\documentclass[preprint,3p,11pt]{elsarticle}
\usepackage{amssymb}

\usepackage[caption=false]{subfig}% Support for small, `sub' figures and tables

\usepackage{microtype}
\usepackage{dcolumn}  % Align table columns on decimal point
\usepackage{color}
\usepackage{bm}       % Bold math
\usepackage{siunitx}    % Format data
\usepackage{booktabs}   % Better delimitations in tables
\usepackage{makecell}   % Modify cells in tables
\usepackage{hyperref}
\usepackage[capitalize]{cleveref}   % Better references
\usepackage{array}
\usepackage{threeparttable}
\usepackage{mleftright}
\usepackage{balance}
\usepackage{textgreek}
\usepackage{longtable}
\usepackage{doi}

\newcommand{\ccite}[1]{\textcolor{blue}{\textbf{\cite{#1}}}}
\newcommand{\aw}{a_{\text{\tiny W}}}		% speed dependence of broadening parameter (math mode required)
\newcommand{\del}{.\kern-\nulldelimiterspace}

\biboptions{numbers,sort&compress}

\journal{J Quant Spectrosc Radiat Transf}

\begin{document}

\LTcapwidth=\textwidth

\begin{frontmatter}

%% Title, authors and addresses

%% use the tnoteref command within \title for footnotes;
%% use the tnotetext command for theassociated footnote;
%% use the fnref command within \author or \address for footnotes;
%% use the fntext command for theassociated footnote;
%% use the corref command within \author for corresponding author footnotes;
%% use the cortext command for theassociated footnote;
%% use the ead command for the email address,
%% and the form \ead[url] for the home page:
%% \title{Title\tnoteref{label1}}
%% \tnotetext[label1]{}
%% \author{Name\corref{cor1}\fnref{label2}}
%% \ead{email address}
%% \ead[url]{home page}
%% \fntext[label2]{}
%% \cortext[cor1]{}
%% \affiliation{organization={},
%%             addressline={},
%%             city={},
%%             postcode={},
%%             state={},
%%             country={}}
%% \fntext[label3]{}

\title{\texorpdfstring{CO$_2$ collision-induced line parameters for the $\nu_3$ band of $^{12}$CH$_4$ measured using a hard-collision speed-dependent line shape and the relaxation matrix formalism}{CO2 collision-induced line parameters for the nu 3 band of 12CH4 measured using a hard-collision speed-dependent line shape and the relaxation matrix formalism}}

%% use optional labels to link authors explicitly to addresses:
%% \author[label1,label2]{}
%% \affiliation[label1]{organization={},
%%             addressline={},
%%             city={},
%%             postcode={},
%%             state={},
%%             country={}}
%%
%% \affiliation[label2]{organization={},
%%             addressline={},
%%             city={},
%%             postcode={},
%%             state={},
%%             country={}}

\author[ulb]{T. Bertin}
\author[ulb]{J. Vander Auwera\texorpdfstring{\corref{cor1}\fnref{jvda}}{}}
\ead{jean.vander.auwera@ulb.be}

\cortext[cor1]{Corresponding author}
\fntext[jvda]{Senior research associate with the F.R.S.-FNRS, Belgium}

\address[ulb]{Spectroscopy, Quantum Chemistry and Atmospheric Remote Sensing (SQUARES), C.P. 160/09, Universit\'e Libre de Bruxelles, 50 avenue F.D. Roosevelt, B-1050 Brussels, Belgium}

%\affiliation[squares]{organization={Spectroscopy, Quantum Chemistry and Atmospheric Remote Sensing (SQUARES), Universit\'e libre de Bruxelles}, addressline={C.P. 160/09, 50 avenue F.D. Roosevelt}, city={Brussels}, postcode={B-1050}, state={}, country={Belgium}}

\def\theadfont{}
%------------------------------------------------------------------------------

\begin{abstract}

Ten high resolution Fourier transform spectra of the pentad region near 3.3 μm of methane diluted in carbon dioxide at total pressures up to 800 hPa have been recorded at $296.5 \, (5)$ K. Including a high resolution spectrum of pure methane at low pressure, these spectra have been analyzed using multi-spectrum fitting techniques. The methane lines were modeled using hard-collision speed-dependent line profiles and line mixing was included in the strongest absorption regions, considering the first order Rosenkranz approximation and the relaxation matrix formalism. CO$_2$ broadening and shift coefficients have been measured, together with the speed dependence of broadening. Results obtained using the two line mixing models are intercompared and compared with previous work.

\end{abstract}

%------------------------------------------------------------------------------

\begin{keyword}

%% keywords here, in the form: keyword \sep keyword
%% PACS codes here, in the form: \PACS code \sep code
%% MSC codes here, in the form: \MSC code \sep code
%% or \MSC[2008] code \sep code (2000 is the default)

High resolution infrared spectroscopy \sep CO$_2$ broadening and shift coefficients \sep Speed dependence of broadening \sep First order line mixing \sep Relaxation matrix formalism

\end{keyword}

\end{frontmatter}

%------------------------------------------------------------------------------

\section{Introduction}\label{sc:introduction}
\renewcommand{\color}[1]{}

%The fairly common presence of methane in the universe \ccite{swain_presence_2008, sromovsky_comparison_2012,  janson_direct_2013, guzman_marmolejo_methane_2015, encrenaz_invited_2022} makes it a critical molecule in chemical \ccite{wogan_abundant_2020} and physical process of stellar and planetary atmospheres. Being an important contributor to the temperature, it has been studied under several conditions \ccite{kondratyev_atmospheric_1985, pavlov_greenhouse_2000, wordsworth_climate_2016a, ramirez_methane_2018}

Methane is rather ubiquitous in planetary atmospheres. In addition to being a strong greenhouse gas on the Earth \ccite{can21a}, it is indeed observed in the atmospheres of the giant planets of the Solar system (\ccite{sin20a, san22a} for example), Titan \ccite{din19a} and exoplanets \ccite{swa08a, mac15a, swa21a}. Although its presence in the CO$_2$ rich atmosphere of Mars is unclear \ccite{gre22a}, the possible contribution of CH$_4$-CO$_2$ collision induced absorption (CIA) to global warming of the atmosphere of early Mars \ccite{wor17a} motivated recent spectroscopic investigations \ccite{tur19a, tra22a}. In particular, the work of Tran \textit{et al.} \ccite{tra22a} was focused on the modeling of absorption spectra of the 3.3 μm region of CH$_4$ diluted in CO$_2$ at total pressures from 3 to 25 bars, recorded using Fourier transform spectroscopy. Following up on that contribution, the present work aimed to characterize the effects of sub-atmospheric CO$_2$ pressures on the shape of lines of the $\nu_3$ band of $^{12}$CH$_4$, observed in that spectral range.

% The Martian atmosphere is constituted of 96\% of carbon dioxide, thus the accuracy in both of these fields relies heavily on the availability of spectroscopic parameters recorded in these conditions. Notably, the lack of information has lead to the use of parameters induced by the pressure of other species \ccite{wordsworth_transient_2017} which have been shown to not be adequate.

An extensive review of the literature reporting measurements of line shape parameters for the $\nu_3$ band of methane can be found in \ccite{ess21a}. Spectroscopic studies dealing with the CH$_4$+CO$_2$ system are rather scarce \ccite{gha05a, fis14a, lyu14a, ess14a, man17a, kor18a, vis19a, ess21a, tra22a}. They are summarized in Table \ref{table:ow}. Using a distributed feedback diode laser spectrometer, Gharavi and Buckley \ccite{gha05a} measured CO$_2$ (as well as self, N$_2$, CO and H$_2$O) pressure broadening coefficients of the R(3) and R(4) manifolds of the $2\nu_3$ band at 296 K and higher temperatures. The measurements were carried out using a Voigt profile, the Lorentz halfwidth of all the components of each manifold being set the same. Also using a Voigt line shape and applying a multispectrum fitting procedure to 12 high resolution Fourier transform spectra recorded at 296 K (3 spectra of pure CH$_4$ and 9 spectra of CH$_4$+CO$_2$ mixtures at total pressures from 22 to 759 hPa), Lyulin \textit{et al.} \ccite{lyu14a} measured CO$_2$ broadening and shift coefficients for \textcolor{red}{respectively} 533 and 386 CH$_4$ lines belonging to several bands observed in the $5550-6140$ cm$^{-1}$ spectral range. In addition to using the Voigt line shape, other studies considered finer effects such as Dicke narrowing \ccite{dic53a}. In particular, Fissiaux \textit{et al.} \ccite{fis14a} compared CO$_2$ broadening coefficients measured using the Voigt, Rautian and Galatry profiles for 28 lines of the $\nu_4$ band of $^{12}$CH$_4$ observed in tunable diode laser spectra recorded at 4 total pressures between 8 and 50 hPa. Vispoel \textit{et al.} \ccite{vis19a} followed a similar approach for 10 lines of the $\nu_3$ band of $^{12}$CH$_4$, relying on spectra recorded at 4 total pressures between 20 and 60 hPa. Using the Voigt and Galatry line shapes, Es-sebbar and Farooq \ccite{ess14a} measured the intensities and CO$_2$ (as well as N$_2$, O$_2$, H$_2$, He and Ar) broadening and narrowing coefficients for 7 lines of the P(11) manifold of the $\nu_3$ band of $^{12}$CH$_4$ observed in 10 spectra recorded at $297\,(1)$ K and total pressures up to about 100 torr using a difference frequency laser system. The CO$_2$ broadening coefficients are unfortunately only provided in a figure. Es-sebbar and Farooq extended these measurements to the P(5), P(7), P(9), P(10), P(12) and P(13) manifolds of the same band, relying on 8 spectra of each manifold recorded at total pressures between 10 and 201 torr \ccite{ess21a}. The broadening coefficients and average narrowing coefficients are reported. \textcolor{red}{Manne \textit{et al.} \ccite{man17a} used a distributed feedback laser to record 8 spectra of the R(3) manifold of $^{12}$CH$_4$ (3 lines) and 9 spectra of the R(4) manifold of $^{13}$CH$_4$ (4 lines) perturbed by CO$_2$ (as well as air and He) at total pressures up to about 150 and 200 hPa, respectively. The Voigt, Rautian, Galatry and speed dependent Voigt profiles, not including or including first order line mixing \ccite{ros75a} with the latter, were used to measure line shape parameters. Most of these are reported in figures, numerical values being only provided for the air, He and CO$_2$ broadening coefficients, their relative speed dependence and the first order line mixing parameters measured with the speed dependent Voigt profile with line mixing.} As already mentioned above, the work of Tran \textit{et al.} \ccite{tra22a} was focused on the modeling of absorption spectra of the 3.3 μm region of CH$_4$ diluted in CO$_2$ at total pressures from 3 to 25 bars, relying on line shape parameters available in the \texttt{HITRAN} database \ccite{gor20a}. To develop a method to determine the concentration of methane during oxy-methane combustion in a mixture of CO$_2$, O$_2$ and Ar, Koroglu \textit{et al.} \ccite{kor18a} measured absorption cross sections of methane mixed with each of these three gases at 296 K and atmospheric pressure in the $3.402-3.405$ μm range using FTIR, and at two wavelengths in the P(8) manifold at temperatures from 700 to 2000 K and pressures from 0.1 to 1.5 atm using a shock tube and a distributed feedback inter-band cascade laser.

\begin{table}[!htb]
\small
\renewcommand{\arraystretch}{1.2}
\centering
\caption{Previous measurements of CO$_2$ collision induced parameters (``Par.'') reported for methane in the literature.} \vspace{12pt}
\label{table:ow}
\begin{tabular}{>{\raggedright\arraybackslash}p{5cm}cccl>{\raggedright\arraybackslash}p{4cm}}
\toprule
Reference & Band & Range / cm$^{-1}$ & \# lines & Par.$^{\dag}$ & Line profile $^{\ddag}$ \\
\midrule
Gharavi \& Buckley (2005) \ccite{gha05a} & $2\nu _3$ & $6046-6058$ & 7 & $b_L^0$ & Voigt \\
Lyulin \textit{et al.} (2014) \ccite{lyu14a} & $2 \nu _3$ & $5550-6140$ & 533 & $b_L^0$, $\delta^0$ & Voigt \\
Fissiaux \textit{et al.} (2014) \ccite{fis14a} & $\nu _4$ & $1241-1369$ & 28 & $b_L^0$ & Voigt, Rautian \& Galatry \\
Vispoel \textit{et al.} (2019) \ccite{vis19a} & $\nu _3$ & $2906-2959$ & 11 & $b_L^0$ & Voigt, Rautian \& Galatry \\
Es-sebbar \& Farooq (2014) \ccite{ess14a} & $\nu _3$ & $2905-2908$ & 9 & $b_L^0$, $\beta^0$ & Voigt \& Galatry \\
Manne \textit{et al.} (2017) \ccite{man17a} & $\nu _3$ & $3057-3058$ & 3 & $b_L^0$, $\aw$ & qsdVoigt with first-order line mixing \\
Es-sebbar \& Farooq (2021) \ccite{ess21a} & $\nu _3$ & $2884-2969$ & 49 & $b_L^0$, $\beta^0$ & Voigt \& Galatry \\
\bottomrule
\\[-6pt]
\multicolumn{6}{l}{$^{\dag}$ $b_L^0$, $\delta^0$, $\beta^0$ and $\aw$ are the broadening, shift, narrowing and speed dependence of broadening coefficients.} \\
\multicolumn{6}{l}{$^{\ddag}$ ``qsd'' stands for ``quadratic speed dependent.''}
\end{tabular}
\end{table}

% Manne \textit{et al.} \ccite{man17a} used a distributed feedback laser to record 8 spectra of the R(3) manifold of $^{12}$CH$_4$ (3 lines) and 9 spectra of the R(4) manifold of $^{13}$CH$_4$ (4 lines) perturbed by CO$_2$ (as well as air and He) at total pressures up to about 150 and 200 hPa, respectively. The line shape parameters measured using the speed dependent Voigt line shape with first order line mixing \ccite{ros75a} are reported.

% The Voigt, Rautian, Galatry and speed dependent Voigt profiles, not including or including first order line mixing \ccite{ros75a} with the latter, were used to retrieve the line shape parameters from spectra recorded at 8 total pressures up to about 150 hPa ($^{12}$CH$_4$) and spectra obtained at 9 total pressures up to about 200 hPa ($^{13}$CH$_4$). Most of the measured the spectroscopic parameters using the speed dependent Voigt line shape with line mixing are reported.

Modeling the effects of higher pressures on the infrared spectrum of methane is challenging, going beyond studies of finer pressure effects such as those referred to here above. As a result of the tetrahedral symmetry of the molecule, methane absorption bands indeed involve manifolds of closely spaced spectral lines that interact by means of inelastic collisions at higher pressures \ccite{bar58a}. This line mixing is frequently modeled using the first order approximation \ccite{ros75a}. However, this ``weak coupling'' approximation is not appropriate for pressures approaching 1 atm \ccite{far21a}, therefore calling for the more appropriate solution relying on the relaxation matrix formalism \ccite{fan63a, lev92a}.

In this context, the present work aims to provide a larger set of parameters characterizing the effects of the pressure of CO$_2$ on the $3.3$ μm region of the spectrum of methane. The present measurements were performed considering hard collision Dicke narrowing and speed dependence of the broadening, both of which being included in the relaxation matrix and first order models developped by Ciury\l{}o and Pine \ccite{ciu00a}. Comparisons of the first order approximation with the relaxation matrix formalism are presented, highlighting the limitations of each model and the differences between the retrieved parameters. Comparisons of the present measurements with previous work are presented as well when possible.

%------------------------------------------------------------------------------

\section{Experimental details}\label{sc:experiments}

% Recording of the CH4+CO2 spectra: 25/02/2020 (jvda2603) -> 11/03/2020 (jvda2632)

Unapodized high resolution spectra of pure methane (Fluka, purity $\geq 99.0$ \%) and methane diluted in carbon dioxide (Aldrich, purity $\geq 99.8$ \%) have been recorded using a Bruker IFS 120 to 125 HR upgrade Fourier transform spectrometer (\texttt{FTS}). The instrument was equipped with a tungsten source, a 1.15 mm entrance aperture diameter, a KBr beamsplitter, a band pass filter limiting the radiation reaching the detector to the $2550-3200$ cm$^{-1}$ range and an InSb detector cooled down to 77 K. The sample was contained in a double jacketed $19.7 \pm 0.2$ cm long stainless steel cell, closed by CaF$_2$ windows and located inside the evacuated spectrometer. The temperature of the cell was stabilized at $296.5 \pm 0.5$ K using a ThermoHaake DC50/K20 thermostat and water as heat exchanger. It was measured using two TSiC 301 sensors (IST Innovative Sensor Technology; stated accuracy of $\pm 0.3$ K in the $10-90$ $^{\circ}$C range) fixed on the outer cell wall. The methane and carbon dioxide pressures were measured using MKS Baratron manometers model 690A of 10 and 1000 Torr full scale ranges, respectively. Their accuracy of reading are conservatively estimated to be equal to 0.5 \%. The total pressures and initial mole fractions of methane are listed in Table \ref{tab:exp_details}.
\begin{table}[!htb]
\caption{Total pressures $P_{tot}$, initial mole fractions of methane $x$ in the CO$_2$ mixtures [$x = P(\mathrm{CH}_4)/P_{tot}$ where $P(\mathrm{CH}_4)$ is the methane pressure measured just after filling the cell] and number of interferograms co-added ($n$). The numbers between parentheses are the uncertainties on the total pressures given in the units of the last quoted digits, estimated as the sum in quadrature of half of the peak-to-peak variations of the pressures measured during the recording of the interferograms and the 0.5 \% accuracy of reading of the pressure gauge. The uncertainty estimated for spectrum S1 is set to 3 \% of the measured pressure to conservatively account for the fact that the measured low pressure is outside the calibrated range of the gauge \ccite{cac22a}.}
\label{tab:exp_details}
\centering
\vspace{1 cm}
\begin{tabular}{lD{.}{.}{5.7}D{.}{.}{6.5}c}
\toprule
\\
\multicolumn{1}{l}{\#} & \multicolumn{1}{c}{$P_{tot}$ / hPa} & \multicolumn{1}{c}{$x \times 10^3$} & \multicolumn{1}{c}{$n$} \\
\\
\midrule
\\
% The uncertainties associated with the variations of the pressures during the recording of the interferograms are listed in "sub-atmospheric/exp_data/pressure_temperature.pxp"
% Recording of the CH4+CO2 spectra: 25/02/2020 (jvda2603) -> 11/03/2020 (jvda2632)
S1  &  0.407\,(12) & 1000.  & 320 \\ % jvda2614_z4_tr: uncertainty = 3% of 0.4071 hPa
S2  &  25.99\,(13) &  9.98  & 204 \\ % jvda2603_z4_tr (MOPD = 300 cm) -> jvda2638_z4_tr (MOPD = 150 cm)
S3  &  51.18\,(26) &  7.34  & 320 \\ % jvda2605_z4_tr
S4  & 100.34\,(50) &  5.11  & 218 \\ % jvda2608_z4_tr
S5  & 201.2\,(1.0) &  5.00  & 320 \\ % jvda2611_z4_tr
S6  & 400.3\,(2.0) &  5.04  & 320 \\ % jvda2617_z4_tr
S7  & 400.6\,(2.0) &  3.81  & 320 \\ % jvda2620_z4_tr
S8  & 601.1\,(3.0) &  2.17  & 320 \\ % jvda2623_z4_tr
S9  & 601.7\,(3.0) &  3.06  & 320 \\ % jvda2626_z4_tr
S10 & 801.7\,(4.0) &  2.20  & 320 \\ % jvda2632_z4_tr
S11 & 802.8\,(4.0) &  2.83  & 320 \\ % jvda2629_z4_tr
\\
\bottomrule
\end{tabular}
\end{table}
All the interferograms were recorded with a maximum optical path difference (\texttt{MOPD}) of 150 cm, corresponding to an approximate spectral resolution (defined as 0.9/\texttt{MOPD}) of 0.006 cm$^{-1}$. The number of interferograms co-added to yield the spectra are also provided in Table \ref{tab:exp_details}. Transmittance spectra were generated using the average of two low resolution (0.1 cm$^{-1}$) empty cell spectra recorded just before and after the corresponding sample spectrum. They were interpolated 4 times.

% ** Recording of the "ILS" spectra:

% - N2O: jvda2589 & jvda2590 (03-05/12/2019) - 2.7 um filter (2150-3500 cm-1), MOPD = 300 cm, iris = 1.00 mm, path = 698 cm
% - CH4: jvda2593 & jvda2595 (09-11/12/2019) - I0093 filter (2550-3200 cm-1), MOPD = 300 cm, iris = 1.00 mm, path = 698 cm
% * STORM: Air conditioning system down from 19/02/2020 onward
% - CH4: jvda2597 (21-22/02/2020) - I0093 filter (2550-3200 cm-1), MOPD = 150 cm, iris = 1.15 mm, path = 19.7 cm
% - CH4: jvda2601 (24-22/02/2020) - I0093 filter (2550-3200 cm-1), MOPD = 300 cm, iris = 1.15 mm, path = 19.7 cm

% ** N2O (3-4/12/2019) - From 'fd_ils.tex' Uncertainty on P = 0.004 hPa and on T = 1.5 K
%
% filename            x_min   x_max   L/cm   T      P/hPa  vmr  iso typ mopd/cm apd iris  foc_len  diam
% jvda2589_z32_sb1.1  2520.0  2600.0  698.0  295.5  0.019  1.0  1.0  T  300.000  1  1.00   418.0   78.0
% jvda2589_z32_sb2.1  3430.0  3510.0  698.0  295.5  0.019  1.0  1.0  T  300.000  1  1.00   418.0   78.0
% jvda2590_z32_sb1.1  2520.0  2600.0  698.0  295.5  0.028  1.0  1.0  T  300.000  1  1.00   418.0   78.0
% jvda2590_z32_sb2.1  3430.0  3510.0  698.0  295.5  0.028  1.0  1.0  T  300.000  1  1.00   418.0   78.0
%
% Temperature was not measured

\begin{figure}[!htb]
\centering
\includegraphics[width = \textwidth]{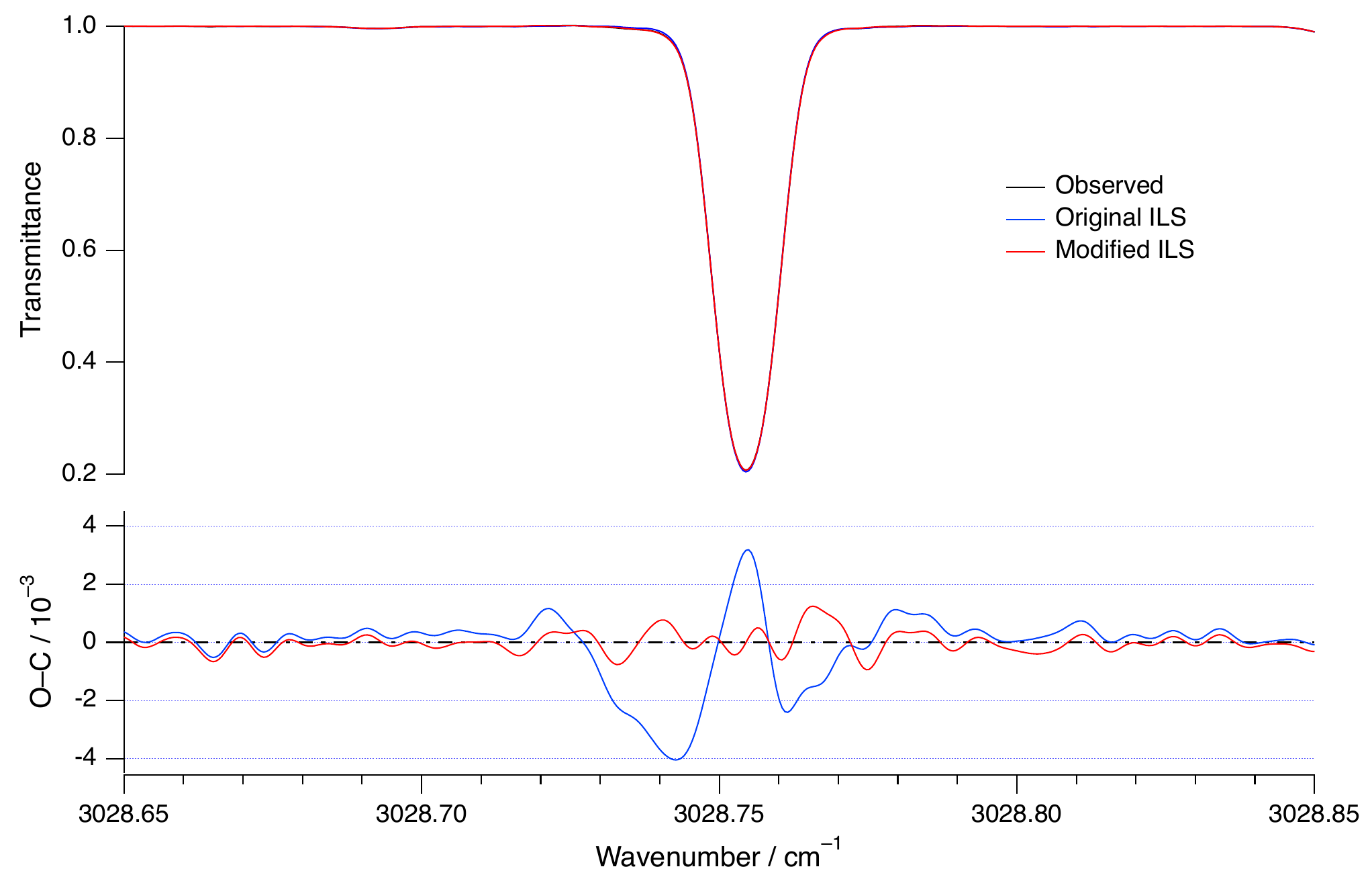}
\caption{Least squares fit of the R(0, $A_1$, 1, 3) line of the $\nu_3$ band of $^{12}$CH$_4$ (spectrum S1 in Table \ref{tab:exp_details}, 296.5 K and 19.7 cm) with the \texttt{ILS} determined using N$_2$O (blue trace) and modified relying on the R(0, $A_1$, 1, 3) and R(1, $F_1$, 1, 10) lines observed in spectrum S1 (red trace).}
\label{fig:ils}
\end{figure}

Knowledge of the instrument line shape (\texttt{ILS}) characterizing the alignment of the \texttt{FTS} during the recording of the CH$_4$+CO$_2$ spectra is required to analyze them. The \texttt{ILS} was determined using a method described in another contribution \ccite{ber23a}. Following other approaches \ccite{has99a, boo19a}, it relies on high resolution spectra of molecules such as N$_2$O or OCS recorded at low pressure in the same instrumental conditions as the sample under study to determine the evolution with optical path difference of a so-called modulation function ($\eta$ in Eq. 1 of \ccite{boo19a}) from which the \texttt{ILS} in the spectral domain can be generated. A N$_2$O spectrum was therefore recorded for that purpose. Unfortunately, a storm broke down the compressor of the air conditioning system of the laboratory (installed on the roof of the building) after the recording of the N$_2$O spectrum and before the start of the measurements on the CH$_4+$CO$_2$ system. As its repairing turned out to require several months to complete, the experiments on CH$_4+$CO$_2$ were conducted without air conditioning. The resulting variations of temperature in the laboratory of about 1.5 K (peak to peak) during the recording of the interferograms did not affect the gaseous sample because the temperature of the cell was stabilized using a thermostat (see above), but had consequences on the alignment of the spectrometer. As a result, the \texttt{ILS} determined using N$_2$O did not correctly model the actual alignment of the \texttt{FTS}, as is shown in Fig. \ref{fig:ils}. The modulation function determined using N$_2$O lines was therefore modified to improve the modeling of the shapes of the isolated R(0, $A_1$, 1, 3) and R(1, $F_1$, 1, 10) lines of the $\nu_3$ band of $^{12}$CH$_4$ observed in spectrum S1 (see Table \ref{tab:exp_details}). These lines are centered near 3028.752 and 3038.499 cm$^{-1}$, respectively. As done in our previous contribution on the $\nu_3$ band of $^{12}$CH$_4$ \ccite{far21a}, the methane lines are identified with $\Delta J (J'',C'',\alpha'', \alpha')$ where $J''$ is the rotational quantum number of the lower level, $C'' = A_1$, $A_2$, $E$, $F_1$ or $F_2$ is its rovibrational symmetry in the molecular symmetry group T$_{\mathrm{d}}$(M) \ccite{bunjen06} and $\alpha$ is a label used to fully identify rotational levels within vibrational polyads \ccite{broetal92a} ($\alpha''$ and $\alpha'$ are associated with the lower and upper levels of the transition, respectively). Although the modeling of these lines obtained using the modified modulation function \textcolor{red}{is} satisfactory, as also shown for R(0, $A_1$, 1, 3) in Fig. \ref{fig:ils}, it is most probably not the best estimate of the actual \texttt{ILS} of the spectrometer. % \textcolor{red}{(see below)}

% R(0) in spectrum S1: Doppler FWHM = 9.3E-3 cm-1, Self broadening (HITRAN) = 0.081 cm-1/atm -> Lor FWHM = 6.6E-5 cm-1

% USELESS as the ILS determined using N2O was not used. The instrument line shape (\texttt{ILS}) of the Bruker IFS 120 to 125 HR upgrade \texttt{FTS} was determined using a method describe in another contribution \ccite{ber23a}. It relied on a high resolution ($\delta_{max} = 300$ cm) spectrum of the \textcolor{red}{$2\nu_1$ (near 2560 cm$^{-1}$) or $\nu_1+\nu_3$ (near 3480 cm$^{-1}$) band} of $^{14}$N$_2^{16}$O (Matheson, unknown purity) recorded 2.5 months before the experiments carried out on the CH$_4$+CO$_2$ system. The instrument was fitted with a tungsten source, a 1 mm entrance aperture diameter, a CaF$_2$ beamsplitter, an optical low pass filter with a cut-off at 3600 cm$^{-1}$ and an InSb detector cooled down to 77 K. The sample was contained in a Pyrex White-type cell \ccite{whi42a} set to provide an absorption path length of $698\,(2)$ cm, closed by CaF$_2$ windows and linked under vacuum to the spectrometer. The spectrum was recorded at room temperature [$295.5 \, (1.5)$ K], stabilized by an air conditioning system. The sample pressure, measured using a MKS Baratron gauge model 390 HA of 10 Torr full scale, was smaller than 2 Pa. The spectrum resulted from the co-addition of 534 interferograms and was interpolated 32 times.

%------------------------------------------------------------------------------

\section{Modeling the absorption spectra}\label{sc:modeling_spectra}

Each of the 11 observed spectra were modeled as the convolution of the \texttt{ILS} with a molecular transmittance $\tau(\tilde{\nu})$ ($\tilde{\nu}$ is the wavenumber, in cm$^{-1}$). The molecular transmittance was generated using the following expression
\begin{equation}\label{eq:molecular_transmittance}
\tau(\tilde{\nu }) = b(\tilde{\nu}) \, \exp \left\{- N \ell \, [8\pi^3 / (3hc)] \, \tilde{\nu} \, \left[1 - \exp \{- hc \tilde{\nu} / (k_B T)\}\right] F(\tilde{\nu }) \right\}
\end{equation}
where $b(\tilde{\nu})$ is the baseline represented by a polynomial expansion, $N$ is the density of absorbing molecules (in molecule/cm$^{3}$), $h$ is the Planck constant, $c$ is the speed of light in vacuum, $k_B$ is the Boltzmann constant, $T$ is the temperature (in K), $\ell$ is the optical path length (in cm) and $F(\tilde{\nu})$ is the ``spectral distribution.''

\begin{figure}[!htb]
\centering
\subfloat[][]{\includegraphics[width = 0.5\linewidth]{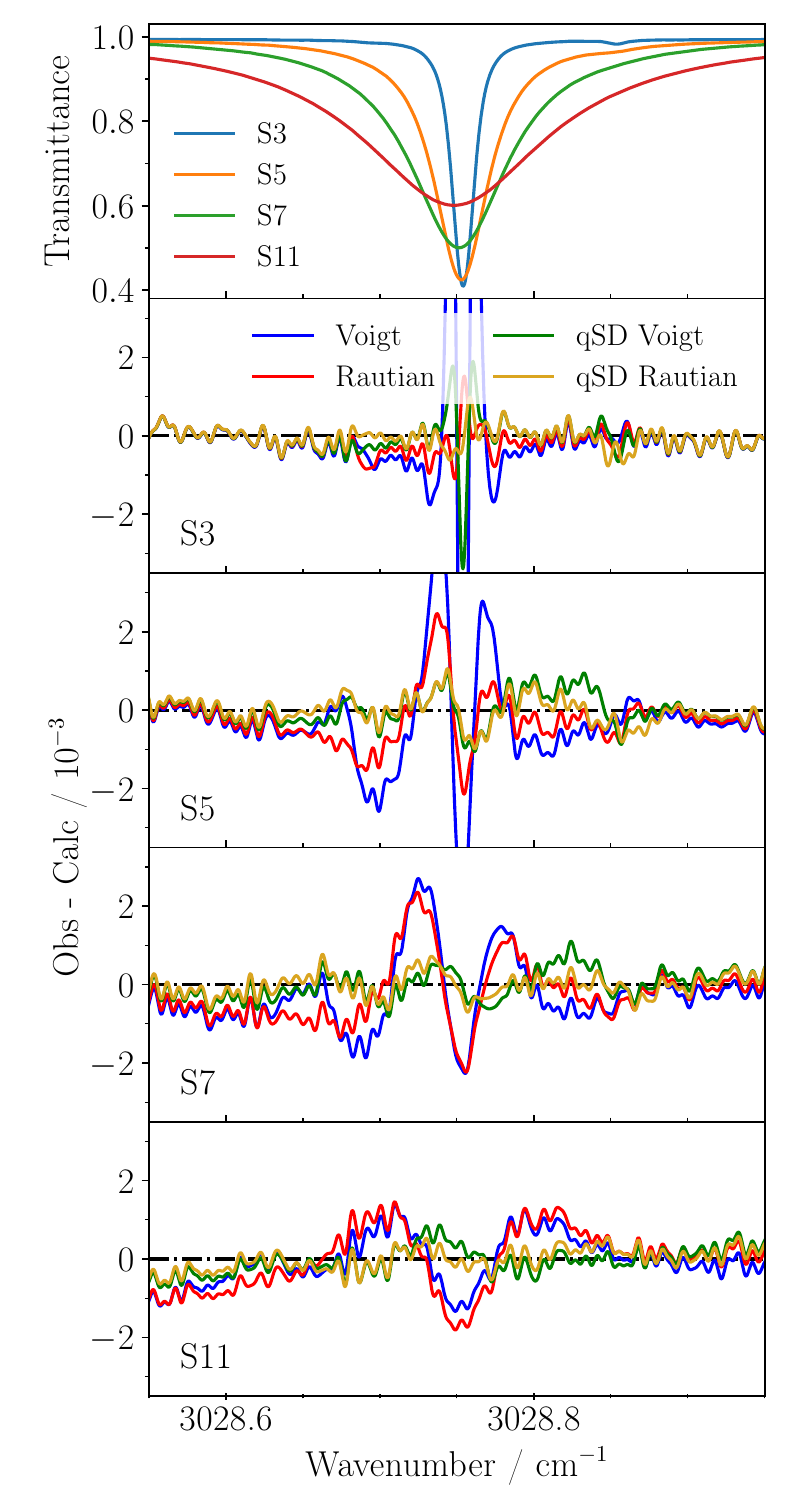}\label{fig:prof:r0}}
\subfloat[][]{\includegraphics[width = 0.5\linewidth]{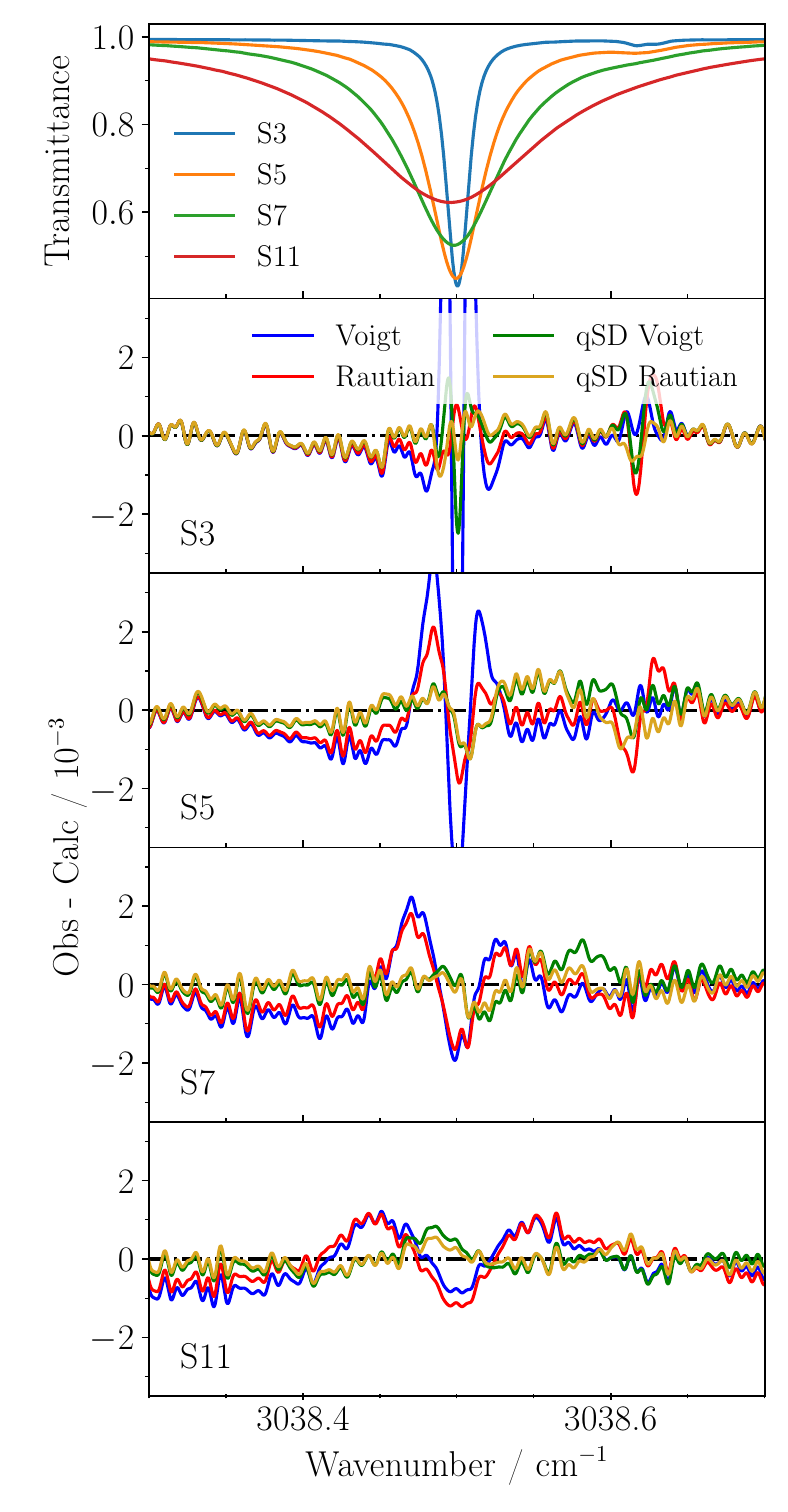}\label{fig:prof:r1}}
\caption{Results of the fit of the R(0, $A_1$, 1, 3) (left) and R(1, $F_1$, 1, 10) (right) lines of the $\nu_3$ band of $^{12}$CH$_4$ perturbed by CO$_2$ using 4 different theoretical line shape models (``qSD'' stands for ``quadratic speed dependent''). Four of the 11 transmittance spectra measured are presented in the top row (296.5 K and 19.7 cm; the pressures and mole fractions are provided in Table \ref{tab:exp_details}). The best-fit residuals obtained for the 4 spectra of the two lines are presented in the lower 4 rows, the color of each trace being associated with a specific line shape model.}
\label{fig:prof}
\end{figure}

In the absence of line mixing, the contributions of the various transitions to the spectral distribution is given by \ccite{har21a}
\begin{equation}\label{eq:isolated_line_spectral_distribution}
F(\tilde{\nu}) = \sum_n \frac{1}{4\pi \varepsilon_0} \, \mu_n^2 \, \rho_n \, g(\tilde{\nu} - \tilde{\nu}_n)
\end{equation}
In this expression, $\varepsilon_0$ is the electric constant and the index $n$ represents a rovibrational transition corresponding to a spectral line, the summation running over all the transitions leading to lines observed in the considered spectral range. For each transition $n$, $\rho_n$ is the relative population of the lower level, $\mu_n$ is the electric dipole transition moment (in D) and $g(\tilde{\nu } - \tilde{\nu}_n)$ is the line shape function centered around the wavenumber $\tilde{\nu}_n = E'_n - E_n''$ where $E'_n$ and $E_n''$ are the energies of the upper and lower levels, respectively. The relative population of the lower level is expressed as % \ccite{}
\begin{equation}\label{eq:rel_pop}
\rho_n = g_n'' \, \frac{\exp \{- hc E_n'' / (k_B T)\}}{Q(T)}
\end{equation}
where $g_n''$ is the degeneracy of the lower level of the transition $n$ and $Q(T)$ is the total internal partition sum. The latter was calculated to be equal to $592.07$ and $1184.2$ at 296.5 K for $^{12}$CH$_4$ and $^{13}$CH$_4$, respectively \ccite{gam21a}. The electric dipole transition moment associated with a given transition $n$ was obtained from the integrated absorption cross section $\sigma_n$ of the corresponding line using the following relation % \ccite{}
\begin{equation}\label{eq:transition_moment} % Lines 823-825 of lm_calc_co2_2015.for
\mu_n^2 = \frac{\sigma_n}{A \, \tilde{\nu}_n \, \rho_n \, \left[1 - \exp \{- hc \tilde{\nu}_n / (k_B T)\}\right]}
\end{equation}
where $A = [8\pi^3 / (3hc)] \times [1/(4\pi \varepsilon_0)] \approx 4.162 \times 10^{-19}$ D$^{-2}$cm$^2$. In the present work, the line shape function used to model the profile of the $^{12}$CH$_4$ lines was chosen to be the hard collision speed dependent profile introduced by Lance \textit{et al.} \ccite{lan97a} because it minimized the residuals obtained for the isolated R(0, $A_1$, 1, 3) and R(1, $F_1$, 1, 10) lines, as shown in Fig. \ref{fig:prof}.
This observation is partially supported by previous work in which narrowing \ccite{ess14a, vis19a, ess21a}, speed dependence \ccite{man17a} or both effects \ccite{pin19a} were included for CO$_2$ or other buffer gases. Hard collisions were considered because CO$_2$ is almost three times as heavy as methane. The hard collision speed dependent profile was computed in its quadratic speed dependence approximation \ccite{roh07a} using the \texttt{FORTRAN} routine ``\texttt{qSDHC.for}'' \ccite{tra13a}. Although it has more physical foundation, the hypergeometric speed dependence model was not considered because it did not improve the residuals but significantly increased computation time. The correlation between velocity- and phase-changing collisions was neglected for the same reason. The three other line shape models considered in Fig. \ref{fig:prof} were computed as follows. The quadratic speed dependent Voigt line shape was calculated using the routine ``\texttt{qSDV.for}'' from Tran \textit{et al.} \ccite{tra13a}, while the Voigt and Rautian profiles were computed using the complex probability (or error) function (see Eq. 4 of \ccite{far21a} for example) calculated using the \texttt{weideman40a()} function provided in the ``\texttt{Py4CAtS}'' package \ccite{sch19a}. The shape of the weaker $^{13}$CH$_4$ lines was modeled with a Voigt profile or a quadratic speed dependent Voigt profile with fixed speed dependence of the broadening (see below). 

In addition to its position $\tilde{\nu}_n$, the hard collision speed dependent profile used to describe the shape of each $^{12}$CH$_4$ line $n$ involves 3 parameters. They are the speed dependent line broadening $\Gamma_n(v)$ ($v$ is the absolute speed of the absorbing molecule), the line shift $\Delta_n$ and the line narrowing $\beta_n$. These 3 parameters are in cm$^{-1}$. Their dependence with pressure, including the speed dependence of the line broadening expressed by the coefficient $\aw$ \ccite{pin19a, lis10a}, was modeled according to:
\begin{eqnarray}
\Gamma_n(v) &=& P_{tot} \left[ b_L^0(\mathrm{CO}_2) (1 - x) + b_L^0(\mathrm{self})\, x \right] \times \left[ 1 + \aw \, \left( \frac{v^2}{v_p^2} - \frac{3}{2}\right) \right] \label{eq:line_broadening} \\
\Delta_n &=& P_{tot} \left[ \delta^0(\mathrm{CO}_2) (1 - x) + \delta^0(\mathrm{self})\, x \right] \label{eq:line_shift} \\[6 pt]
\beta_n &=& \beta^0 \, P_{tot} \label{eq:line_narrowing}
\end{eqnarray}
where $x = P(\mathrm{CH}_4)/P_{tot}$ is the methane mole fraction, $P_{tot}$ is the total pressure (in atm), $v_p$ is the most probable speed of the methane molecules, and $b_L^0(\mathrm{CO}_2)$, $b_L^0(\mathrm{self})$, $\delta^0(\mathrm{CO}_2)$, $\delta^0(\mathrm{self})$ and $\beta^0$ are the CO$_2$ broadening, self broadening, CO$_2$ shift, self shift and narrowing coefficients (in cm$^{-1}$atm$^{-1}$) of the methane lines, respectively. As it is defined in Eq. \ref{eq:line_broadening}, the speed dependence of the line broadening $\aw$ is unitless and is assumed to be the same for CO$_2$ and self broadening. Narrowing is also assumed to be independent of the broadening gas.

At higher pressures, observed spectra can \textit{a priori} not be modeled by the sum of isolated lines anymore because of the rise of collisional line mixing effects \ccite{bar58a}. The use of Eq. \ref{eq:isolated_line_spectral_distribution} would indeed lead to characteristic signatures in the residuals \ccite{har21a}. Several models of these effects exist \ccite{bar58a, kol58a, fan63a, ben66a, gor66a, gor71a, ros75a, lam77a, smi81a, ciu00a}. Within the Rosenkranz first order approximation \ccite{ros75a}, Eq. \ref{eq:isolated_line_spectral_distribution} can still be used to model observed spectra with the line shape function involved given by \ccite{gen87a, pin97a, pin97b}
\begin{equation}\label{eq:line-mixing_rosenkranz}
g_{lm}^n(\tilde{\nu} - \tilde{\nu}_n) = \mathrm{Re}\{w(\tilde{\nu} - \tilde{\nu}_n)\} + Y_n^0 \, P_{tot} \, \mathrm{Im}\{w(\tilde{\nu} - \tilde{\nu}_n)\}
\end{equation}
where $w(\tilde{\nu} - \tilde{\nu}_n)$ is the complex normalized spectral shape of the isolated line, its real ($\mathrm{Re}\{\}$) and imaginary ($\mathrm{Im}\{\}$) parts being conveniently provided by the routines ``\texttt{qSDHC.for}'' and ``\texttt{qSDV.for}'' \ccite{tra13a} for the quadratic speed dependent hard collision (\textit{i.e.} Rautian) and Voigt profiles, respectively. $Y_n^0$ (in atm$^{-1}$) is the first order line mixing parameter, expressing the coupling of line $n$ with neighboring lines. This approach has drawbacks \ccite{har21a} and is \textit{a priori} limited to weak line coupling and small overlapping. Ciury\l{}o and Pine \ccite{ciu00a} derived a general form of the spectral distribution for hard collision speed dependent profiles affected by line mixing:
\begin{eqnarray}
F (\tilde{\nu }) &=& \frac{1}{\pi} \, \mathrm{Re} \mleft\{ \bm{\mu} ^T \mleft[ \bm{1} - \bm{G}(\tilde {\nu }) \bm{\beta}\mright]^{-1} \bm{G}(\tilde{\nu }) \bm{\rho } \bm{\mu } \mright\} \label{eq:rel_mat} \\
% F (\tilde{\nu }) &=& \frac{1}{\pi } \mathrm{Re} \mleft\{ \bm{\mu} ^T \mleft[ \bm{1} - \bm{G}(\tilde {\nu }) \mleft( \bm{\beta} - \bm{C} \mright)\mright]^{-1} \bm{G}(\tilde{\nu }) \bm{\rho } \bm{\mu } \mright\}, \\
\bm{G}(\tilde{\nu}) &=& \int f_M(\bm{v}) \times \mleft[ \bm{W}(v) + \bm{\beta} - i\tilde{\nu}\bm{1} + i\bm{\tilde{\nu}_0} + \frac{i}{2\pi c}(\bm{k}^T \bm{v})\bm{1} \mright]^{-1} \mathrm{d}^3\bm{v} \nonumber
% \label{eq:general}
\end{eqnarray}
Considering that $N$ lines are involved, $\bm{G}(\tilde{\nu})$ is a $N \times N$ matrix, $\bm{\mu}$ is a $N$-dimensional vector of the electric dipole transition moments, $\bm{1}$ is the $N \times N$ unit matrix and $\bm{\tilde{\nu}_0}$, $\bm{\rho}$ and $\bm{\beta}$ are $N \times N$ diagonal matrices of the line centers, relative populations ($\bm{\rho}_{nn} = \rho_n$ given in Eq. \ref{eq:rel_pop}) and narrowing parameters ($\bm{\beta}_{nn} = \beta_n$ given in Eq. \ref{eq:line_narrowing}), respectively. $\bm{\rho}$ and $\bm{\beta}$ are considered to be speed independent \ccite{rau67a}. $\bm{k}$ is the $N$-dimensional wave vector of amplitude $|\bm{k}| = 2 \pi \tilde{\nu }$, $\bm{v}$ is a $N$-dimensional vector of the absolute velocity of the absorbing molecule and $\bm{W}$ is the $N \times N$ relaxation matrix. Its diagonal elements are equal to $\bm{W}_{nn} = \Gamma _n(v) + i\Delta _n$, where $\Gamma_n(v)$ and $\Delta_n$ are as given in Eqs. \ref{eq:line_broadening} and \ref{eq:line_shift}, respectively. Its off-diagonal elements are $\bm{W}_{mn} = -W_{mn}P_{tot}$, where $W_{mn}$ is the speed independent line mixing parameter that characterizes the coupling between lines $n$ and $m$. The off-diagonal elements are connected by the detailed balance equation, \textit{i.e.} $W_{mn} \, \rho_n = W_{nm} \, \rho_m$ \ccite{har21a}. All the elements of $\bm{W}$ are in cm$^{-1}$. $f_M(\bm{v})$ is the Maxwell-Boltzmann velocity distribution \ccite{ciu00a}. % $\bm{C}$ is the $N \times N$ correlated relaxation matrix of elements characterizing collisions that changes both the velocity and the phase or state of the molecule. It is equal to $0$ in the uncorrelated case, resulting in a slightly simpler form of \cref{eq:general}:
The first-order line-mixing parameters $Y_n^0$ introduced in Eq. \ref{eq:line-mixing_rosenkranz} are related to the off-diagonal elements of the relaxation matrix by \ccite{gen87a}:
\begin{equation}\label{eq:y_from_w}
Y_n^0 = 2 \sum_{m \neq n} \frac{\mu_m}{\mu_n} \, \frac{\bm{W}_{mn}}{\tilde{\nu}_n - \tilde{\nu}_m}
\end{equation}

To analyze the recorded spectra in the first order line mixing approximation (Eqs. \ref{eq:molecular_transmittance}, \ref{eq:isolated_line_spectral_distribution} and \ref{eq:line-mixing_rosenkranz}), the multi-spectrum analysis software \ccite{tud12a, dan14a} already used to study the spectra of the $\nu_3$ band of methane perturbed by air \ccite{far21a} was used. A multi-spectrum analysis software was specifically developped during this work (\url{https://github.com/TBSpectroscopy/rm-fit}) to analyze the recorded spectra using the relaxation matrix formalism (Eqs. \ref{eq:molecular_transmittance} and \ref{eq:rel_mat}). The program implements a Trust Region Reflective least squares fitting algorithm to adjust a synthetic spectrum to each of the 11 observed spectra, relying on the function \texttt{least\_squares()} provided in the \texttt{optimize} package of the \texttt{SciPy} library (\texttt{https://scipy.org}) \ccite{vir20a}. Mostly written in Python, the software relies on the \texttt{FORTRAN} and Python routines mentionned above to generate the line shapes \ccite{tra13a, sch19a} and the \texttt{Eigen} C++ library (\texttt{https://eigen.tuxfamily.org}) to perform the matrix operations.

%------------------------------------------------------------------------------

\section{Analysis}\label{sc:analysis}

% 13/02/2024: Range update to match the results presented, i.e. covering P(13) to R(13) and nothing more. H2O lines included are listed in 'article_supmat/appendix/linelists/d2b_linelists-uncertainties/h2o_included.txt'.

The multispectrum analysis of the measured spectra S1 to S11 was carried out in the range \textcolor{red}{$2880-3150$} cm$^{-1}$, considering lines of $^{12}$CH$_4$ stronger than $10^{-23}$ cm$^{-1}$/(molecule cm$^{-2}$) at 296 K. These lines belong to the $\nu_3$, $\nu_2+\nu_4$, $2\nu_2$ and $\nu_3+\nu_4-\nu_4$ bands. Lines of $^{13}$CH$_4$ were also included when observed. As absorption of water vapor present in the evacuated spectrometer was observed in the spectra, \textcolor{red}{77 lines} of H$_2^{16}$O stronger than $10^{-22}$ cm$^{-1}$/(molecule cm$^{-2}$) at 296 K had to be included in the range \textcolor{red}{$2966 - 3150$} cm$^{-1}$. A Voigt function was used to model the shape of the $^{13}$CH$_4$ and H$_2^{16}$O lines, the parameters of which being taken from the \texttt{HITRAN} database \ccite{gor20a} (downloaded from \texttt{https://hitran.org} in February 2021). The CO$_2$ broadening and shift coefficients of the stronger $^{13}$CH$_4$ lines were fitted when needed. They were otherwise left fixed to their air counterparts.

As is apparent from section \ref{sc:modeling_spectra}, the modeling of each $^{12}$CH$_4$ line required 8 parameters, \textit{i.e.} $\tilde{\nu}_n$, $\sigma_n$ or $\mu_n$, $b_L^0(\mathrm{CO}_2)$, $b_L^0(\mathrm{self})$, $\delta^0(\mathrm{CO}_2)$, $\delta^0(\mathrm{self})$, $\beta^0$ and $\aw$. Inclusion of first order line mixing added the $Y_n^0$ parameter, while use of the relaxation matrix formalism resulted in the addition of speed independent off-diagonal elements $W_{mn}$, the number of which depended upon the number of interacting lines involved in the considered spectral range. Line mixing couples lines of the same $A$, $E$ and $F$ symmetry \ccite{pin97a}. This selection rule was explicitly taken into account in the relaxation matrix formalism, but could not in the first order approximation because the corresponding program did not allow enforcing it \ccite{far21a}. Coupling between the same $1 \leftrightarrow 1$ or $2 \leftrightarrow 2$ symmetries ($A_2 \leftrightarrow A_2$ for example) was considered despite the fact that it is known to be small \ccite{pie99a, pin00a}. In the present work, line mixing was only considered for lines of the $\nu_3$ band of $^{12}$CH$_4$.

\begin{figure}[!htbp]
\centering
\includegraphics[width = \textwidth]{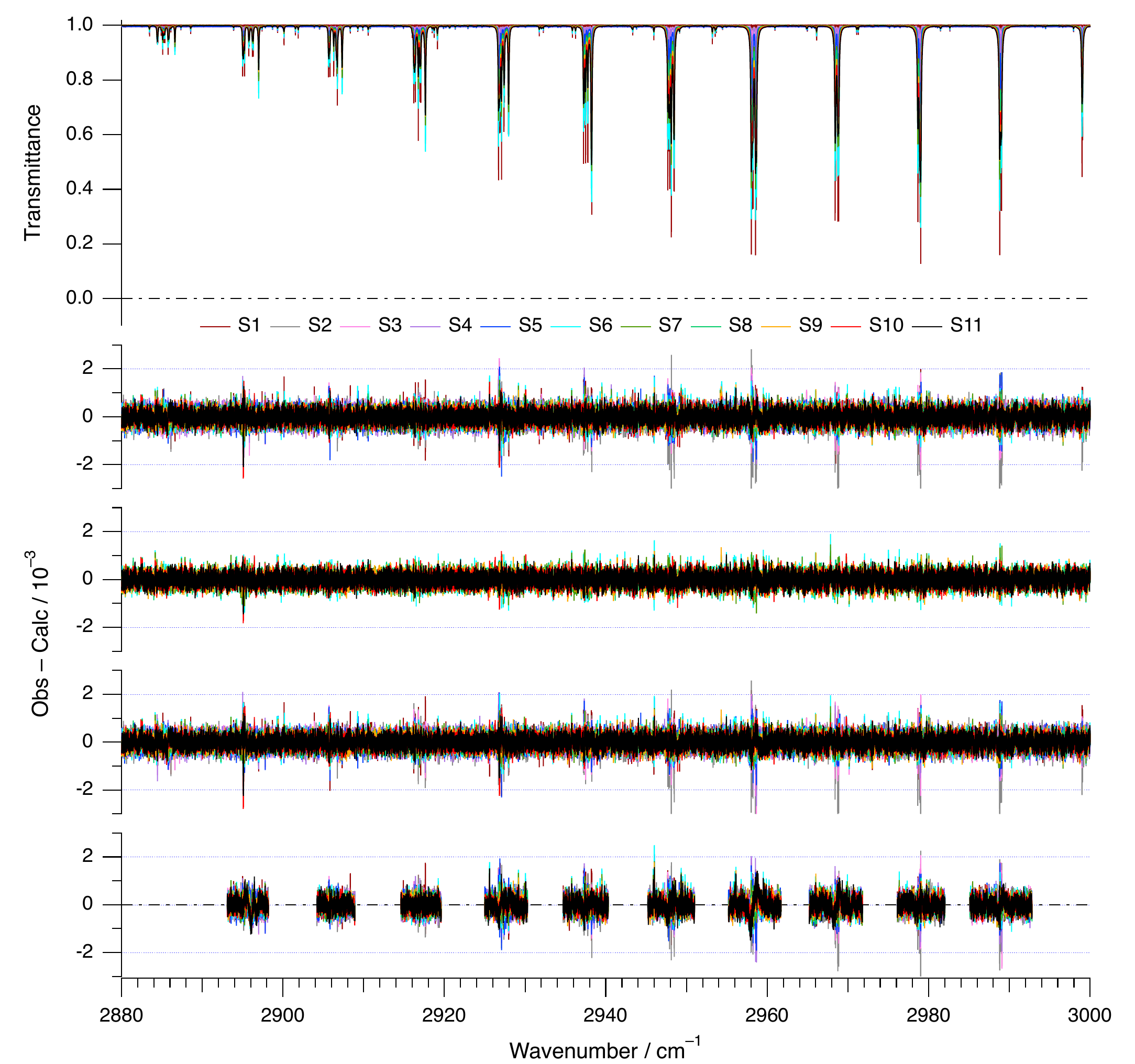}
\caption{Transmittance spectra of the P branch of the $\nu _3$ band of methane (top panel) and corresponding residuals obtained for the fits F1, F2, F3 and F4 (from top to bottom).}
\label{fig:fit_P}
\end{figure}

\begin{figure}[!htbp]
\centering
\includegraphics[width = \textwidth]{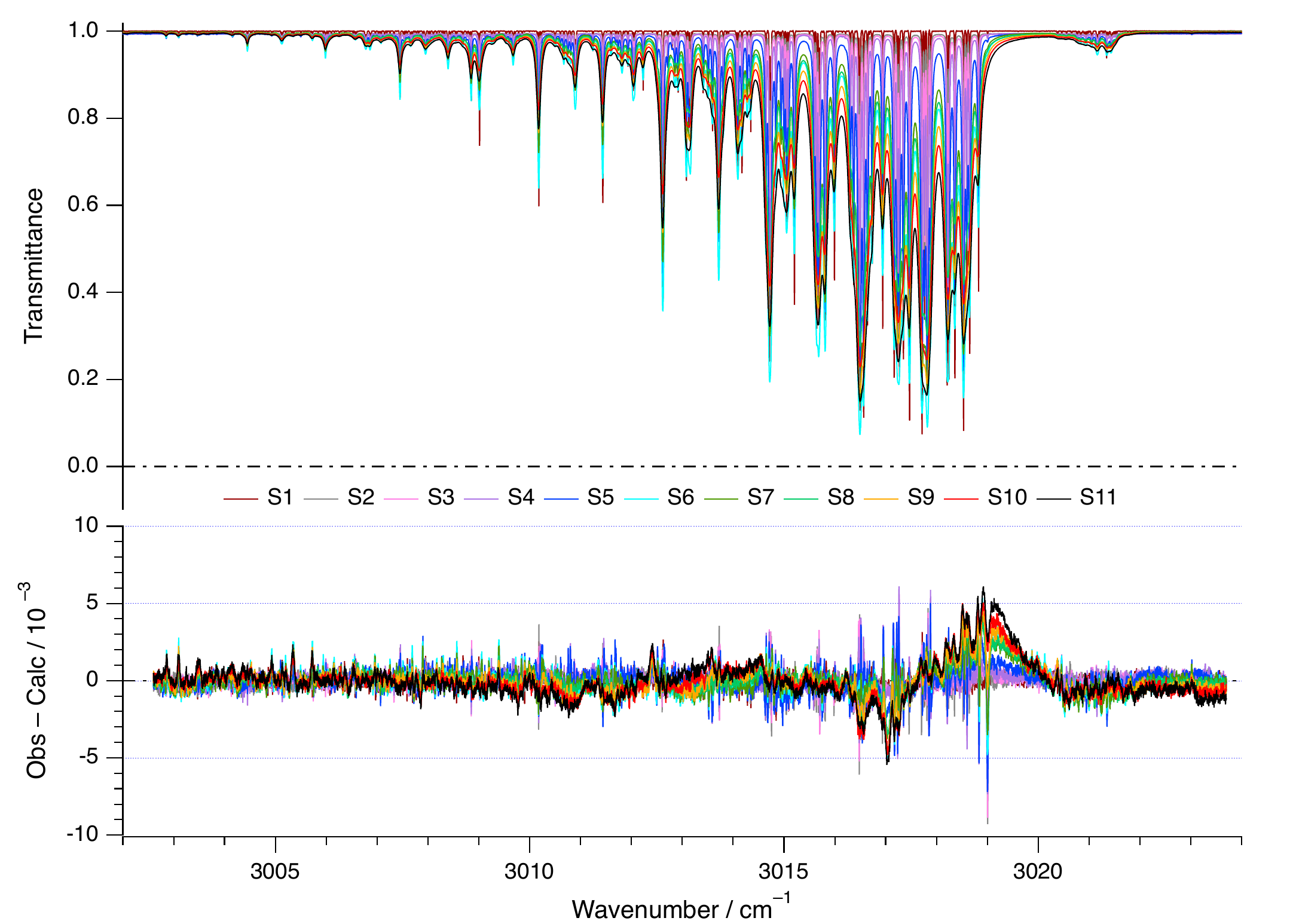}
\caption{Transmittance spectra of the Q branch of the $\nu _3$ band of methane (top panel) and corresponding residuals obtained for fit F4.}
\label{fig:fit_Q}
\end{figure}

\begin{figure}[!htbp]
\centering
\includegraphics[width = \textwidth]{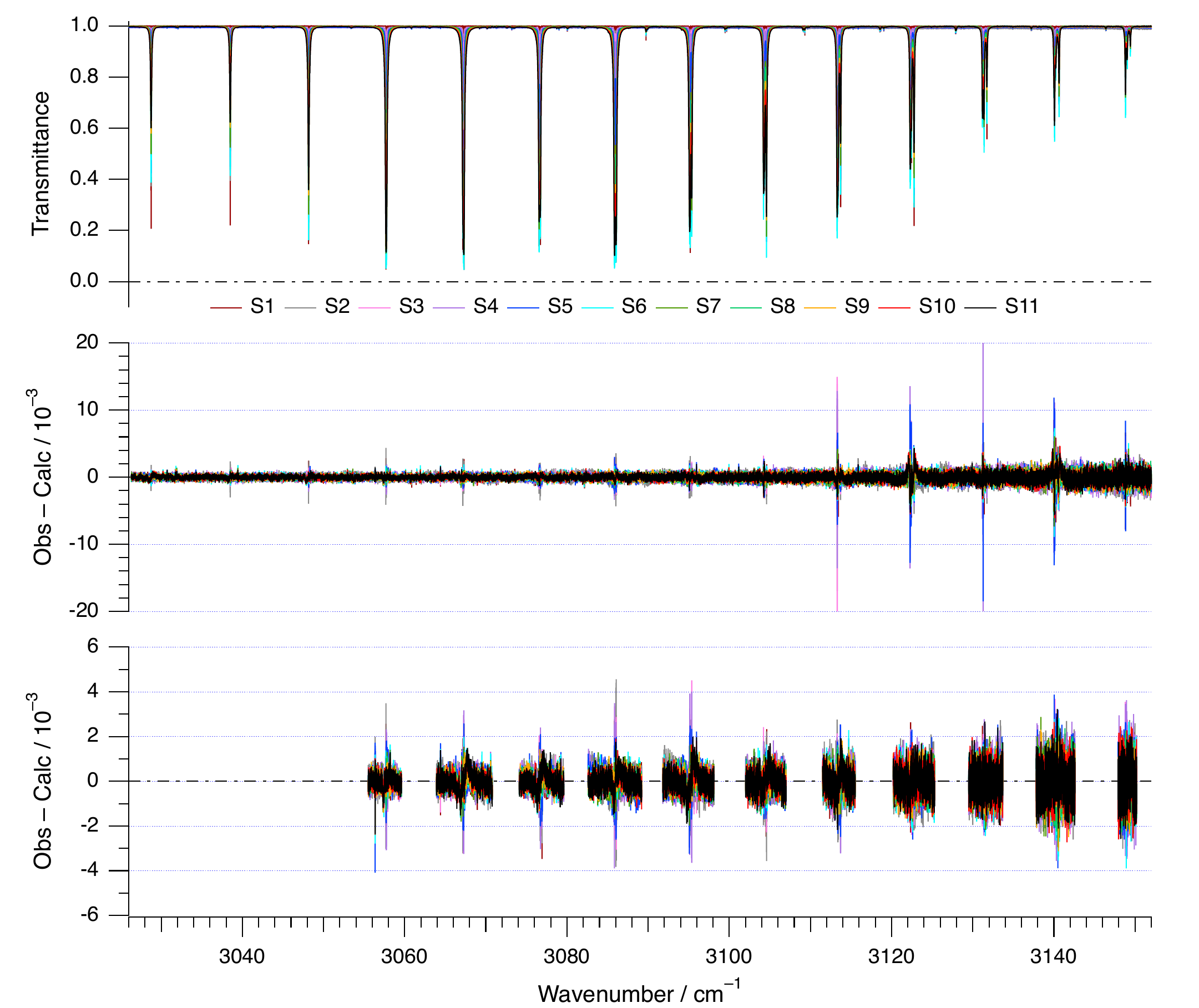}
\caption{Transmittance spectra of the R branch of the $\nu _3$ band of methane (top panel) and corresponding residuals obtained for the fits F1 (middle panel) and F4 (bottom panel).}
\label{fig:fit_R}
\end{figure}

\begin{figure}[!htbp]
\centering
\includegraphics[width = \linewidth]{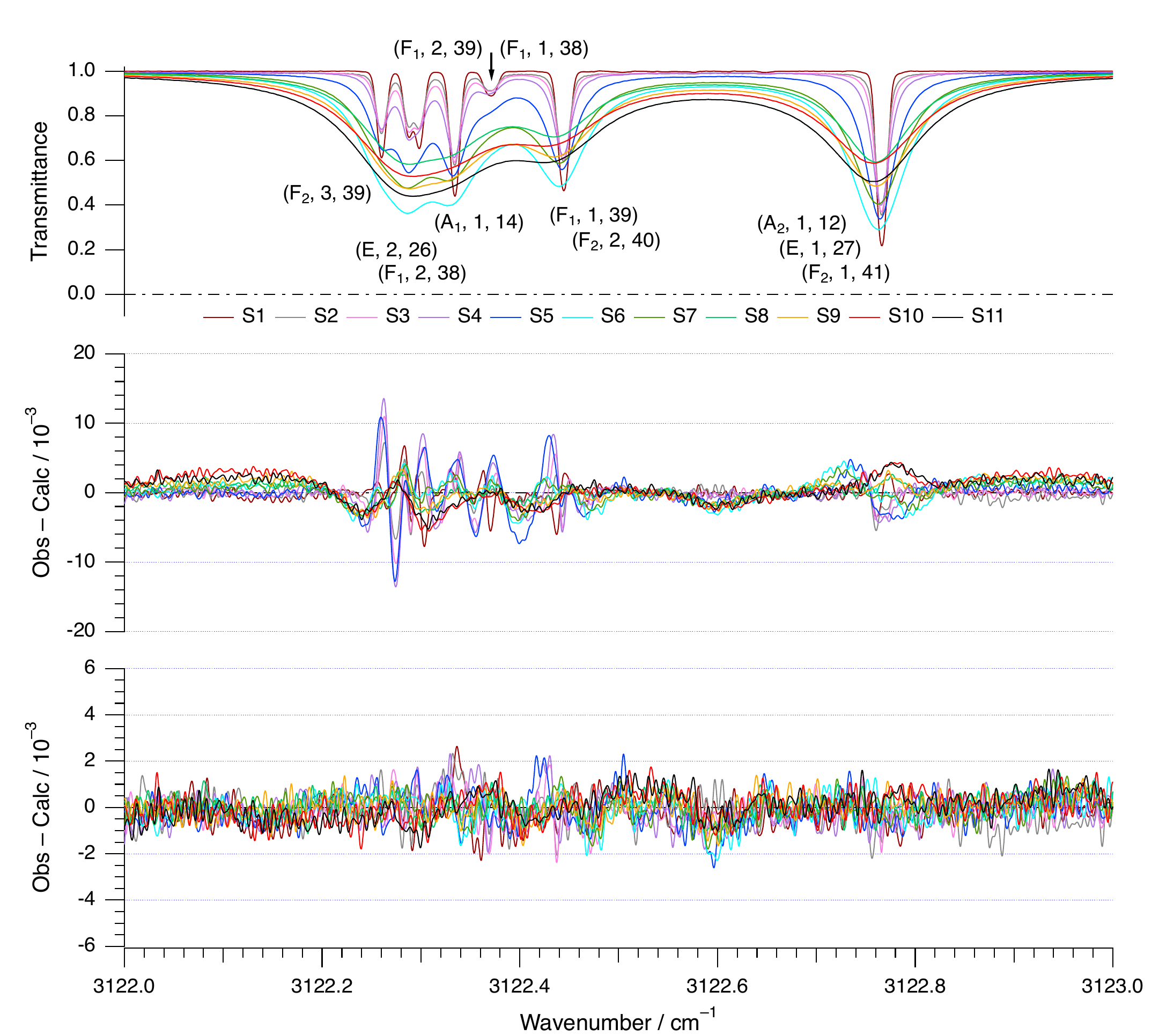}
\caption{Same as Fig. \ref{fig:fit_R}, showing the R(10) manifold. In the upper panel, the lines are identified with $(C'',\alpha'', \alpha')$ (see text for the meaning of the symbols).}
\label{fig:fit_rosen_cyuri}
\end{figure}

The positions $\tilde{\nu}_n$, intensities $\sigma_n$, assignments, air broadening $b_L^0(\mathrm{air})$ and air shift $\delta^0(\mathrm{air})$ coefficients of the $^{12}$CH$_4$ lines were taken from the \texttt{HITRAN} database \ccite{gor20a} (downloaded from \texttt{https://hitran.org} in December 2021). When fixed, $b_L^0(\mathrm{CO}_2)$ and $\delta^0(\mathrm{CO}_2)$ were therefore assumed to be equal to their air counterparts. \textcolor{red}{As in \ccite{tra22a}, $b_L^0(\mathrm{CO}_2)$ could have been set to $1.3 \times b_L^0(\mathrm{air})$. The spectrum of the strongest methane line the CO$_2$ broadening of which was set to the air broadening available in HITRAN in the present work was calculated at the lowest and highest total pressures (S2 and S11 in Table \ref{tab:exp_details}), with $b_L^0(\mathrm{CO}_2)$ set to $b_L^0(\mathrm{air})$ (spectra identified as $S_{calc}^{1.0}$ hereafter) and $1.3 \times b_L^0(\mathrm{air})$ (spectra identified as $S_{calc}^{1.3}$ hereafter). The peak amplitudes of the two ratios $S_{calc}^{1.3}/S_{calc}^{1.0}$ are smaller (lowest total pressure) or similar (highest total pressure) to the noise level observed in the residuals presented in Figs. \ref{fig:fit_P} to \ref{fig:fit_rosen_cyuri}. Therefore, the impact on the reported results of fixing the CO$_2$ broadening of the weak lines to the air broadening available in HITRAN is believed to be very small if not negligible.} The self broadening and self shift coefficients of the $^{12}$CH$_4$ lines were fixed to the values obtained as follows. The self broadening coefficients measured using a Voigt profile for several bands of $^{12}$CH$_4$ \ccite{pre05a, lep06a, smi10a, lyu11a, smi14a, has17a} were fitted to $b_L^0(\mathrm{self}) = \exp \{a + b \, m^2 + c \, |m|^3 + d \, m^4 \}$ where $a$, $b$, $c$ and $d$ were adjustable parameters and $m = -J''$, $J''$ and $J''+1$ for P, Q and R branch lines, respectively. Separate fits were performed for lines originating from levels with the $A_1$, $A_2$, $F_1$, $F_2$ and $E$ symmetries. As the self broadening coefficients calculated for the $A_1$, $A_2$, $F_1$ and $F_2$ symmetries with the fitted values of the $a$, $b$, $c$ and $d$ parameters were within $\pm 2$ \% of their averages, only these averages were considered. The two sets of calculated self broadening coefficients thus determined are listed in Table \ref{tab:b0_self}.
\begin{table}[htb!]
\caption{Values of the self broadening coefficients determined through fits of literature data \ccite{pre05a, lep06a, smi10a, lyu11a, smi14a, has17a} (see text for details). $m = -J''$, $J''$ and $J''+1$ for P, Q and R branch lines, respectively. ``Unc'' are the standard deviations, expressed as a percentage of the corresponding values.}
\label{tab:b0_self}
\centering
\vspace{1 cm}
\begin{tabular}{D{.}{.}{2.0}cccD{.}{.}{2.1}}
\toprule
\\
\multicolumn{1}{c}{$|m|$} & \multicolumn{1}{c}{$A$ \& $F$} & \multicolumn{1}{c}{Unc} & \multicolumn{1}{c}{$E$} & \multicolumn{1}{c}{Unc} \\
\\
\midrule
\\
  1 & 0.0812 & 6.0 & 0.0791 & 8.2 \\
  2 & 0.0812 & 6.0 & 0.0790 & 8.2 \\
  3 & 0.0811 & 6.0 & 0.0786 & 8.2 \\
  4 & 0.0807 & 6.0 & 0.0779 & 8.3 \\
  5 & 0.0800 & 6.1 & 0.0769 & 8.4 \\
  6 & 0.0789 & 6.2 & 0.0755 & 8.6 \\
  7 & 0.0775 & 6.3 & 0.0737 & 8.8 \\
  8 & 0.0758 & 6.4 & 0.0715 & 9.0 \\
  9 & 0.0736 & 6.6 & 0.0691 & 9.4 \\
 10 & 0.0712 & 6.8 & 0.0665 & 9.7 \\
 11 & 0.0686 & 7.1 & 0.0637 &10.  \\
 12 & 0.0659 & 7.4 & 0.0609 &11.  \\
 13 & 0.0630 & 7.7 & 0.0582 &11.  \\
 14 & 0.0602 & 8.1 & 0.0557 &12.  \\
 15 & 0.0575 & 8.5 & 0.0535 &12.  \\
 16 & 0.0551 & 8.8 & 0.0516 &13.  \\
 17 & 0.0529 & 9.2 & 0.0503 &13.  \\
 18 & 0.0510 & 9.5 & 0.0495 &13.  \\
 19 & 0.0497 & 9.8 & 0.0494 &13.  \\
\\
\bottomrule
\end{tabular}
\end{table}
The self shift coefficient of all the methane lines considered was fixed to $\delta^0(\mathrm{self}) = -0.0090 \, (13)$ cm$^{-1}$atm$^{-1}$. This value is the average of the 66 self shift coefficients measured at 296 K by Pine \ccite{pin19a} in the Q branch of the $\nu_3$ band of $^{12}$CH$_4$ ($J=1$ to $13$) using a Rautian line shape with first order line mixing, each value being weighted with the square of the inverse of its uncertainty of measurement. Three of the 4 self broadening coefficients measured in the R branch of the $\nu_2 + \nu_4$ band of $^{12}$CH$_4$ by Mondelain \textit{et al.} \ccite{mon05a} agree with this average within experimental uncertainties. The parameters $\beta^0$, $Y_n^0$ and $W_{mn}$ for $m \neq n$ were set to zero when not fitted. The same applies to $\aw$, except for fits F3 and F4 (see below). In addition to the line parameters, one (P and R branches) to three (Q branch) baseline [$b(\tilde{\nu})$ in Eq. \ref{eq:molecular_transmittance}] parameters were fitted for each spectrum included in the analysis, together with the methane mole fraction $x = P(\mathrm{CH}_4)/P_{tot}$ associated with spectra S2 to S11. %\textcolor{red}{[what about $x$(H$_2$O)?]}

% SELF BROADENING: see 'sub-atmospheric/literature/self_broadening/'. Unweigthed fits performed because they led to more consistent results than weighted fits. SELF SHIFT: see 'sub-atmospheric/literature/self_shift/'

The multispectrum analysis of the measured spectra was carried out 4 different ways, identified as fits F1 to F4 from now on. The first 3 fits relied on the weak line mixing first order approximation (Eq. \ref{eq:line-mixing_rosenkranz}), while fit F4 relied on the relaxation matrix formalism (Eq. \ref{eq:rel_mat}). \textcolor{red}{Table \ref{table:analyses_fit} provides an overview of the 4 fits. Note that the parameters of a limited number of lines indicated as ``fixed'' in this Table were actually fitted to improve the residuals.
\newcolumntype{R}{>{$}r<{$}} % math-mode version of "r" column type
\newcolumntype{L}{>{$}l<{$}} % math-mode version of "l" column type
\begin{table}[!htbp]
\renewcommand*{\arraystretch}{2}
\centering
\caption{Summary of the 4 fits (F1 to F4) applied to analyze the 11 spectra of Table \ref{tab:exp_details}. Fits F1, F3 and F4 included the 11 spectra listed in Table \ref{tab:exp_details}, while fit F2 considered only spectra S6 to S11. The branches studied in each fit are indicated in the second header row of columns ``F1'' to ``F4.'' The observed line shapes were modeled using the quadratic speed dependent Rautian profile. Line mixing was modeled using the first order approximation (fits F1 to F3) or the relaxation matrix formalism (fit F4). $I$ is the line intensity in cm$^{-1}$/(molecule cm$^{-2}$) at 296 K; only lines with $I > 10^{-23}$ were considered. In the first three rows of columns ``F1'' to ``F4,'' the number of lines with at least one fitted parameter is given first, followed by the count of lines without fitted parameters.}
\label{table:analyses_fit}
\small
\vspace{1 cm}
\begin{threeparttable}
\begin{tabular}{ccccc}
\toprule
$^{12}$CH$_4$ line  & F1 & F2 & F3 & F4 \\[-9pt]
parameters  & P \& R & P & P & P, Q \& R \\
\midrule
 $I < 5 \times 10^{-22}$ & 138/2147 & 79/975 & 80/974 & 61/1598 \\[-6pt]
 $5 \times 10^{-22} \leq I < 5 \times 10^{-21}$ & 239/1 & 100/0 & 100/0 & 173/1 \\[-6pt]
 $I \geq 5 \times 10^{-21}$ & 147/3 & 70/0 & 70/0 & 213/2 \\[-3pt]
 $b_L^0(\mathrm{CO}_2)^{\dag}$ & fit/$b_L^0(\mathrm{air})$ & fit/$b_L^0(\mathrm{air})$ & fit/$b_L^0(\mathrm{air})$ & fit/$b_L^0(\mathrm{air})$ \\[-6pt]
 $\delta^0(\mathrm{CO}_2)^{\dag}$ & fit/$\delta^0(\mathrm{air})$ & fit/$\delta^0(\mathrm{air})$ & fit/$\delta^0(\mathrm{air})$ & fit/$\delta^0(\mathrm{air})$ \\[-6pt]
 $Y_n^0$ or $W_{mn}$ ${}^{\ddag}$ & fit/0 & fit/0 & fit/0 & fit/0 \\[-6pt]
 $\beta^0$ ${}^{\ddag}$ & fit/0 & 0/0 & fit/0 & fit/0 \\[-6pt]
 $\aw$ ${}^{\ddag}$ & fit/0 & fit/0 & 0.110/0.110 & fit/0.110 \\
\bottomrule
\end{tabular}
\begin{tablenotes}
    \item[$^{\dag}$] ``fit'' applies to all the fitted lines.
    \item[$^{\ddag}$] ``fit'' only applies to $^{12}$CH$_4$ lines with $I \geq 5 \times 10^{-21}$; $Y_n^0$ or $W_{mn}$ were non zero for the $\nu_3$ band only.
\end{tablenotes}
\end{threeparttable}
\end{table}
As mentioned in the introduction, this work aimed to analyze spectra recorded at total pressures approaching 1 atm using the relaxation matrix formalism (\textit{i.e.} ``fit F4''), more appropriate than the first order line mixing approximation. However, the spectra were also analyzed using the latter (identified as ``fit F1''), as it made it possible to compare both approaches and results of this work with literature. This work actually started with fit F1. These measurements showed that the speed dependence of line broadening $\aw$ did not exhibit a rotational dependence (see below). To determine an average value of $\aw$ without ``interferences'' from Dicke narrowing, ``fit F2'' was performed relying on spectra involving total pressures larger than 400 hPa to allow use of the speed dependent Voigt line shape model. Fit F2 was also limited to the P branch as it is less congested and first order line mixing applies. In ``fit F3,'' the P branch was remeasured using the same line shape as fits F1 and F4, \textit{i.e.} the speed dependent hard collision profile, with $\aw$ fixed to the average value obtained in fit F2 to check its impact on the measured line shape parameters.} In all fits, the $J''$ manifolds in the P and R branches were measured individually as they are well separated from each other. The couplings of lines involving different $J$ levels were therefore ignored in these branches. Because of its high line density, the Q branch was only analyzed in its entirety, in the range $3002.6-3023.7$ cm$^{-1}$, using the relaxation matrix formalism (fit F4). To avoid having to deal with large matrices, lines separated by more than 1 cm$^{-1}$ were considered to not be coupled and inter-$J$ couplings were limited to $\Delta J = \pm 1$ and $J \leq 4$. The first restriction was quite effective at reducing the sizes of the matrices and prevented unrealistically large line mixing parameters. The second restriction is justified by the increase of the energy gap separating rotational levels, larger than $50$ cm$^{-1}$ between $J = 4$ and $J = 5$.

% 05/04/2024 ("..., lines separated by more than 1 cm$^{-1}$ were considered to not be coupled..."): In 'article_revised/response_to_review/line_mixing/offdiag_W/line_coupling.pxp', the figure shows that 7 separations are larger than 1 cm-1 (but smaller to ~2.5 cm-1). I guess Thibault included them because it improved the residuals.

% Fits F1, F3 and F4 included the 11 spectra listed in Table \ref{tab:exp_details}, while fit F2 considered only spectra S6 to S11 (total pressures $\geq$ 400 hPa).

The lack of calibration of the wavenumber scales of the spectra was accounted for by multiplying the line positions read from \texttt{HITRAN} by the factor $(1+\mathrm{C})$. In fits F1 to F3, the positions of the lines in all the analyzed manifolds of the P and R branches were kept fixed to the \texttt{HITRAN} values and C was fitted. For fit F4, C was only fitted for the P(8) to P(12), Q(8) to Q(13) and R(0) to R(8) manifolds of the $\nu_3$ band, involving line positions measured using sub-Doppler spectroscopy \ccite{abe13a}. The other manifolds were then measured with C set to the average of the values thus obtained, fitting the line positions. This ``calibration'' was possible because the 11 spectra were recorded one after the other, with the same instrumental conditions. During the analysis, some of the line positions available in \texttt{HITRAN} were found to be off by a few $10^{-2}$ cm$^{-1}$ and therefore fitted. In the Q branch, line positions were determined in a separate fit of spectrum S1. Note that there seems to be a typo in the position of the R(4, $A_1$, 1, 7) line provided in \texttt{HITRAN} as it reads 3067.300026 cm$^{-1}$ instead of the value of 3067.300224 cm$^{-1}$ reported by Abe \textit{et al.} \ccite{abe13a}.

\begin{figure}[!htb]
\centering
\includegraphics[width = 0.5\linewidth]{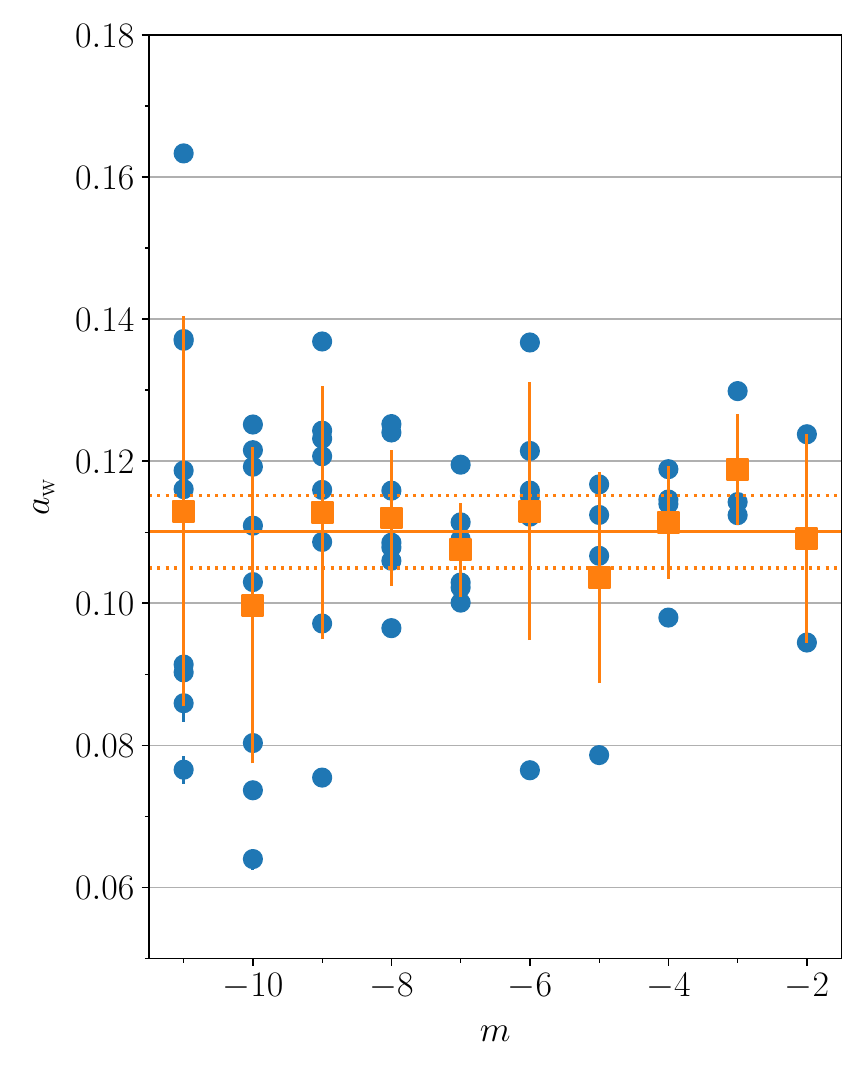}
\caption{Speed dependence of line broadening coefficient $\aw$ measured in the P branch of the $\nu_3$ band of $^{12}$CH$_4$ in fit F2 ($m = -J''$). The blue circle are the measurements, the orange squares present the averages for each manifold (the error bars display the corresponding standard deviations) and the horizontal orange line shows the weighted average of the latter.}
\label{fig:sd_average}
\end{figure}

As indicated in Table \ref{table:analyses_fit}, the speed dependence of the line broadening was set to $\aw = 0.110$ for fits F3 and F4. This value was obtained as follows. Figure \ref{fig:sd_average} presents the values of $\aw$ measured in fit F2 and the corresponding averages for each manifold.
These averages do not exhibit any rotational dependence. They were therefore averaged, each value being multiplied by the inverse of the square of its standard deviation, yielding $\aw = 0.110 \, (5)$. The results presented in Fig. \ref{fig:sd_average} and their average are very similar to values reported in the literature, for example for methane \ccite{pin19a, dev15a}, carbon monoxide \ccite{dev12a}, carbon dioxide \ccite{devi07a, bui14a} and acetylene \ccite{oku23a}. % \textcolor{red}{(although $\aw \approx 0.22$)}

Figures \ref{fig:fit_P}, \ref{fig:fit_Q} and \ref{fig:fit_R} present the residuals obtained at the end of the fits performed in the present work for the P, Q and R branches, respectively. The sharp signatures observed for the lower pressures spectra (S2 and S3) are believed to arise from a still imperfect modeling of the \texttt{ILS} and a slight saturation of the stronger lines. The use of the relaxation matrix resulted in a clear reduction of the residuals for the more congested manifolds, \textit{i.e.} R(9) to R(13) above 3110 cm$^{-1}$ and, to a lesser extent, P(8) to P(11) below 2930 cm$^{-1}$. Shown in Fig. \ref{fig:fit_rosen_cyuri}, the residuals obtained for the R(10) manifold are very revealing in that regard, where the first order approximation fails to represent the couplings involving the close lying lines of $F$ symmetry observed between 3122.2 and 3122.5 cm$^{-1}$. However, no improvement in the residuals is observed for the more widespread lower $J$ manifolds of the P and R branches, the signatures being even slightly larger in some cases [for R(6) and R(7), for example]. Figure \ref{fig:fit_Q} shows that the residuals exhibit a rather strong signature in the low $J$ region of the Q branch. It is interpreted as resulting from the sum of many weak interactions that were neglected in the present modeling because of the above-mentioned restrictions, which had no consequence in the more widespread P and R branches. These restrictions made the general relaxation matrix equation manageable. In this case however, a hybrid method using both the relaxation matrix and first order line mixing should probably be used, as was done by Pine \ccite{pin19a}.

All the lines involving at least 1 fitted parameter in fits F1 and F4 are provided as supplementary material. % \textcolor{red}{We consider the parameters from F4 to be our final results. In addition, for the regions not covered by F4 (see Figs. \ref{fig:fit_P} to \ref{fig:fit_R}), the parameters from F1 are considered to be the final results. About 760 broadening, 220 relative speed-dependence, 250 Dicke narrowing and 660 shift parameters were measured across 800 lines from both $^{12}$CH$_4$ and $^{13}$CH$_4$ between $2846$ and $3186$ cm$^{-1}$.}

\section{Results}\label{sc:results}

The CO$_2$ broadening coefficients $b_L^0(\mathrm{CO}_2)$, speed dependence of the broadening coefficient $\aw$, CO$_2$ shift coefficients $\delta^0$ and first order line mixing parameters $Y_n^0$ measured in this work for the $\nu_3$ band of $^{12}$CH$_4$ are presented in Figs. \ref{fig:y0} and \ref{fig:sd} to \ref{fig:lmfo} ($m = -J''$, $J''$ and $J'' + 1$ for P, Q and R branch lines, respectively).
As already highlighted, the 4 fits analyzed the P branch, fits F1 and F4 also dealt with the R branch and the Q branch was only studied in fit F4.

% Overall, the spread of the results presented in Figs. \ref{fig:y0} and \ref{fig:sd} to \ref{fig:delta} is rather small in the P and Q branches and up to $m=9$ in the R branch and is somewhat larger for $m \geq 10$. The lower precision of measurement for $m \geq 10$ is attributed to \textcolor{red}{...} The reduction of the signal-to-noise ratio in the transmittance spectra as the upper limit of the band pass filter used is approached (see the residuals presented in Fig. \ref{fig:fit_R}) probably also contributes.

% However, the lines from the R(3)--R(10) and the whole Q branch still present some disparities, which has two origins. The first is the saturated absorption of the most intense manifolds (see Fig. \ref{fig:fit_R}) getting out of the validity range of Beer-Lambert law. The second is the overlapping of the lines, both of which are stronger in the $R$ and $Q$ branches, with $Q$ being the most extreme. By itself, the overlapping makes the determination harder, but it is worsened by the uncertainty coming from the line-mixing. Other differences will be discussed for each parameter.

The uncertainties $\Delta \zeta$ on these measured parameters $\zeta$, illustrated with error bars in the figures, were estimated using the following sum in quadrature
\begin{equation}\label{eq:errors}
\Delta \zeta = \sqrt{ (\delta \zeta)^2 + (\delta T)^2 + (\delta P_{tot})^2 + [\delta P(\mathrm{CO}_2)]^2 + (\delta_{ils})^2 }
\end{equation}
In this expression, $\delta \zeta$ is the $1\sigma$ precision of measurement of the parameters, as provided by the least squares fitting algorithm. The uncertainty on the sample temperature was estimated to be $\delta T \approx 0.2$ \%, from the sum in quadrature of the accuracy of measurement of the sensors (\textit{i.e.} $\pm 0.3$ K) and the uncertainty on the measured temperature itself (\textit{i.e.} $\pm 0.5$ K). The uncertainty on the measured total pressure is $\delta P_{tot} \approx 0.5$ \% (see Table \ref{tab:exp_details}). The uncertainties on the CO$_2$ pressures $\delta P(\mathrm{CO}_2)$ include the uncertainties on the measured mole fractions of methane and the presence of impurities. They do not apply to the narrowing and line mixing parameters, involving the total pressure only (see Eqs. \ref{eq:line_narrowing} and \ref{eq:line-mixing_rosenkranz}). The line shape exhibits similar behavior with respect to the Dicke narrowing parameter and the speed dependence of broadening. The average covariance of the fit with respect to these two parameters, estimated to be equal to 10 \% of their standard deviations, was therefore also considered for these two parameters. \textcolor{red}{The contribution to the uncertainties on the measured parameters of uncertainties in the modeling of the instrument line shape (\texttt{ILS}) is identified by $\delta_{ils}$ in Eq. \ref{eq:errors}. It was estimated using a method similar to that described by Loos \textit{et al.} \ccite{loo17a}. A few manifolds were remeasured with the modulation function (from which the \texttt{ILS} is generated in the spectral domain) modified to minimize the sum of the square of the residuals obtained for each of them. The RMS of the differences between the values of the parameters obtained using the initial and modified \texttt{ILS} were used as the contributions $\delta_{ils}$ to their uncertainties}. They amount to $0.4$\% for the broadening coefficients, $0.9$\% for the shift coefficients, $2.0$\% for the speed dependence of the broadening and $5.7$\% for the narrowing coefficients. As already observed \ccite{far21a, loo17a}, the impact of uncertainties in the \texttt{ILS} was observed to be larger for smaller values of the line mixing parameters. $\delta_{ils}$ was therefore set to 3.4 and 3.2 \% for $W_{mn} > 0.001$ cm$^{-1}$atm$^{-1}$ and $Y^0_n > 0.03$ atm$^{-1}$, respectively. Smaller values of these 2 parameters were respectively assigned uncertainties $\delta_{ils} = 15$ and 18 \%. Contrary to our previous contribution on the $\nu_3$ band of methane \ccite{far21a}, self broadening and self shift were considered in the present work. In that previous work, the neglect of self broadening was conservatively estimated to contribute 0.3 \% to the uncertainties of the measured air broadening coefficients. Involving similar experimental conditions and noting that $b_L^0(\mathrm{CO}_2) \simeq 1.2 \times b_L^0(\mathrm{air})$ on average \ccite{ess21a}, the impact on $b_L^0(\mathrm{CO}_2)$, $\aw$ and $\beta^0$ of uncertainties on the self broadening coefficients listed in Table \ref{tab:b0_self} was included in the present work. The value of $\delta^0(\mathrm{self}) = -0.0090 \, (13)$ cm$^{-1}$atm$^{-1}$ to which the self shift coefficients of all the methane lines were fixed to is similar to the CO$_2$ shift coefficients measured in this work (see Fig. \ref{fig:delta}).
Even though Eq. \ref{eq:line_shift} and the mole fractions listed in Table \ref{tab:exp_details} show that the uncertainty of 14 \% on $\delta^0(\mathrm{self})$ contributes less than 0.2 \% to the measured CO$_2$ shift coefficients, it was included in $\Delta \zeta$.

\subsection{\texorpdfstring{CO$_2$ broadening coefficients}{CO2 broadening coefficients}}\label{sc:broadening}

\begin{figure}[!htb]
\centering
\includegraphics[width = \linewidth]{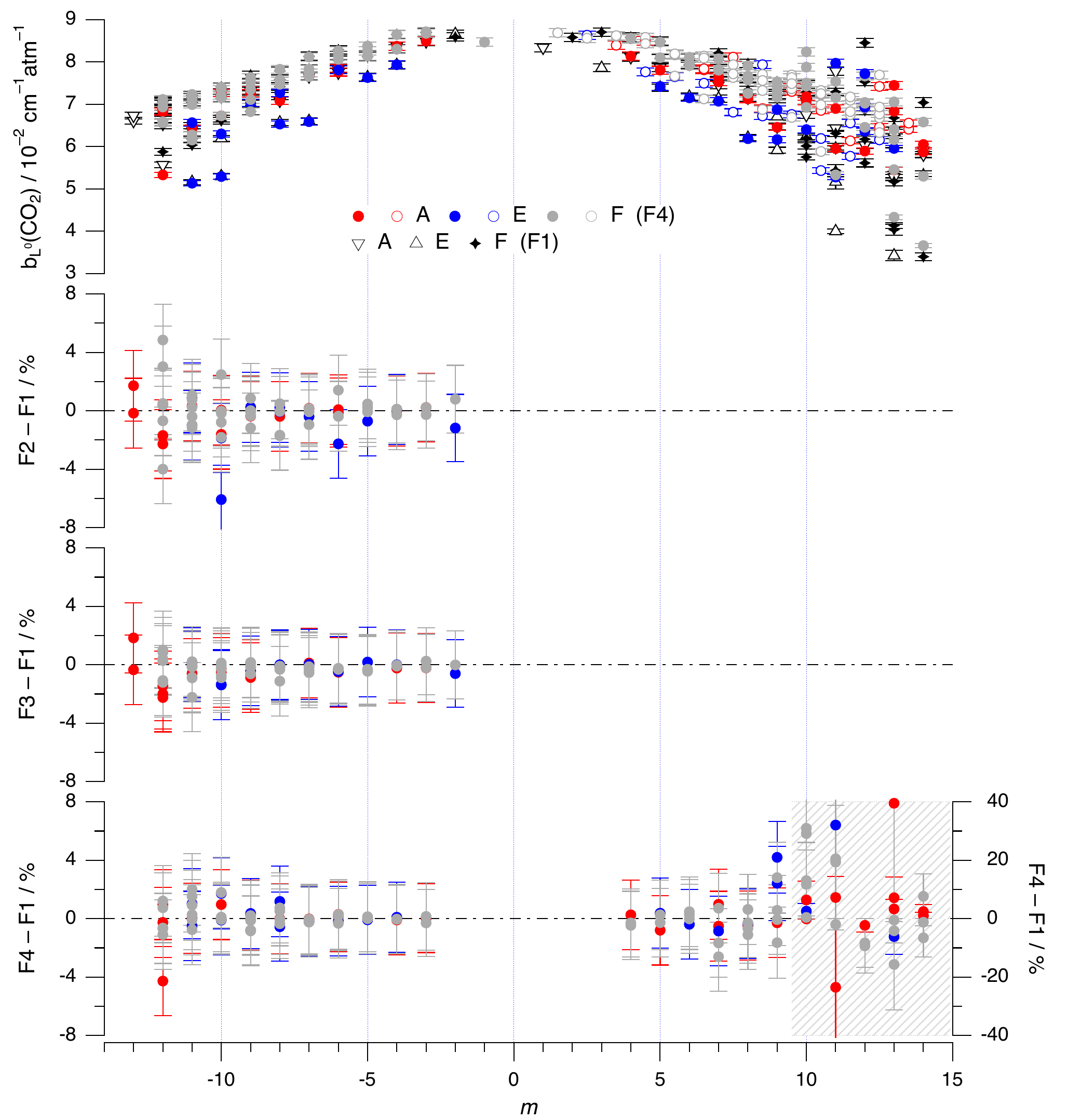}
\caption{CO$_2$ broadening coefficient measured at 296.5 K in the P ($m = -J''$), Q ($m = J'' + 0.5$) and R ($m = J'' + 1$) branches of the $\nu _3$ band of $^{12}$CH$_4$ with line mixing modeled using the relaxation matrix formalism (fit F4; colored symbols, filled for P and R branch lines and open for Q branch lines) and the first order approximation (fit F1, black symbols) (upper panel); differences with the CO$_2$ broadening coefficients measured in fit F1 of the values retrieved in fits F2, F3 and F4 are presented in the lower 3 panels. In the lower panel, the points over the hatched area refer to the right axis and three points, corresponding to differences of about $12$ \% at $m=-12$, $54$ \% at $m=11$ and $71$ \% at $m=13$, are off scale. The error bars represent the estimated uncertainties (see section \ref{sc:results} for details).}
\label{fig:y0}
\end{figure}
The upper panel of Fig. \ref{fig:y0} presents the CO$_2$ broadening coefficients measured in this work in the P, R and Q (fit F4 only)  branches of the $\nu _3$ band of $^{12}$CH$_4$ with line mixing modeled using the relaxation matrix formalism (fit F4) or the first order approximation (fit F1). Differences of the CO$_2$ broadening coefficients measured in fits F2, F3 and F4 with the values retrieved from fit F1 are presented in the lower 3 panels. Most of these differences are smaller than 1 \% and their averages are $-0.22 \, (1.4)$ \%, $-0.34 \, (0.7)$ \% and $0.18 \, (1.6)$ \% for fits F2, F3 and F4 ($m < 10$), respectively (the numbers between parentheses are the corresponding standard deviations). The largest difference of about 12 \% is observed for the P(12, $F_1$, 45, 1) line. The differences are somewhat larger for $m \geq 10$. The corresponding manifolds involve closely lying lines and the residuals obtained for these in fit F1 exhibit signatures (see Fig. \ref{fig:fit_R}) indicating that the first order line mixing model failed to reproduce satisfactorily the observed spectra. The resulting broadening coefficients retrieved from fit F1 may therefore be impacted by this imperfect modeling, leading to the large differences observed in the lower panel of Fig. \ref{fig:y0}.

% Second sentence removed: "Some strongly blended lines ($\Delta \tilde{\nu } < 0.01$ cm$^{-1}$) appeared with indistinguishable parameters and are not included."

% As the variations may simply be due to the larger noise level in the corresponding spectral range, it is not clear whether or not the use of the relaxation matrix did improve the measurement of this parameter.

Overall, the $b_L^0(\mathrm{CO}_2)$ parameter is well decorrelated from the other parameters. The upper panel of Fig. \ref{fig:y0} shows that the CO$_2$ broadening coefficients of lines of $E$ symmetry tend to be smaller than for lines of $A$ and $F$ species, which do not exhibit any marked difference. This has been observed previously for different collision partners \ccite{bal86a, mar93a, gab97a, far21a} but the origin of such a behavior has not been clearly identified \ccite{fox90a}. As is commonly observed for methane, all three branches exhibit the same $J''$ dependence, \textit{i.e.} a decrease with increasing $|m|$ of the CO$_2$ broadening coefficients following a maximum near $|m| = 3$. This behavior has been attributed to the increase of the energy separation between the rotational levels, leading to a reduction of the collisional transfer probabilities. % The difference of the measured $Q(1)$ is caused by the algorithm trying to reduce the residuals coming most probably from some remaining line mixing. 

\begin{figure}[!htb]
\centering
\includegraphics[width = \linewidth]{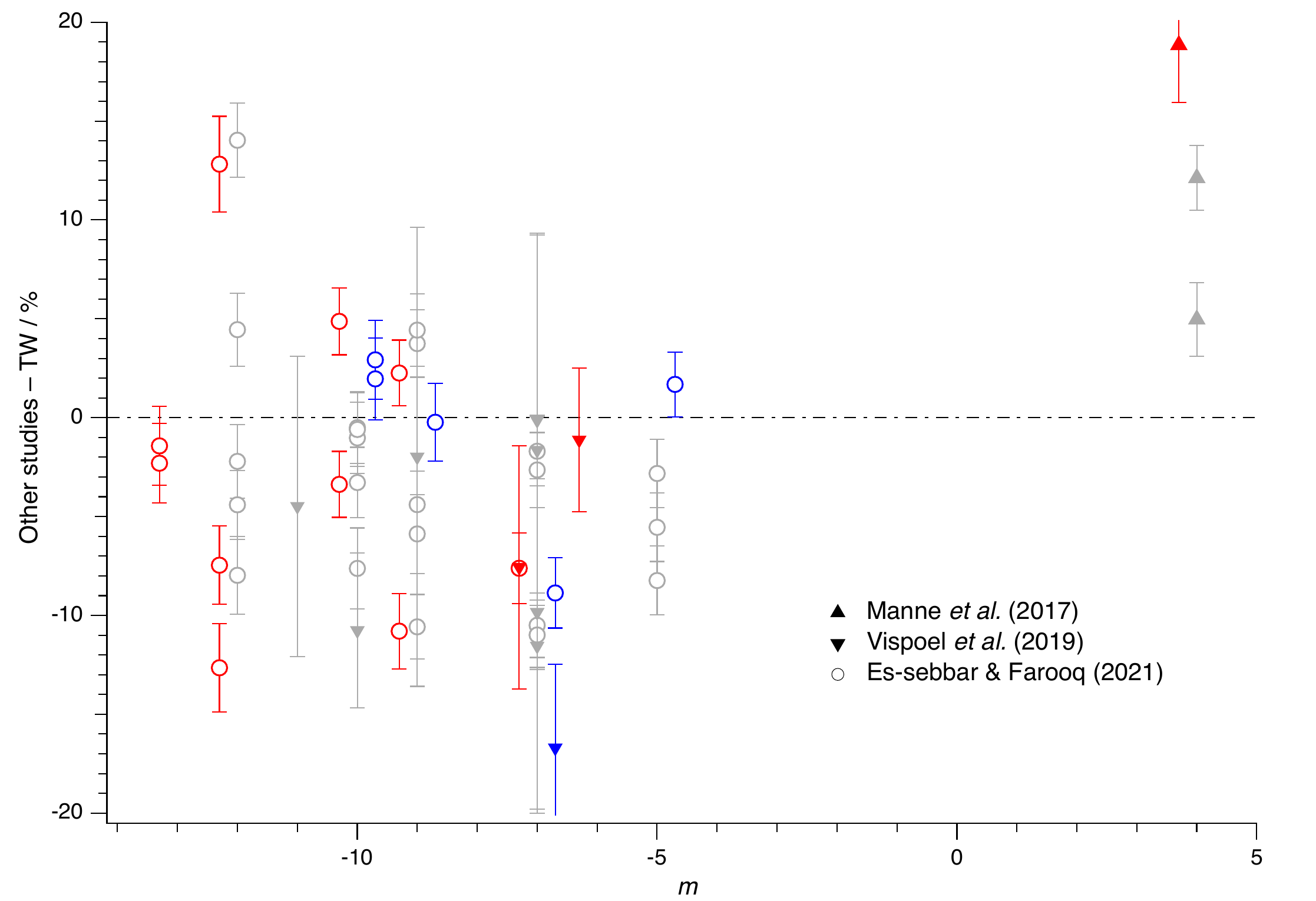}
\caption{Differences with this work (TW) of CO$_2$ broadening coefficients for the $\nu_3$ band of $^{12}$CH$_4$ measured in previous work (Manne \textit{et al.} \ccite{man17a}, Vispoel \textit{et al.} \ccite{vis19a} and Es-sebbar and Farooq \ccite{ess21a}). Red, blue and grey symbols correspond to the $A$, $E$ and $F$ symmetries, respectively. To improve readability, the differences for the $A$ and $E$ symmetries are provided for $m-0.3$ and $m+0.3$, respectively.}
\label{fig:br_lit}
\end{figure}

Table \ref{table:ow_y0} and Fig. \ref{fig:br_lit} compare the CO$_2$ broadening coefficients measured in this work with all the measurements reported for the $\nu_3$ band of $^{12}$CH$_4$ \ccite{ess21a, vis19a, man17a}.
\begin{table}[!htbp]
\small
\centering
\caption{\textcolor{red}{Comparison of CO$_2$ broadening coefficients (in $10^{-2}$ cm$^{-1}$atm$^{-1}$) reported previously for the $\nu_3$ band of \textnormal{$^{12}$CH$_4$} and measured in this work (TW).} The numbers between parentheses are the uncertainties, given in the units of the last digits quoted.} \vspace{12pt}
\label{table:ow_y0}
\begin{tabular}{llrrS[table-format=1.7, table-number-alignment=right]S[table-format=1.7, table-number-alignment=right]S[table-format=1.7, table-number-alignment=right]S[table-format=1.7, table-number-alignment=right]}
\toprule
 $J''$   & $C''$   & $\alpha'$   & $\alpha''$   & {\thead[tc]{\ccite{man17a}\\(qsdVoigt)$^{\dag}$}}   & {\thead[tc]{\ccite{vis19a}\\(Rautian)}}   & {\thead[tc]{\ccite{ess21a}\\(Galatry)}} & {TW} \\
\midrule
 $P(13)$ & $A_2$ & $18$ & $1$ &            &            & 6.456(53) & 6.608(79)   \\
 $P(13)$ & $A_1$ & $16$ & $1$ &            &            & 6.636(54) & 6.732(80)   \\
 $P(12)$ & $A_1$ & $15$ & $2$ &            &            & 5.968(71) & 6.849(82)   \\
 $P(12)$ & $F_1$ & $43$ & $3$ &            &            & 7.514(45) & 6.633(80)   \\
 $P(12)$ & $F_2$ & $45$ & $3$ &            &            & 7.245(45) & 7.012(84)   \\
 $P(12)$ & $A_2$ & $14$ & $1$ &            &            & 6.425(55) & 6.876(82)   \\
 $P(12)$ & $F_2$ & $46$ & $2$ &            &            & 6.547(55) & 7.061(84)   \\
 $P(12)$ & $F_1$ & $44$ & $2$ &            &            & 6.729(39) & 7.084(85)   \\
 $P(12)$ & $F_2$ & $47$ & $1$ &            &            & 6.428(43) & 6.494(78)   \\
 $P(12)$ & $A_1$ & $16$ & $1$ &            &            & 6.016(65) & 5.570(68)   \\
 $P(11)$ & $F_1$ & $42$ & $2$ &            & 6.840(459) &           & 7.166(86)   \\
 $P(10)$ & $F_2$ & $37$ & $3$ &            &            & 7.118(45) & 7.146(85)   \\
 $P(10)$ & $E$   & $24$ & $2$ &            &            & 5.449(42) & 5.299(63)   \\
 $P(10)$ & $F_1$ & $36$ & $2$ &            &            & 7.164(44) & 7.409(88)   \\
 $P(10)$ & $A_1$ & $14$ & $1$ &            &            & 7.025(34) & 7.274(87)   \\
 $P(10)$ & $F_1$ & $37$ & $1$ &            & 6.560(200) & 7.276(45) & 7.351(88)   \\
 $P(10)$ & $F_2$ & $38$ & $2$ &            &            & 7.165(48) & 7.212(86)   \\
 $P(10)$ & $A_2$ & $12$ & $1$ &            &            & 7.058(34) & 6.665(80)   \\
 $P(10)$ & $F_2$ & $39$ & $1$ &            &            & 6.217(58) & 6.611(79)   \\
 $P(10)$ & $E$   & $25$ & $1$ &            &            & 6.425(55) & 6.194(75)   \\
 $P( 9)$ & $A_2$ & $12$ & $1$ &            &            & 7.426(34) & 7.320(87)   \\
 $P( 9)$ & $F_2$ & $32$ & $2$ &            &            & 7.629(37) & 7.411(88)   \\
 $P( 9)$ & $F_1$ & $33$ & $3$ &            &            & 6.825(34) & 7.696(92)   \\
 $P( 9)$ & $A_1$ & $10$ & $1$ &            &            & 6.547(52) & 7.321(87)   \\
 $P( 9)$ & $F_1$ & $34$ & $2$ &            &            & 7.198(38) & 7.513(90)   \\
 $P( 9)$ & $E$   & $23$ & $1$ &            &            & 7.021(54) & 7.012(84)   \\
 $P( 9)$ & $F_2$ & $33$ & $1$ &            & 6.970(740) & 7.426(45) & 7.115(85)   \\
 $P( 9)$ & $F_1$ & $35$ & $1$ &            &            & 6.428(54) & 6.837(82)   \\
 $P( 7)$ & $F_1$ & $26$ & $2$ &            & 6.980(679) & 6.929(34) & 7.744(92)   \\
 $P( 7)$ & $E$   & $17$ & $1$ &            & 5.490(200) & 6.006(38) & 6.603(79)   \\
 $P( 7)$ & $F_2$ & $24$ & $2$ &            & 7.180(590) & 7.225(45) & 8.136(97)   \\
 $P( 7)$ & $A_2$ & $10$ & $1$ &            & 7.080(380) & 7.076(45) & 7.660(91)   \\
 $P( 7)$ & $F_2$ & $25$ & $1$ &            & 7.820(640) & 7.623(55) & 7.835(93)   \\
 $P( 7)$ & $F_1$ & $27$ & $1$ &            & 7.530(750) & 7.526(44) & 7.666(91)   \\
 $P( 6)$ & $A_2$ & $6$  & $1$ &            & 7.760(190) &           & 7.825(93)   \\
 $P( 5)$ & $F_1$ & $18$ & $2$ &            &            & 7.685(44) & 8.125(97)   \\
 $P( 5)$ & $E$   & $12$ & $1$ &            &            & 7.763(34) & 7.640(91)   \\
 $P( 5)$ & $F_2$ & $17$ & $1$ &            &            & 8.142(44) & 8.368(100)  \\
 $P( 5)$ & $F_1$ & $19$ & $1$ &            &            & 7.578(45) & 8.258(99)   \\
 $R( 3)$ & $A_2$ & $7$  & $1$ & 9.670(139) &            &           & 8.114(97)   \\
 $R( 3)$ & $F_2$ & $15$ & $1$ & 8.970(60)  &            &           & 8.584(100)  \\
 $R( 3)$ & $F_1$ & $16$ & $1$ & 9.600(40)  &            &           & 8.587(100)  \\
\bottomrule
\\[-6pt]
\multicolumn{6}{l}{$^{\dag}$ ``qsd'' stands for ``quadratic speed dependent.''}
\end{tabular}
\end{table}
As Es-sebbar and Farooq \ccite{ess21a} and Vispoel \textit{et al.} \ccite{vis19a} used several line shape models, the measurements performed using the profile closest to the one used in the present work were selected. Considering the uncertainties, the CO$_2$ broadening coefficients from Vispoel \textit{et al.} \ccite{vis19a} generally agree although some discrepancies are observed. They are however systematically lower than the present measurements. Those from Es-sebbar and Farroq \ccite{ess21a} are closer to the present results although they also tend to be lower than the present results. This trend towards lower values of these two studies most probably results from the fact that the present analysis included both Dicke narrowing and speed dependence while these analyses only considered the former. The closer agreement of the present measurements with the results of Es-sebbar and Farroq \ccite{ess21a} may originate from the fact that the use of the Galatry profile leads to broadening coefficients larger than the Rautian line shape model \ccite{mon05a, vis19a}. The CO$_2$ broadening coefficients measured by Vispoel \textit{et al.} \ccite{vis19a} using the Galatry profile are indeed on average 2.6 \% higher than measured using the Rautian profile. The measurements from Manne \ccite{man17a} show the opposite effect by not considering the Dicke narrowing. % Even so, the measurements seems quite different but this can probably be attributed to our bigger pressure range having more constraint on the fit.

\begin{figure}[!htb]
\centering
\includegraphics[width = 0.5\linewidth]{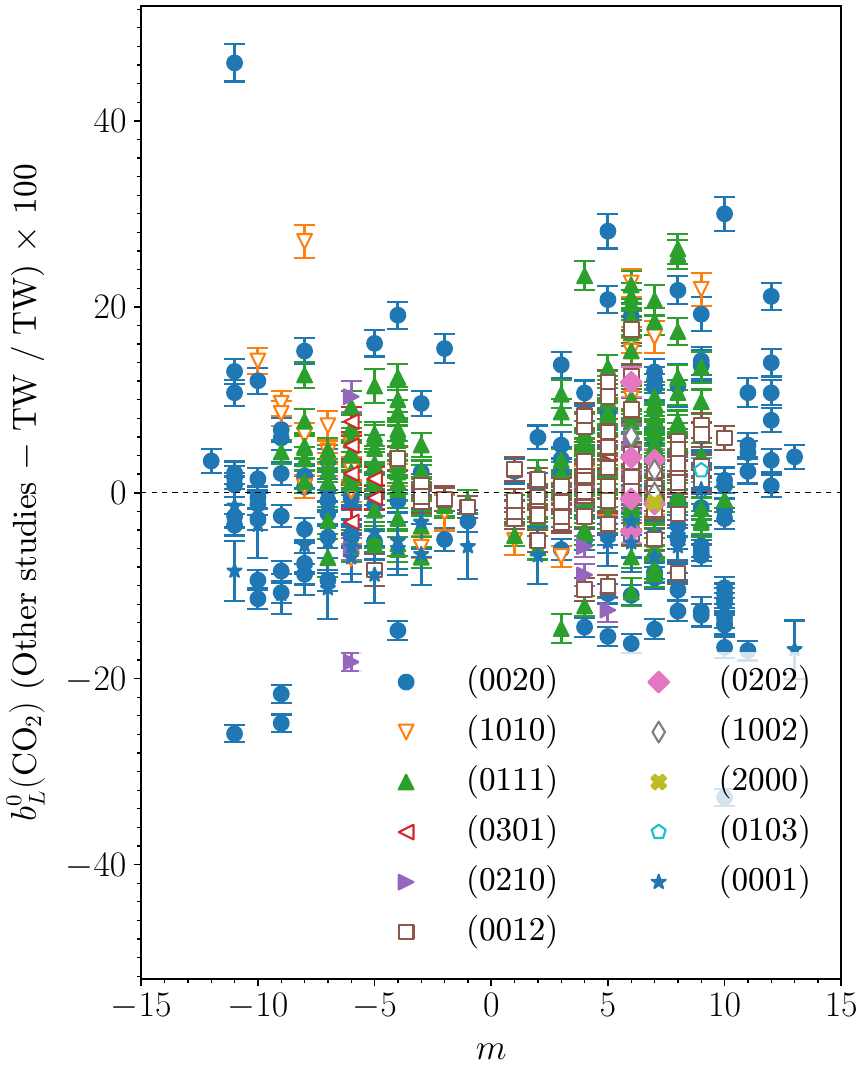}
\caption{Differences with this work (TW) of CO$_2$ broadening coefficients reported for various cold bands of $^{12}$CH$_4$. The bands are identified with the upper vibrational level involved, \textit{i.e.} $(v_1v_2v_3v_4)$ where $v_i$ is the vibrational quantum number associated with the mode of vibration $i$. All the measurements are from Lyulin \textit{et al.} \ccite{lyu14a}, except for the $\nu_4$ band \ccite{fis14a}.}
\label{fig:y0_allbands}
\end{figure}

Figure \ref{fig:y0_allbands} compares the present measurements for the $\nu_3$ band with measurements reported for the $\nu_4$ band \ccite{fis14a} and 10 cold bands observed in the $5550-6140$ cm$^{-1}$ range, involving changes of the vibrational quantum numbers up to 4 \ccite{lyu14a}. No obvious vibrational dependence can be put forward.
 % Differences between our measurements and those from other bands (taken from references listed in \cref{table:ow}) are shown in \ref{fig:y0_allbands}. Transitions emerging from the same lower levels are compared here but do not present a clear sign of vibrational dependence. Furthermore, none of the references used speed-dependence, so an accurate answer would be impossible.

\subsection{Speed dependence of broadening}\label{sec:sd_broadening}

\begin{figure}[!htb]
\centering
\includegraphics[width = \linewidth]{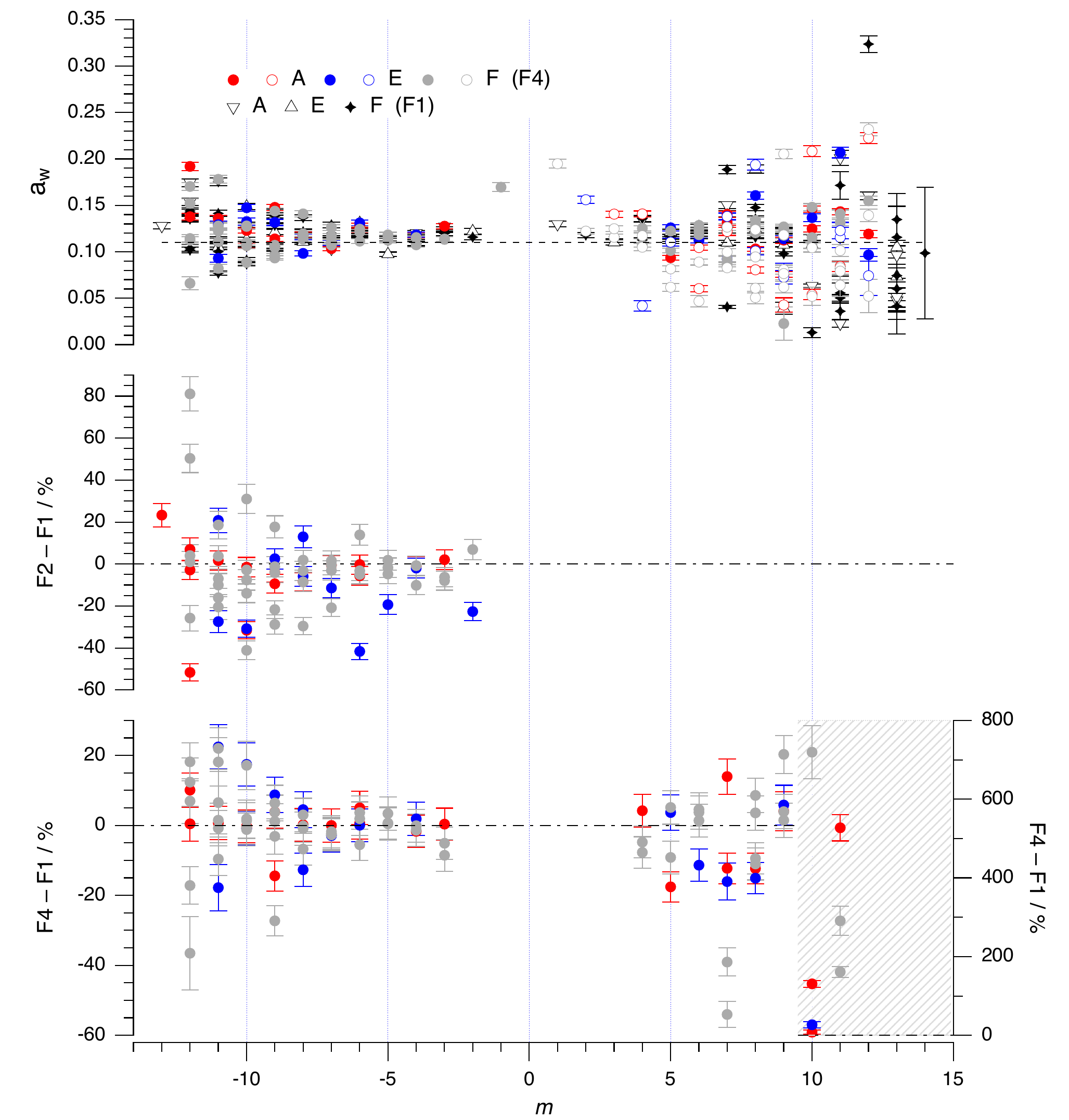} \\
\caption{Speed-dependence of the broadening measured at 296.5 K in the P ($m = -J''$), Q ($m = J''$) and R ($m = J'' + 1$) branches of the $\nu _3$ band of $^{12}$CH$_4$ with line mixing modeled using the relaxation matrix formalism (fit F4; colored symbols, filled for P and R branch lines and open for Q branch lines) and the first order approximation (fit F1, black symbols) (upper panel); differences with the speed dependence coefficients measured in fit F1 of the values retrieved in fits F2 and F4 are presented in the lower 2 panels. In the lower panel, the points over the hatched area refer to the right axis. The error bars represent the estimated uncertainties (see section \ref{sc:results} for details).}
\label{fig:sd}
\end{figure}
The speed dependence of line broadening coefficients $\aw$ measured in this work in fits F1 and F4 are presented in the upper panel of Fig. \ref{fig:sd}, together with the average value $\aw = 0.110$ determined using a quadratic speed dependent Voigt profile to fit spectra at total pressures from 400 to 803 hPa (fit F2; see section \ref{sc:analysis}). Differences of the $\aw$ coefficients measured in fits F2 and F4 with the values retrieved from fit F1 are presented in the lower 2 panels. The averages of these differences are $-4.2 \, (20.)$ \% and $1.3 \, (21.)$ \% for fits F2 and F4 ($m < 10$), respectively. The differences presented in the lower panel of Fig. \ref{fig:sd} are significantly larger for $m \geq 10$. These large differences may result from the imperfect modeling of the corresponding congested manifolds provided by the first order line mixing model (fit F1), as already discussed in section \ref{sc:broadening}. However, the speed dependence coefficients measured in the Q branch in fit F4 exhibit a spread similar to the parameters retrieved in the R branch in fits F1 and F4 (see upper panel of Fig. \ref{fig:sd}) that may indicate that \textcolor{red}{this spread has a physical origin}.

Although the $\aw$ coefficients presented in the upper panel of Fig. \ref{fig:sd} seem to be on average larger in the P branch than in the R and Q branches, the spread of the data prevents any definite conclusion on a possible rotational dependence to be drawn. Additionally, the data do not exhibit any obvious dependence on the tetrahydral symmetry, unlike the broadening coefficients. As already indicated, the range of $\aw$ values obtained in this work is similar to what is reported in the literature for methane and other molecules \ccite{pin19a, dev15a, dev12a, devi07a, bui14a, oku23a}. The few values measured for $|m| \leq 2$ in the Q branch are high compared to neighboring data. This is likely due to the combined effects of the lower intensity of the corresponding lines and correlations between speed dependence, narrowing and residual line mixing.

The $\aw$ coefficients determined in fit F2 [from which the average value $\aw = 0.110\,(5)$ was determined] are on average $4-5$ \% smaller than in the other 2 fits, as highlighted by the negative average differences mentioned here above. This may indicate that narrowing was underestimated in fits F1 and F4 and speed dependence consequently overestimated, implying that the two parameters are still correlated with the range of pressures used in the present work. This point is further discussed in section \ref{sc:narrowing}.

% The values obtained in the $R$ branch show slightly less dispersions when using the relaxation-matrix model, with a mean difference between $F1$ and $F4$ of about $7\%$. This seems to indicate that some of the misrepresented line mixing was filled with a combination of narrower and wider lines. However, each parameter did not converge properly as the uncertainty grew more and more at higher energies, so this has to be interpreted with caution. 

Table \ref{table:ow_sd} compares the speed dependence of broadening coefficients $\aw$ measured in fits F1 and F4 with values reported by Manne \textit{et al.} \ccite{man17a}, retrieved using a quadratic speed dependent Voigt profile with first order line mixing. The latter are about an order of magnitude smaller than determined in the present work.
\begin{table}[!htb]
\small
\centering
\caption{Speed dependence of the broadening $\aw$ (unitless) for the R(3) manifold of the $\nu_3$ band of $^{12}$CH$_4$ measured in fits F1 and F4 of this work (TW) and reported by Manne \textit{et al.} \ccite{man17a}. The numbers between parentheses are the uncertainties, given in the units of the last digits quoted.} \vspace{12pt}
\label{table:ow_sd}
% \input{tables/sd_comp.tex}
% \begin{tabular}{lccS[table-format=1.7, table-number-alignment=right]S[table-format=1.7, table-number-alignment=right]}
\begin{tabular}{lccD{.}{.}{1.6}D{.}{.}{2.9}D{.}{.}{2.9}}
\toprule
$C''$ & $\alpha'$ & $\alpha''$ & \multicolumn{1}{c}{\ccite{man17a}} & \multicolumn{1}{c}{TW (F4)} & \multicolumn{1}{c}{TW (F1)} \\
\midrule
 $A_2$ & $7$  & $1$ & 0.012\,(7) & 0.1407\,(32) & 0.1350\,(31) \\
 $F_2$ & $15$ & $1$ & 0.008\,(7) & 0.1253\,(29) & 0.1358\,(32) \\
 $F_1$ & $16$ & $1$ & 0.01\,(3)  & 0.1074\,(25) & 0.1128\,(26) \\
\bottomrule
\end{tabular}
\end{table}
The origin of this discrepancy is unclear. It may come from the different ranges of total pressures used in the two studies, \textit{i.e.} up to 803 hPa in the present work and up to 200 hPa in \ccite{man17a}. The rather large uncertainties characterizing the values reported in \ccite{man17a} possibly indicate that the range of total pressures was too low.

\subsection{Dicke narrowing}\label{sc:narrowing}

The Dicke narrowing coefficients $\beta^0$ were measured for several lines of the P and R branches of the $\nu_3$ band of $^{12}$CH$_4$ in fits F1 and F3, and for the 3 branches in fit F4. They seemed to exhibit a rotational dependence, whereas $\beta^0$ is expected to be constant for the uncorrelated hard collision line shape model \ccite{rau67a} because collisional broadening and the Doppler effect can be considered as statistically independent. Although it may only be apparent, resulting from the increasing dispersion of the measured values that follows the increasing number of lines per manifold as $J$ becomes larger, the observation of such a rotational dependence was interpreted as resulting from the correlation between the measured Dicke narrowing coefficients and speed dependence of broadening, even though their contributions to the observed spectra dominate in different pressure regimes (\textit{i.e.} intermediate pressures for the former and higher pressures for the latter). Fit F3 provided another indication of the existence of such a correlation. Indeed, it involved decorrelating $\beta^0$ and $\aw$ by fixing the latter. The spread of the measured narrowing coefficients was found to be similar to fits F1 and F4, but any resemblance of a $J''$ dependence was absent.

The rough average of the Dicke narrowing coefficients measured in the present work is $\langle \beta^0 \rangle \approx 0.005$ cm$^{-1}$atm$^{-1}$. To the best of our knowledge, narrowing coefficients were only reported twice for the $\nu_3$ band of methane perturbed by CO$_2$, from measurements carried out using the Galatry profile in spectra recorded at $297\,(1)$ K. Performed in the P(11) manifold observed at total pressures up to 101 torr, these measurements yielded an average narrowing coefficient of $\langle \beta^0 \rangle = 0.02724\,(43)$ cm$^{-1}$atm$^{-1}$ \ccite{ess14a}. A more precise average of $\langle \beta^0 \rangle = 0.02934\,(18)$ cm$^{-1}$atm$^{-1}$ was recently reported by the same authors, from a more extensive study of the P branch relying on spectra recorded at total pressures between 4 and 201 torr  \ccite{ess21a}.

The rather small Dicke narrowing coefficients obtained in the present work together with the possible correlation with speed dependence were interpreted as an indication that the present measurements did not rely on enough spectra corresponding to total pressures below about 200 hPa. Indeed, Fig. \ref{fig:prof} shows that Dicke narrowing effects are largest in that range of pressures. Hence, the narrowing coefficients measured in the present work should probably be considered as fitting parameters only and their values are therefore not reported.

\subsection{\texorpdfstring{CO$_2$ shift coefficients}{CO2 shift coefficients}}

\begin{figure}[!htb]
\centering
\includegraphics[width = 0.5\linewidth]{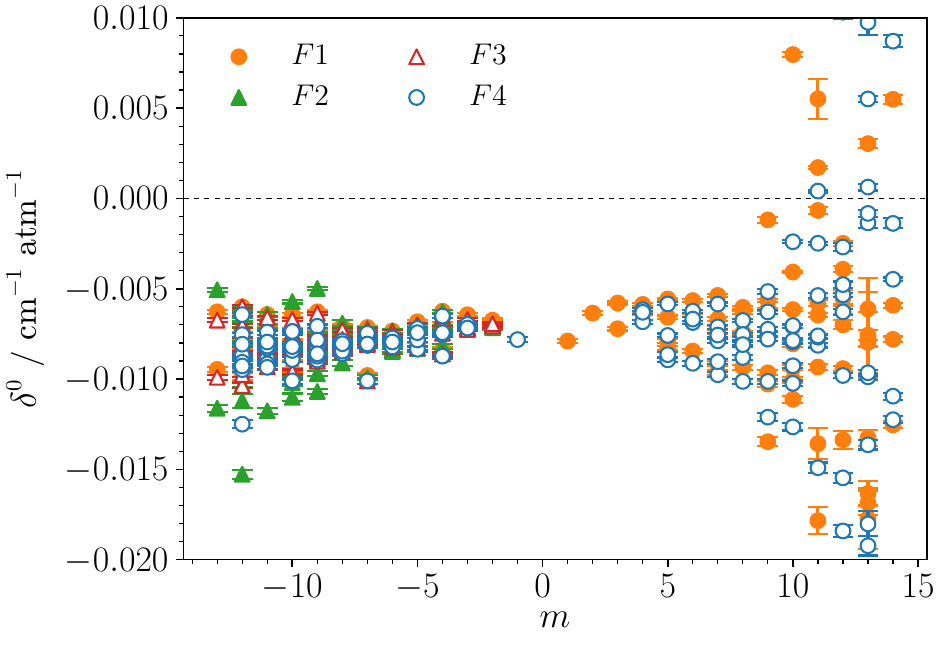}\label{fig:delta:delta} \\
\includegraphics[width = 0.5\linewidth]{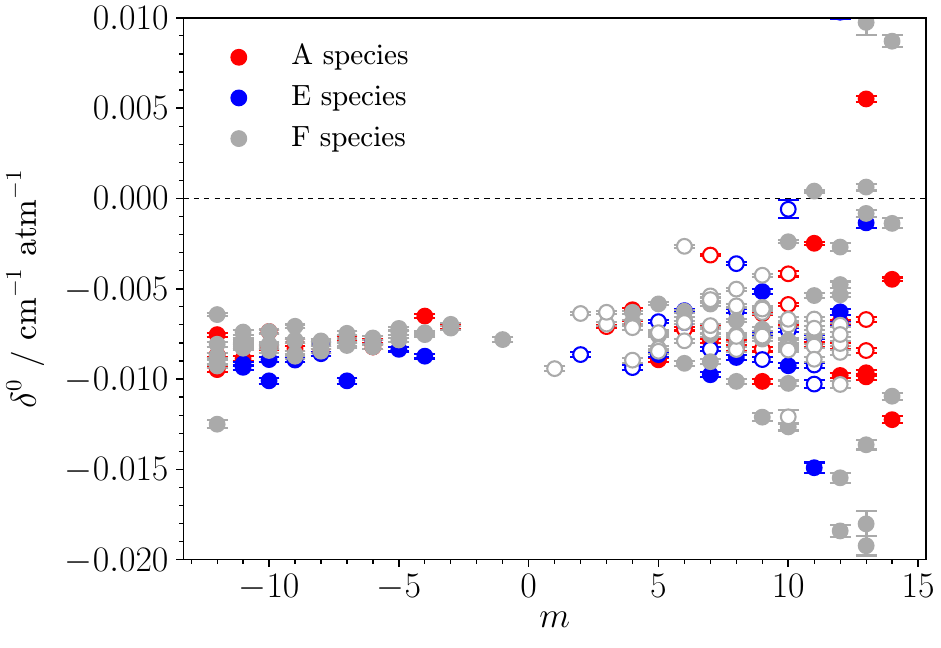} \label{fig:delta:delta_rm}
\caption{CO$_2$ shift coefficients measured at 296.5 K in the P ($m = -J''$) and R ($m = J'' + 1$) branches of the $\nu_3$ band of $^{12}$CH$_4$ through the 4 fits described in Table \ref{table:analyses_fit} (upper panel) and results obtained using the relaxation matrix formalism (fit F4) for the P, Q ($m = J''$, open symbols) and R branches (lower panel).}
\label{fig:delta}
\end{figure}
The CO$_2$ shift coefficients measured in fits F1 to F4 are presented in the upper panel of Fig. \ref{fig:delta}. To the best of our knowledge, this is the first time that these coefficients are measured for the $\nu_3$ band of methane. The spread of the reported shift parameters is larger in the R branch, particularly for $m \geq 10$. As already noted for the speed dependence coefficients and because it is observed for fits F1 and F4, it is unclear whether this trend is real or results from some imperfect modeling of line mixing in the corresponding congested manifolds.

The absolute value of the shift coefficients resulting from fit F1 clearly appear to be systematically smaller than those obtained from the other 3 fits, by about $4-5$ \%. The origin of this difference is unclear. It could come from correlations between the CO$_2$ shift and first order line mixing coefficients, but none was observed. Figure \ref{fig:delta} also shows that the more limited range of pressures characterizing fit F2 had visible consequences on the precision of measurement of the shift coefficients. This observation highlights the sensitivity of this parameter to the experimental information available.

% THIBAULT sur ``The origin of this difference is unclear'' (15/12/2023): Je suis quasiment sûr que c'est un problème de sensibilité aux conditions initiales du x_calibration_factor

% The residual shift of the line intensity originating from neglected line-mixing can easily be misinterpreted as pressure shift by the algorithm. This resulted in an averaged value shifted downward in $F4$ compared to $F1$. It is more pronounced in the $R$ branch ($4\%$ in $P$ and $7\%$ in $R$) as the couplings are stronger. The resulting better symmetry between the branches tends to confirm the improvement from using a more complete line-mixing profile.

% Ma question: "Cette explication ne fonctionne pas pour F2 et F3. Pourquoi les shifts de F2 et F3 sont plus grands en valeur absolue que F1?"

% THIBAULT (10/12/2023): "Je n'ai pas de réponse définitive pour ça mais étrangement, on trouve l'opposé dans la figure 11. Les gamma 2 sont plus grand pour F1 (et F4) que F2, ce qui semble confirmer le problème de corrélation avec le narrowing. Mon hypothèse sur le shift est que la largeur de raie plus grande de F2 et F3 (due à gamma 2 plus petit) masque une partie du mélange de raie, ce qui "corrige" le fit du shift."

The CO$_2$ shift coefficients resulting from fit F4 with distinction between $A$, $E$ and $F$ species are presented in the lower panel of Fig. \ref{fig:delta}. In the P branch, the lines of $E$ symmetry seem to undergo more shift than the other lines. This posssible trend is however not visible in the other branches, preventing any definite conclusion to be drawn.

% The Q branch like the first-order R, has a slightly higher average than P. The cause, as was mentioned above, is the wrong representation of line-mixing, by the approximation in the previous case, and remaining couplings here.

\subsection{First order line mixing model}

\begin{figure}[!htb]
\centering
% *** Figure generated in 'manuscript/supmat/supmat.pxp' 'response_to_review/line_mixing/mixed_lines/mixed-y_lines.pxp' ***
\includegraphics[width = \linewidth]{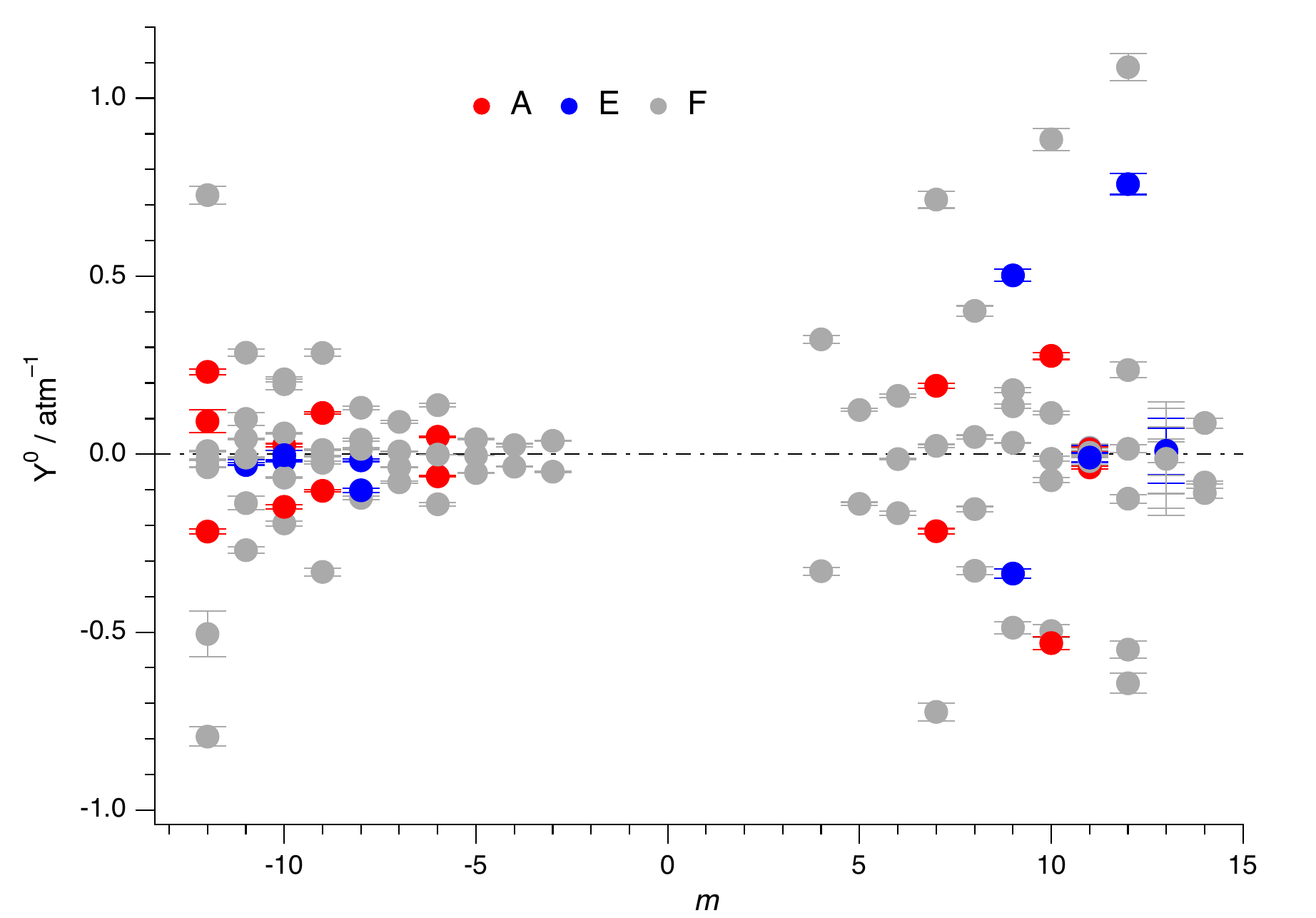}
\caption{First order line mixing coefficients measured at 296.5 K in the P ($m = -J''$) and R ($m = J'' + 1$) branches of the $\nu_3$ band of $^{12}$CH$_4$ in fit F1.}
\label{fig:lmfo}
\end{figure}
\textcolor{red}{Table \ref{table:y0} lists the first order line mixing coefficients $Y_n^0$ measured in fit F1 for 111 lines of the $\nu_3$ band of $^{12}$CH$_4$. They are presented in Fig. \ref{fig:lmfo}. Table \ref{table:y0} includes two additional $Y_n^0$ values, associated with lines of the $\nu_2+\nu_4$ ($F_2$) band and also fitted to improve the residuals. The $Y_n^0$ coefficients of all the other lines were set to zero.} Fits F2 and F3 yield very similar results for the P branch and are therefore not presented. As also observed for the $\nu_3$ band of $^{12}$CH$_4$ perturbed by air \ccite{far21a}, most line mixing coefficients are in the range from $-0.5$ to $0.5$ atm$^{-1}$.
\small
\begin{longtable}{lcD{.}{.}{2.3}D{.}{.}{1.8}D{.}{.}{4.8}c}
\caption{First order line mixing coefficients $Y^0_n$(F1) and $Y^0_n$(F4) (in atm$^{-1}$) for the $\nu_3$ band of $^{12}$CH$_4$ measured in fit F1 and calculated using Eq. \ref{eq:y_from_w}, respectively. The number provided between parentheses after each value is the estimated uncertainty, in the units of the last digit quoted. Its absence for $Y^0_n$(F4) indicates that it is larger than 0.1 or the value of $Y^0_n$(F4). The identification, position $\tilde{\nu}_n$ (in cm$^{-1}$) and intensity $I_n$ [in $10^{-21}$ cm$^{-1}$/(molecule cm$^{-2}$)] of the lines are from the \texttt{HITRAN} database \ccite{gor20a}. Two lines (identified by $\dag$) belong to the $\nu_2+\nu_4$ ($F_2$) band. In the column labeled ``$W_{mn}$,'' the first (\textit{resp.} second) pair of digits $i$:$j$ refers to the $A_1 \leftrightarrow A_2$, $E \leftrightarrow E$ and $F_1 \leftrightarrow F_2$ [\textit{resp.} $(A \ \mathrm{or} \ F)_1 \leftrightarrow (A \ \mathrm{or} \ F)_1$ and $(A \ \mathrm{or} \ F)_2 \leftrightarrow (A \ \mathrm{or} \ F)_2$] couplings: $i$ is the number of couplings considered in fit F4 and $j$ is the total number of possible couplings involving line $n$ in the $J''$ manifold it belongs to.}
\label{table:y0}\\
\toprule
Line $n$ & \multicolumn{1}{c}{$\tilde{\nu}_n$} & \multicolumn{1}{c}{$I_n$} & \multicolumn{1}{c}{$Y^0_n$(F1)} & \multicolumn{1}{c}{$Y^0_n$(F4)} & \multicolumn{1}{c}{$W_{mn}$} \\
\midrule
\endfirsthead
\caption[]{(Continued)}\\
\toprule
Line $n$ & \multicolumn{1}{c}{$\tilde{\nu}_n$} & \multicolumn{1}{c}{$I_n$} & \multicolumn{1}{c}{$Y^0_n$(F1)} & \multicolumn{1}{c}{$Y^0_n$(F4)} & \multicolumn{1}{c}{$W_{mn}$} \\
\midrule
\endhead
\multicolumn{6}{r}{{Continued on next page}}
\endfoot
\endlastfoot
P(12, $A_1$, 2, 15)         & 2894.9961 &  11.09  &  0.2314\,(79)   &   0.2123\,(78)  & 1:1/0:1 \\  %  8.2 %
P(12, $F_1$, 3, 43)         & 2895.0560 &   6.618 &  0.7277\,(250)  &   0.6446\,(238) & 1:3/0:2 \\  % 11.4 %
P(12, $F_2$, 3, 45)         & 2895.1288 &   6.561 & -0.7930\,(270)  &  -0.6502\,(240) & 1:3/0:2 \\  % 18.0 %
P(12, $A_2$, 1, 14)         & 2895.2333 &  11.02  & -0.2172\,(74)   &  -0.2138\,(79)  & 1:2/0:0 \\  %  1.6 %
P(12, $F_2$, 2, 46)         & 2895.7573 &   6.439 &  0.0089\,(20)   &                 &         \\
P(12, $F_1$, 2, 44)         & 2896.2026 &   6.521 & -0.0367\,(15)   &                 &         \\
P(12, $F_2$, 1, 47)         & 2896.2999 &   6.523 & -0.0158\,(28)   &                 &         \\
P(12, $F_1$, 1, 45)         & 2896.9801 &   6.696 & -0.5051\,(640)  &                 &         \\
P(12, $A_1$, 1, 16)         & 2896.9902 &  11.16  &  0.0922\,(320)  &                 &         \\
P(11, $F_1$, 3, 41)         & 2905.6337 &  11.32  &  0.2852\,(97)   &   0.2626        & 3:3/2:2 \\  %  7.9 %
P(11, $E$, 2, 27)           & 2905.6975 &   7.486 & -0.0191\,(34)   &                 &         \\
P(11, $F_2$, 3, 39)         & 2905.8137 &  11.40  & -0.2694\,(91)   &  -0.2514        & 3:3/2:2 \\  %  6.7 %
P(11, $F_1$, 2, 42)         & 2906.2825 &  11.06  &  0.0435\,(15)   &   0.0459        & 3:3/2:2 \\  %  5.5 %
P(11, $E$, 1, 28)           & 2906.5888 &   7.389 & -0.0303\,(18)   &                 &         \\
P(11, $F_2$, 2, 40)         & 2906.6476 &  11.10  & -0.0093\,(19)   &  -0.0230        & 3:3/2:2 \\  % ****  
P(11, $F_2$, 1, 41)         & 2907.3213 &  11.41  & -0.1372\,(190)  &   0.0271        & 3:3/2:2 \\  % ****  
P(11, $F_1$, 1, 43)         & 2907.3361 &  11.41  &  0.0991\,(180)  &  -0.0587        & 3:3/2:2 \\  % ****  
P(10, $F_2$, 3, 37)         & 2916.2016 &  18.36  &  0.2107\,(72)   &   0.1972\,(73)  & 1:2/0:2 \\  %  6.4 %
P(10, $E$, 2, 24)           & 2916.3018 &  12.07  & -0.0190\,(33)   &                 &         \\
P(10, $F_1$, 2, 36)         & 2916.3963 &  17.87  & -0.1944\,(66)   &  -0.1914\,(79)  & 3:3/0:1 \\  %  1.5 %
P(10, $A_1$, 1, 14)         & 2916.7538 &  29.62  &  0.0257\,(44)   &   0.0269\,(10)  & 1:1/0:0 \\  %  4.7 %
P(10, $F_1$, 1, 37)         & 2916.9662 &  17.70  &  0.0584\,(20)   &   0.0595\,(23)  & 3:3/0:1 \\  %  1.9 %
P(10, $F_2$, 2, 38)         & 2917.0661 &  17.88  & -0.0659\,(23)   &  -0.0643\,(24)  & 1:2/1:2 \\  %  2.3 %
P(10, $A_2$, 1, 12)         & 2917.6291 &  30.55  & -0.1477\,(69)   &  -0.0261\,(10)  & 1:1/0:0 \\  % ****  
P(10, $F_2$, 1, 39)         & 2917.6526 &  18.31  &  0.1971\,(150)  &  -0.0057\,(3)   & 1:2/0:2 \\  % ****  
P(10, $E$, 1, 25)           & 2917.6628 &  12.22  & -0.0027\,(140)  &                 &         \\
P(9, $A_2$, 1, 12)          & 2926.7002 &  45.83  &  0.1161\,(39)   &   0.1020\,(38)  & 1:1/0:0 \\  % 12.1 %
P(9, $F_2$, 2, 32)          & 2926.7830 &  27.36  &  0.2844\,(97)   &   0.2954\,(109) & 2:3/0:1 \\  %  3.9 %
P(9, $F_1$, 3, 33)          & 2926.8851 &  27.09  & -0.3304\,(110)  &  -0.2984\,(110) & 1:2/0:2 \\  %  9.7 %
P(9, $A_1$, 1, 10)          & 2927.0762 &  45.46  & -0.1029\,(35)   &  -0.1029\,(38)  & 1:1/0:0 \\  %  0.0 %
P(9, $F_1$, 2, 34)          & 2927.3726 &  26.82  &  0.0123\,(21)   &   0.0196\,(7)   & 1:2/1:2 \\  % ****  
P(9, $F_2$, 1, 33)          & 2927.9321 &  27.65  & -0.0060\,(20)   &  -0.0028        & 2:3/0:1 \\  % ****  
P(9, $F_1$, 1, 35)          & 2927.9636 &  27.65  & -0.0240\,(44)   &  -0.0163\,(31)  & 1:2/0:2 \\  % ****  
P(8, $F_2$, 2, 30)          & 2937.2346 &  38.80  &  0.1307\,(45)   &   0.1215\,(45)  & 2:2/0:1 \\  %  7.0 %
P(8, $E$, 2, 19)            & 2937.3082 &  25.71  & -0.0173\,(30)   &   0.0005        & 1:1/0:0 \\  % ****  
P(8, $F_1$, 2, 28)          & 2937.4950 &  38.53  & -0.1229\,(42)   &  -0.1173\,(54)  & 2:2/0:1 \\  %  4.6 %
P(8, $F_2$, 1, 31)          & 2937.7672 &  38.17  &  0.0154\,(26)   &   0.0179\,(18)  & 2:2/1:1 \\  % 16.2 %
P(8, $E$, 1, 20)            & 2938.1926 &  26.06  & -0.1015\,(55)   &  -0.0005        & 1:1/0:0 \\  % ****  
P(8, $F_1$, 1, 29)          & 2938.2154 &  39.10  &  0.0407\,(32)   &  -0.0225\,(9)   & 1:2/0:1 \\  % ****  
P(7, $F_1$, 2, 26)          & 2947.6680 &  51.28  &  0.0908\,(31)   &   0.0938\,(52)  & 2:2/1:1 \\  %  3.2 %
P(7, $F_2$, 2, 24)          & 2947.9121 &  50.69  & -0.0787\,(27)   &  -0.0711\,(150) & 2:2/1:1 \\  %  9.7 %
P(7, $F_2$, 1, 25)          & 2948.4214 &  51.49  &  0.0079\,(14)   &   0.0111        & 2:2/1:1 \\  % ****  
P(7, $F_1$, 1, 27)          & 2948.4741 &  51.71  & -0.0359\,(13)   &  -0.0344\,(131) & 2:2/1:1 \\  %  4.2 %
P(6, $A_1$, 1, 8)           & 2958.0173 & 105.2   &  0.0488\,(17)   &   0.0545\,(20)  & 1:1/0:0 \\  % 11.6 %
P(6, $F_1$, 1, 21)          & 2958.1200 &  62.73  &  0.1378\,(47)   &   0.1355\,(49)  & 1:2/0:0 \\  %  1.7 %
P(6, $F_2$, 2, 22)          & 2958.2329 &  62.26  & -0.1407\,(48)   &  -0.1337\,(54)  & 1:1/1:1 \\  %  5.0 %
P(6, $A_2$, 1, 6)           & 2958.5364 & 105.7   & -0.0623\,(21)   &  -0.0543\,(20)  & 1:1/0:0 \\  % 12.8 %
P(6, $F_2$, 1, 23)          & 2958.6508 &  63.15  & -0.0009\,(3)    &  -0.0027\,(4)   & 1:1/0:1 \\  % ****  
P(5, $F_1$, 2, 18)          & 2968.4033 &  70.85  &  0.0423\,(14)   &   0.0495\,(18)  & 1:1/0:1 \\  % 16.8 %
P(5, $F_2$, 1, 17)          & 2968.7362 &  71.41  & -0.0530\,(18)   &  -0.0469\,(22)  & 2:2/0:0 \\  % 11.5 %
P(5, $F_1$, 1, 19)          & 2968.8552 &  71.30  & -0.0035\,(6)    &  -0.0022\,(4)   & 1:1/0:1 \\  % ****  
R(7, $F_2$, 1, 25)          & 2971.0753 &   1.956 &  0.0257\,(47)   &                 &         \\  
R(7, $F_1$, 1, 26)          & 2971.2826 &   1.995 & -0.0824\,(33)   &                 &         \\  
P(4, $F_2$, 1, 15)          & 2978.6505 &  72.12  &  0.0256\,(44)   &   0.0271\,(10)  & 1:1/0:0 \\  %  6.0 %
P(4, $F_1$, 1, 13)          & 2978.9199 &  72.54  & -0.0351\,(12)   &  -0.0270\,(10)  & 1:1/0:0 \\  % 23.1 %
P(3, $F_2$, 1, 9)           & 2988.9325 &  64.33  &  0.0375\,(13)   &   0.0524\,(19)  & 1:1/0:0 \\  % ****  
P(3, $F_1$, 1, 11)          & 2989.0335 &  64.53  & -0.0493\,(17)   &  -0.0522\,(19)  & 1:1/0:0 \\  %  6.0 % 
R(3, $F_2$, 1, 15)          & 3057.7265 & 126.0   &  0.3227\,(110)  &   0.3250\,(123) & 1:1/0:0 \\  %  0.7 %
R(3, $F_1$, 1, 16)          & 3057.7607 & 126.0   & -0.3284\,(110)  &  -0.3250\,(123) & 1:1/0:0 \\  %  1.0 %
R(4, $F_2$, 1, 20)          & 3067.1642 & 127.4   &  0.1242\,(42)   &   0.1127\,(41)  & 1:1/0:0 \\  %  9.2 %
R(4, $F_1$, 1, 18)          & 3067.2611 & 126.6   & -0.1392\,(47)   &  -0.1135\,(41)  & 1:1/0:0 \\  % 18.5 %
R(5, $F_1$, 2, 23)          & 3076.5496 & 117.9   &  0.1638\,(56)   &   0.1627\,(59)  & 2:2/0:1 \\  %  0.6 %
R(5, $F_2$, 1, 21)          & 3076.6770 & 116.6   & -0.1656\,(56)   &  -0.1529\,(78)  & 2:2/0:1 \\  %  7.6 %
R(5, $F_1$, 1, 24)          & 3076.7252 & 117.0   & -0.0138\,(24)   &  -0.0116\,(18)  & 2:2/0:1 \\  % 16.2 %
R(6, $A_1$, 1, 10)          & 3085.8322 & 169.3   &  0.1922\,(66)   &   0.1712\,(63)  & 1:2/0:0 \\  % 10.9 %
R(6, $F_1$, 1, 25)          & 3085.8607 & 101.3   &  0.7153\,(240)  &   0.6799\,(251) & 2:3/0:0 \\  %  5.0 %
R(6, $F_2$, 2, 26)          & 3085.8935 & 101.1   & -0.7238\,(250)  &  -0.6518\,(238) & 1:1/1:2 \\  %  9.9 %
R(6, $A_2$, 1, 8)           & 3086.0308 & 165.9   & -0.2162\,(74)   &  -0.1747\,(64)  & 1:1/0:1 \\  % 19.2 %
R(6, $F_2$, 1, 27)          & 3086.0717 & 100.0   &  0.0234\,(41)   &  -0.0298\,(14)  & 1:1/1:2 \\  % ****  
R(7, $F_1$, 2, 29)          & 3095.0607 &  81.50  &  0.4024\,(140)  &   0.3673\,(147) & 2:3/0:2 \\  %  8.7 %
R(7, $F_2$, 2, 28)          & 3095.1307 &  80.95  & -0.3273\,(110)  &  -0.2724\,(134) & 2:3/0:2 \\  % 16.8 %
R(7, $F_2$, 1, 29)          & 3095.3511 &  79.61  &  0.0488\,(51)   &   0.0377\,(120) & 2:3/0:2 \\  % 22.6 %
R(7, $F_1$, 1, 30)          & 3095.3710 &  79.83  & -0.1543\,(71)   &  -0.1365\,(106) & 2:3/0:2 \\  % 11.6 %
R(8, $F_2$, 2, 33)          & 3104.2054 &  61.33  &  0.1801\,(66)   &   0.4570\,(172) & 3:3/0:1 \\  % ****  
R(8, $E$, 2, 21)            & 3104.2206 &  40.81  &  0.5026\,(170)  &   0.0535\,(38)  & 1:2/0:0 \\  % ****  
R(8, $F_1$, 2, 31)          & 3104.2840 &  60.75  & -0.4873\,(170)  &  -0.4056\,(246) & 2:2/1:2 \\  % 16.8 %
R(8, $F_2$, 1, 34)          & 3104.3365 &  60.81  &  0.0323\,(13)   &  -0.0012        & 3:3/0:1 \\  % ****  
R(8, $E$, 1, 22)            & 3104.5690 &  39.67  & -0.3352\,(130)  &  -0.0551\,(40)  & 1:2/0:0 \\  % ****  
R(8, $F_1$, 1, 32)          & 3104.5749 &  59.56  &  0.1345\,(61)   &  -0.0557\,(75)  & 2:2/1:2 \\  % ****  
R(9, $A_2$, 1, 13)          & 3113.2615 &  72.54  &  0.2761\,(100)  &   0.3680\,(135) & 1:2/0:0 \\  % ****  
R(9, $F_2$, 2, 34)          & 3113.2798 &  43.35  & -0.4968\,(180)  &   1.826\,(757)  & 2:6/0:2 \\  % ****  
R(9, $F_1$, 3, 36)          & 3113.3012 &  41.72  &  0.8844\,(310)  &  -1.757\,(779)  & 1:3/1:5 \\  % ****  
R(9, $A_1$, 1, 12)          & 3113.3803 &  71.52  & -0.5304\,(180)  &  -0.3733\,(137) & 1:1/0:1 \\  % 29.6 %
R(9, $F_1$, 2, 37)          & 3113.4174 &  41.33  &  0.1158\,(54)   &  -0.0653\,(261) & 1:3/2:5 \\  % ****  
R(9, $F_2$, 1, 35)          & 3113.7073 &  41.82  & -0.0133\,(66)   &   3.758         & 2:6/0:2 \\  % ****  
R(9, $F_1$, 1, 38)          & 3113.7119 &  41.80  & -0.0730\,(66)   &  -3.836         & 1:3/0:5 \\  % ****  
R(10, $F_2$, 3, 39)         & 3122.2575 &  28.65  &  0.0183\,(84)   &   0.8407\,(317) & 2:4/0:3 \\  % ****  
R(10, $E$, 2, 26)           & 3122.2853 &  19.13  &  0.0140\,(100)  &                 &         \\
R(10, $F_1$, 2, 38)         & 3122.2962 &  23.98  &  0.0137\,(150)  &  -0.6307\,(297) & 1:4/1:3 \\  % ****  
R(10, $A_1$, 1, 14)         & 3122.3316 &  48.37  &  0.0135\,(69)   &   0.0447\,(28)  & 1:2/0:0 \\  % ****  
R(10, $F_1$, 1, 39)         & 3122.4393 &  23.62  &  0.0044\,(130)  &  13.3           & 1:4/2:3 \\  % ****  
R(10, $F_2$, 2, 40)         & 3122.4440 &  28.16  &  0.0027\,(77)   & -11.3           & 2:4/0:3 \\  % ****  
R(10, $A_2$, 1, 12)         & 3122.7625 &  46.27  & -0.0371\,(40)   &  -0.0471\,(30)  & 1:1/0:1 \\  % 26.9 %
R(10, $E$, 1, 27)           & 3122.7647 &  18.36  & -0.0096\,(130)  &                 &         \\
R(10, $F_2$, 1, 41)         & 3122.7639 &  27.57  & -0.0194\,(110)  &  -0.1704\,(68)  & 1:4/0:3 \\  % ****  
R(11, $E$, 2, 28)           & 3131.1704 &   9.139 &  0.7590\,(290)  &                 &         \\
R(11, $F_1$, 3, 43)         & 3131.1737 &  17.12  &  1.0870\,(380)  &   1.310\,(507)  & 2:5/0:2 \\  % 20.5 %
R(11, $F_2$, 3, 41)         & 3131.2065 &  10.46  & -0.6434\,(280)  &  -1.395\,(811)  & 2:3/0:4 \\  % ****  
R(11, $F_1$, 2, 42)         & 3131.2426 &  17.37  & -0.5490\,(240)  &  -0.4394\,(236) & 2:5/0:2 \\  % 20.0 %
R(11, $F_2$, 2, 41)         & 3131.3472 &   7.526 &  0.2366\,(220)  &                 &         \\
R(11, $F_2$, 1, 43)         & 3131.7352 &  17.25  &  0.0152\,(110)  &  85.6           & 1:3/0:4 \\  % 23.8 %
R(11, $F_1$, 1, 44)         & 3131.7362 &  16.97  & -0.1249\,(120)  & -87.1           & 2:5/0:2 \\  % ****  
R(12, $F_1$, 3, 45)         & 3140.0447 &   6.262 &  0.0099\,(670)  &   3.336         & 2:6/0:4 \\  % ****  
R(12, $F_2$, 3, 48)         & 3140.0660 &   8.062 &  0.0098\,(330)  &  15.3           & 2:5/0:5 \\  % ****  
R(12, $F_2$, 2, 46)         & 3140.0686 &   9.040 &  0.0095\,(1200) & -16.0           & 2:5/0:5 \\  % ****  
R(12, $E$, 2, 30)           & 3140.0836 &   7.090 &  0.0095\,(910)  &                 &         \\
R(12, $F_1$, 2, 44)         & 3140.2204 &   6.394 & -0.0123\,(1600) &                 &         \\
R(12, $F_2$, 1, 47)         & 3140.2230 &   7.250 & -0.0116\,(1400) &                 &         \\
R(12, $E$, 1, 31)           & 3140.6222 &   6.764 &  0.0080\,(660)  &                 &         \\
R(12, $F_1$, 1, 46)         & 3140.6256 &   9.661 & -0.0132\,(470)  &                 &         \\
R(13, $F_2$, 2, 47)         & 3148.8237 &   5.224 &  0.0877\,(140)  &                 &         \\
R(13, $F_2$, 3, 49)         & 3148.8299 &   5.522 & -0.1091\,(140)  &                 &         \\
R(13, $F_2$, 1, 48)         & 3149.4217 &   5.222 & -0.0793\,(37)   &                 &         \\
\bottomrule
\end{longtable}
\normalsize

Since the elements of the relaxation matrix verify the detailed balance relation ensuring that the total population is conserved, the first order line mixing coefficients $Y_n^0$ must satisfy the following sum rule \ccite{har21a}:
\begin{equation}\label{eq:y_sum-rule}
\sum_n \rho_n(T) \, \mu_n^2 \, Y_n^0 = 0
\end{equation}
where $T$ is the temperature, $\rho_n$ is the relative population of the lower level of the transition corresponding to line $n$ (see Eq. \ref{eq:rel_pop}) and the sum runs over all the lines of a given symmetry in a given P, Q or R manifold. As the relative populations $\rho_n$ are approximately the same for all transitions in the same multiplet, the average of the coefficients measured for given $m$ and symmetry is close to zero. The first order line mixing coefficients associated with close lying doublets of lines even appear in pairs of similar absolute values and opposite signs, as can be clearly seen in Fig. \ref{fig:lmfo} for lines of $F$ symmetry.

In the P branch, large line mixing coefficients were obtained for high $J$ lines. Most probably, these result from the combined effects of the increasing density of lines in the corresponding multiplets and the decreasing intensity of these lines. The spread of the line mixing coefficients is significantly larger in the R branch as a result of its more coupled nature, leading to the failure of the first order line mixing model as observed in Figs. \ref{fig:fit_R} and \ref{fig:fit_rosen_cyuri}. Interestingly, the axial symmetry of the coefficients starts to break where signatures in the residuals start to appear (at $m = 10$ in Fig. \ref{fig:fit_R}).

Table \ref{table:ow_lmfo} compares the first order line mixing coefficients measured in the present work with the only data reported in the literature \ccite{man17a}. As the couplings are weak in the R(3) manifold, the first order approximation works very well and leads to good agreement. Contrary to the present work, Manne \textit{et al.} \ccite{man17a} did consider the $10$ times smaller line mixing for the $A_2$ line but this had little impact.
\begin{table}[!htbp]
\centering
\caption{First order line mixing coefficients $Y_n^0$ (atm$^{-1}$) for the R(3) manifold of the $\nu_3$ band of $^{12}$CH$_4$ measured in fit F1 of this work (TW) and reported by Manne \textit{et al.} \ccite{man17a}. The numbers between parentheses are the uncertainties, given in the units of the last digits quoted.} \vspace{6pt}
\label{table:ow_lmfo}
\begin{tabular}{lccD{.}{.}{1.5}D{.}{.}{3.8}}
\toprule
$C''$ & $\alpha'$ & $\alpha''$ & \multicolumn{1}{c}{\ccite{man17a}} & \multicolumn{1}{c}{TW} \\
\midrule
$A_2$ & $7$  & $1$ &  0.03\,(12)  &  0.0\,(\mathrm{fixed}) \\
$F_2$ & $15$ & $1$ &  0.39\,(9)  &  0.323\,(11) \\
$F_1$ & $16$ & $1$ & -0.35\,(12) & -0.328\,(11) \\
\bottomrule
\end{tabular}\end{table}

% More differences can be seen with the speed-dependent Voigt profile, but this is again because of the lower pressure range. Beside that, the results are consistent, as the determination of Dicke narrowing does not have much correlation to that of line mixing. The results from $F1$ with distinction between $A$, $E$, and $F$ species are presented in \cref{fig:delta:delta_rm}.

\subsection{Relaxation matrix formalism}

\textcolor{red}{Involving lines of the $\nu_3$ band of $^{12}$CH$_4$, the 114 off-diagonal relaxation matrix coefficients measured in this work are listed in Table \ref{table:wnn}. 109 coefficients couple lines stronger than $10^{-20}$ cm$^{-1}$/(molecule cm$^{-2}$). The 5 remaining parameters were considered as they improved the residuals. They involved 7 close-lying ($|\Delta \tilde{\nu}| \leq 0.1$ cm$^{-1}$) weaker ($6.6 \times 10^{-21} <$ intensity $< 9.0 \times 10^{-21}$) lines of the P, Q and R branches with $J''=12$.}
\scriptsize
\begin{longtable}{llD{.}{.}{1.7}D{.}{.}{1.3}cllD{.}{.}{1.7}D{.}{.}{1.3}}
\caption{\small Off-diagonal relaxation matrix coefficients $W_{mn}$ (in $10^{-2}$ cm$^{-1}$atm$^{-1}$) measured \textcolor{red}{in this work (column ``TW'') for the $\nu_3$ band of $^{12}$CH$_4$. The estimated uncertainty is provided between parentheses after each value, in the units of the last digit quoted. Zero indicates that the measured value was smaller than its estimated uncertainty. Off-diagonal coefficients not listed were fixed to zero. Column ``Lit.'' lists the values calculated for CH$_4$-air in \ccite{tra06a}, multiplied by 1.3 (as done in \ccite{tra22a}). The identifications of the lines are from the \texttt{HITRAN} database \ccite{gor20a}.}}
\label{table:wnn}\\
\toprule
Line $m$ & Line $n$ & \multicolumn{1}{c}{TW} & \multicolumn{1}{c}{Lit.} & & Line $m$ & Line $n$ & \multicolumn{1}{c}{TW} & \multicolumn{1}{c}{Lit.} \\
\cmidrule(r){1-4}\cmidrule(l){6-9}
\endfirsthead
\caption[]{\small (Continued)}\\
\toprule
Line $m$ & Line $n$ & \multicolumn{1}{c}{TW} & \multicolumn{1}{c}{Lit.} & & Line $m$ & Line $n$ & \multicolumn{1}{c}{TW} & \multicolumn{1}{c}{Lit.} \\
\cmidrule(r){1-4}\cmidrule(l){6-9}
\endhead
\multicolumn{9}{r}{{\small Continued on next page}}
\endfoot
\endlastfoot
P(12, $A_1$, 2, 15) & P(12, $A_2$, 1, 14) & 2.527\,(93)  & 3.151 & & R(3, $F_2$, 1, 15)  & R(3, $F_1$, 1, 16)  & 0.556\,(21)  & 0.449 \\  %  P12A1015002  P12A2014001  %  Q10F2037002  Q10F1039001
P(12, $F_1$, 3, 43) & P(12, $F_2$, 3, 45) & 2.358\,(87)  & 2.680 & & R(4, $F_2$, 1, 20)  & R(4, $F_1$, 1, 18)  & 0.548\,(20)  & 0.380 \\  %  P12F1043003  P12F2045003  %  Q10F1040002  Q10F2038003
P(11, $F_1$, 3, 41) & P(11, $F_2$, 3, 39) & 2.350\,(100) & 3.259 & & R(5, $F_1$, 2, 23)  & R(5, $F_2$, 1, 21)  & 1.042\,(38)  & 1.529 \\  %  P11F1041003  P11F2039003  %  Q10A2014001  Q10A1013001
P(11, $F_1$, 3, 41) & P(11, $F_1$, 2, 42) & 0.0          & 0.002 & & R(5, $F_1$, 2, 23)  & R(5, $F_1$, 1, 24)  & 0.0          & 0.020 \\  %  P11F1041003  P11F1042002  %  Q09F1033001  Q09F2035001
P(11, $F_1$, 3, 41) & P(11, $F_2$, 2, 40) & 0.0          & 0.329 & & R(5, $F_2$, 1, 21)  & R(5, $F_1$, 1, 24)  & 0.028\,(4)   & 0.082 \\  %  P11F1041003  P11F2040002  %  Q09F1035003  Q09F2036002
P(11, $F_1$, 3, 41) & P(11, $F_2$, 1, 41) & 0.0          & 0.008 & & R(6, $F_1$, 1, 25)  & R(6, $F_2$, 2, 26)  & 1.070\,(39)  & 1.254 \\  %  P11F1041003  P11F2041001  %  Q08F2030001  Q08F1032002
P(11, $F_1$, 3, 41) & P(11, $F_1$, 1, 43) & 0.0          & 0.047 & & R(6, $F_1$, 1, 25)  & R(6, $F_2$, 1, 27)  & 0.312\,(15)  & 0.474 \\  %  P11F1041003  P11F1043001  %  Q08F1032002  Q08F2031002
P(11, $F_2$, 3, 39) & P(11, $F_1$, 2, 42) & 0.0          & 0.064 & & R(6, $F_2$, 2, 26)  & R(6, $F_2$, 1, 27)  & 0.0          & 0.070 \\  %  P11F2039003  P11F1042002  %  Q08E_021001  Q08E_022002
P(11, $F_2$, 3, 39) & P(11, $F_2$, 2, 40) & 0.0          & 0.039 & & R(6, $A_1$, 1, 10)  & R(6, $A_2$, 1, 8)   & 1.717\,(63)  & 2.010 \\  %  P11F2039003  P11F2040002  %  Q09A1013001  Q09A2011001
P(11, $F_2$, 3, 39) & P(11, $F_2$, 1, 41) & 0.0          & 0.006 & & R(7, $F_1$, 2, 29)  & R(7, $F_2$, 2, 28)  & 0.971\,(37)  & 1.677 \\  %  P11F2039003  P11F2041001  %  Q07F1026001  Q07F2028001
P(11, $F_2$, 3, 39) & P(11, $F_1$, 1, 43) & 0.0          & 0.280 & & R(7, $F_1$, 2, 29)  & R(7, $F_2$, 1, 29)  & 1.332\,(61)  & 0.905 \\  %  P11F2039003  P11F1043001  %  Q07F2029002  Q07F1027002
P(11, $F_1$, 2, 42) & P(11, $F_2$, 2, 40) & 0.0          & 0.341 & & R(7, $F_2$, 2, 28)  & R(7, $F_1$, 1, 30)  & 0.074\,(34)  & 0.162 \\  %  P11F1042002  P11F2040002  %  Q06F2023002  Q06F1025001
P(11, $F_1$, 2, 42) & P(11, $F_2$, 1, 41) & 0.0          & 0.472 & & R(7, $F_2$, 1, 29)  & R(7, $F_1$, 1, 30)  & 0.130\,(8)   & 0.048 \\  %  P11F1042002  P11F2041001  %  Q05F1019001  Q05F2021001
P(11, $F_1$, 2, 42) & P(11, $F_1$, 1, 43) & 0.0          & 0.002 & & R(8, $F_2$, 2, 33)  & R(8, $F_1$, 2, 31)  & 1.805\,(67)  & 2.357 \\  %  P11F1042002  P11F1043001  %  Q05F2021001  Q05F1020002
P(11, $F_2$, 2, 40) & P(11, $F_2$, 1, 41) & 0.0          & 0.007 & & R(8, $F_2$, 2, 33)  & R(8, $F_2$, 1, 34)  & 0.0          & 0.005 \\  %  P11F2040002  P11F2041001  %  Q04F1017001  Q04F2016001
P(11, $F_2$, 2, 40) & P(11, $F_1$, 1, 43) & 0.0          & 0.028 & & R(8, $F_2$, 2, 33)  & R(8, $F_1$, 1, 32)  & 0.0          & 0.335 \\  %  P11F2040002  P11F1043001  %  Q03F1012001  Q04F2016001
P(11, $F_2$, 1, 41) & P(11, $F_1$, 1, 43) & 0.0          & 0.002 & & R(8, $F_1$, 2, 31)  & R(8, $F_2$, 1, 34)  & 0.146\,(11)  & 0.083 \\  %  P11F2041001  P11F1043001  %  Q03F1012001  Q03F2014001
P(10, $F_2$, 3, 37) & P(10, $F_1$, 2, 36) & 1.946\,(72)  & 2.721 & & R(8, $F_1$, 2, 31)  & R(8, $F_1$, 1, 32)  & 0.0          & 0.047 \\  %  P10F2037003  P10F1036002  %  Q02F2008001  Q03F2014001
P(10, $F_1$, 2, 36) & P(10, $F_1$, 1, 37) & 0.0          & 0.075 & & R(8, $F_2$, 1, 34)  & R(8, $F_1$, 1, 32)  & 0.657\,(46)  & 0.136 \\  %  P10F1036002  P10F1037001  %  Q01F1005001  Q02F2008001
P(10, $F_1$, 2, 36) & P(10, $F_2$, 2, 38) & 0.376\,(15)  & 0.198 & & R(8, $E$, 2, 21)    & R(8, $E$, 1, 22)    & 0.945\,(68)  & 1.100 \\  %  P10F1036002  P10F2038002  %  R03F2015001  R03F1016001
P(10, $F_1$, 1, 37) & P(10, $F_2$, 2, 38) & 0.266\,(10)  & 0.192 & & R(9, $A_2$, 1, 13)  & R(9, $A_1$, 1, 12)  & 2.201\,(81)  & 2.773 \\  %  P10F1037001  P10F2038002  %  R04F2020001  R04F1018001
P(10, $F_1$, 1, 37) & P(10, $F_2$, 1, 39) & 0.199\,(11)  & 0.097 & & R(9, $F_2$, 2, 34)  & R(9, $F_1$, 3, 36)  & 1.857\,(70)  & 2.035 \\  %  P10F1037001  P10F2039001  %  R05F1023002  R05F2021001
P(10, $A_1$, 1, 14) & P(10, $A_2$, 1, 12) & 1.159\,(43)  & 0.756 & & R(9, $F_2$, 2, 34)  & R(9, $F_1$, 2, 37)  & 0.862\,(81)  & 0.454 \\  %  P10A1014001  P10A2012001  %  R05F1023002  R05F1024001
P(9, $A_2$, 1, 12)  & P(9, $A_1$, 1, 10)  & 1.926\,(71)  & 2.918 & & R(9, $F_1$, 3, 36)  & R(9, $F_1$, 2, 37)  & 0.080\,(65)  & 0.037 \\  %  P09A2012001  P09A1010001  %  R05F2021001  R05F1024001
P(9, $F_2$, 2, 32)  & P(9, $F_1$, 3, 33)  & 1.516\,(56)  & 2.142 & & R(9, $F_1$, 2, 37)  & R(9, $F_2$, 1, 35)  & 1.108\,(41)  & 0.157 \\  %  P09F2032002  P09F1033003  %  R06F1025001  R06F2026002
P(9, $F_2$, 2, 32)  & P(9, $F_1$, 2, 34)  & 0.0          & 0.477 & & R(9, $F_2$, 1, 35)  & R(9, $F_1$, 1, 38)  & 0.884\,(43)  & 0.007 \\  %  P09F2032002  P09F1034002  %  R06F1025001  R06F2027001
P(9, $F_1$, 2, 34)  & P(9, $F_2$, 1, 33)  & 0.541\,(20)  & 0.166 & & R(10, $A_1$, 1, 14) & R(10, $A_2$, 1, 12) & 0.992\,(63)  & 0.655 \\  %  P09F1034002  P09F2033001  %  R06F2026002  R06F2027001
P(9, $F_2$, 1, 33)  & P(9, $F_1$, 1, 35)  & 0.026\,(5)   & 0.007 & & R(10, $F_2$, 3, 39) & R(10, $F_1$, 2, 38) & 1.196\,(46)  & 2.358 \\  %  P09F2033001  P09F1035001  %  R06A1010001  R06A2008001
P(8, $F_2$, 2, 30)  & P(8, $F_1$, 2, 28)  & 1.588\,(59)  & 2.111 & & R(10, $F_2$, 3, 39) & R(10, $F_1$, 1, 39) & 2.756\,(100) & 0.721 \\  %  P08F2030002  P08F1028002  %  R07F1029002  R07F2028002
P(8, $F_2$, 2, 30)  & P(8, $F_2$, 1, 31)  & 0.0          & 0.004 & & R(10, $F_1$, 2, 38) & R(10, $F_2$, 2, 40) & 0.307\,(25)  & 0.171 \\  %  P08F2030002  P08F2031001  %  R07F1029002  R07F2029001
P(8, $F_1$, 2, 28)  & P(8, $F_2$, 1, 31)  & 0.070\,(12)  & 0.075 & & R(10, $F_1$, 1, 39) & R(10, $F_2$, 2, 40) & 2.890\,(110) & 0.166 \\  %  P08F1028002  P08F2031001  %  R07F2028002  R07F1030001
P(8, $F_2$, 1, 31)  & P(8, $F_1$, 1, 29)  & 0.511\,(21)  & 0.122 & & R(10, $F_1$, 1, 39) & R(10, $F_2$, 1, 41) & 2.986\,(120) & 0.084 \\  %  P08F2031001  P08F1029001  %  R07F2029001  R07F1030001
P(8, $E$, 2, 19)    & P(8, $E$, 1, 20)    & 0.0          & 0.986 & & R(11, $F_1$, 3, 43) & R(11, $F_2$, 3, 41) & 2.260\,(85)  & 2.396 \\  %  P08E_019002  P08E_020001  %  R08F2033002  R08F1031002
P(7, $F_1$, 2, 26)  & P(7, $F_2$, 2, 24)  & 1.151\,(45)  & 1.391 & & R(11, $F_1$, 3, 43) & R(11, $F_2$, 2, 42) & 3.443\,(150) & 0.242 \\  %  P07F1026002  P07F2024002  %  R08F2033002  R08F2034001
P(7, $F_1$, 2, 26)  & P(7, $F_2$, 1, 25)  & 0.0          & 0.750 & & R(11, $F_2$, 3, 41) & R(11, $F_1$, 2, 42) & 1.033\,(42)  & 0.047 \\  %  P07F1026002  P07F2025001  %  R08F2033002  R08F1032001
P(7, $F_1$, 2, 26)  & P(7, $F_1$, 1, 27)  & 0.0          & 0.008 & & R(11, $F_1$, 2, 42) & R(11, $F_2$, 2, 42) & 0.0          & 0.251 \\  %  P07F1026002  P07F1027001  %  R08F1031002  R08F2034001
P(7, $F_2$, 2, 24)  & P(7, $F_2$, 1, 25)  & 0.200\,(140) & 0.086 & & R(11, $F_2$, 2, 42) & R(11, $F_1$, 1, 44) & 1.936\,(72)  & 0.021 \\  %  P07F2024002  P07F2025001  %  R08F1031002  R08F1032001
P(7, $F_2$, 2, 24)  & P(7, $F_1$, 1, 27)  & 0.440\,(160) & 0.135 & & R(11, $F_2$, 1, 43) & R(11, $F_1$, 1, 44) & 4.236\,(160) & 0.001 \\  %  P07F2024002  P07F1027001  %  R08F2034001  R08F1032001
P(7, $F_2$, 1, 25)  & P(7, $F_1$, 1, 27)  & 0.050\,(16)  & 0.040 & & R(12, $A_1$, 2, 16) & R(12, $A_2$, 1, 15) & 3.807\,(140) & 2.532 \\  %  P07F2025001  P07F1027001  %  R08E_021002  R08E_022001
P(6, $A_1$, 1, 8)   & P(6, $A_2$, 1, 6)   & 1.412\,(52)  & 1.878 & & R(12, $A_2$, 1, 15) & R(12, $A_1$, 1, 17) & 1.979\,(75)  & 0.175 \\  %  P06A1008001  P06A2006001  %  R09A2013001  R09A1012001
P(6, $F_1$, 1, 21)  & P(6, $F_2$, 2, 22)  & 0.768\,(28)  & 1.172 & & R(12, $F_1$, 3, 45) & R(12, $F_2$, 3, 48) & 0.0          & 2.154 \\  %  P06F1021001  P06F2022002  %  R09F2034002  R09F1036003
P(6, $F_2$, 2, 22)  & P(6, $F_2$, 1, 23)  & 0.057\,(9)   & 0.066 & & R(12, $F_1$, 3, 45) & R(12, $F_2$, 2, 46) & 3.320\,(150) & 0.237 \\  %  P06F2022002  P06F2023001  %  R09F2034002  R09F1037002
P(5, $F_1$, 2, 18)  & P(5, $F_2$, 1, 17)  & 0.820\,(30)  & 1.150 & & R(12, $F_2$, 3, 48) & R(12, $F_2$, 2, 46) & 1.881\,(99)  & 0.016 \\  %  P05F1018002  P05F2017001  %  R09F1036003  R09F1037002
P(5, $F_2$, 1, 17)  & P(5, $F_1$, 1, 19)  & 0.013\,(2)   & 0.062 & & & & & \\  %  P05F2017001  P05F1019001  %  R09F1037002  R09F2035001
P(4, $F_2$, 1, 15)  & P(4, $F_1$, 1, 13)  & 0.364\,(13)  & 0.566 & & & & & \\  %  P04F2015001  P04F1013001  %  R09F2035001  R09F1038001
P(3, $F_2$, 1, 9)   & P(3, $F_1$, 1, 11)  & 0.264\,(10)  & 0.567 & & & & & \\  %  P03F2009001  P03F1011001  %  R10A1014001  R10A2012001
Q(12, $A_2$, 1, 17) & Q(12, $A_1$, 2, 15) & 3.217\,(120) & 1.540 & & & & & \\  %  Q12A2017001  Q12A1015002  %  R10F2039003  R10F1038002
Q(12, $F_2$, 3, 46) & Q(12, $F_1$, 3, 47) & 1.706\,(64)  &       & & & & & \\  %  Q12F2046003  Q12F1047003  %  R10F2039003  R10F1039001
Q(11, $F_1$, 1, 40) & Q(11, $F_2$, 1, 42) & 1.776\,(74)  & 0.001 & & & & & \\  %  Q11F1040001  Q11F2042001  %  R10F1038002  R10F2040002
Q(11, $F_1$, 2, 41) & Q(11, $F_2$, 3, 44) & 0.287\,(100) & 0.026 & & & & & \\  %  Q11F1041002  Q11F2044003  %  R10F1039001  R10F2040002
Q(11, $F_1$, 2, 41) & Q(11, $F_1$, 3, 42) & 0.667\,(210) & 0.001 & & & & & \\  %  Q11F1041002  Q11F1042003  %  R10F1039001  R10F2041001
Q(11, $F_2$, 3, 44) & Q(11, $F_1$, 3, 42) & 2.060\,(130) & 1.324 & & & & & \\  %  Q11F2044003  Q11F1042003  %  R11F1043003  R11F2041003
Q(10, $F_2$, 2, 37) & Q(10, $F_1$, 1, 39) & 0.263\,(12)  & 0.075 & & & & & \\  %  R11F1043003  R11F2042002  %  R12A1016002  R12A2015001
Q(10, $F_1$, 2, 40) & Q(10, $F_2$, 3, 38) & 1.862\,(69)  & 1.058 & & & & & \\  %  R11F2041003  R11F1042002  %  R12A2015001  R12A1017001
Q(10, $A_2$, 1, 14) & Q(10, $A_1$, 1, 13) & 0.0          & 0.294 & & & & & \\  %  R11F1042002  R11F2042002  %  R12F1045003  R12F2048003
Q(9, $F_1$, 1, 33)  & Q(9, $F_2$, 1, 35)  & 0.244\,(14)  & 0.003 & & & & & \\  %  R11F2042002  R11F1044001  %  R12F1045003  R12F2046002
Q(9, $F_1$, 3, 35)  & Q(9, $F_2$, 2, 36)  & 1.267\,(48)  & 0.927 & & & & & \\  %  R11F2043001  R11F1044001  %  R12F2048003  R12F2046002
Q(8, $F_2$, 1, 30)  & Q(8, $F_1$, 2, 32)  & 0.019\,(6)   & 0.035 & & & & & \\
Q(8, $F_1$, 2, 32)  & Q(8, $F_2$, 2, 31)  & 0.435\,(19)  & 0.998 & & & & & \\
Q(8, $E$, 1, 21)    & Q(8, $E$, 2, 22)    & 3.169\,(130) & 0.466 & & & & & \\
Q(9, $A_1$, 1, 13)  & Q(9, $A_2$, 1, 11)  & 1.262\,(47)  & 1.263 & & & & & \\
Q(7, $F_1$, 1, 26)  & Q(7, $F_2$, 1, 28)  & 0.061\,(11)  & 0.019 & & & & & \\
Q(7, $F_2$, 2, 29)  & Q(7, $F_1$, 2, 27)  & 1.091\,(40)  & 0.661 & & & & & \\
Q(6, $F_2$, 2, 23)  & Q(6, $F_1$, 1, 25)  & 0.940\,(35)  & 0.586 & & & & & \\
Q(5, $F_1$, 1, 19)  & Q(5, $F_2$, 1, 21)  & 0.266\,(11)  & 0.037 & & & & & \\
Q(5, $F_2$, 1, 21)  & Q(5, $F_1$, 2, 20)  & 1.360\,(51)  & 0.696 & & & & & \\
Q(4, $F_1$, 1, 17)  & Q(4, $F_2$, 1, 16)  & 1.248\,(46)  & 0.473 & & & & & \\
Q(3, $F_1$, 1, 12)  & Q(4, $F_2$, 1, 16)  & 0.322\,(13)  & 0.052 & & & & & \\
Q(3, $F_1$, 1, 12)  & Q(3, $F_2$, 1, 14)  & 0.584\,(22)  & 0.466 & & & & & \\
Q(2, $F_2$, 1, 8)   & Q(3, $F_2$, 1, 14)  & 0.425\,(16)  & 0.176 & & & & & \\
Q(1, $F_1$, 1, 5)   & Q(2, $F_2$, 1, 8)   & 0.029\,(5)   & 0.373 & & & & & \\
\bottomrule
\end{longtable}
\normalsize

% 05/04/2024: The 2 sentences "Table \ref{table:wnn} shows that the values of the off-diagonal relaxation matrix coefficients in the Q and R branches are on average larger than in the P branch. This is consistent with the results of the first order line mixing model." were removed as I do not agree. The figure 'Wmn vs mJ' in 'article_revised/response_to_review/line_mixing/offdiag_W/line_coupling.pxp' shows that they are similar.

\textcolor{red}{With the signal to noise ratio characterizing the spectra analyzed in the present work, it was noticed that the Rosenkranz approximation failed when the manifold included an off-diagonal coefficient larger than about $0.015$ cm$^{-1}$atm$^{-1}$. As in previous work \ccite{pin00a, far21a}, line-mixing between $A_1 \leftrightarrow A_1$, $A_2 \leftrightarrow A_2$, $F_1 \leftrightarrow F_1$ and $F_2 \leftrightarrow F_2$ was observed to be smaller than for the other allowed transitions. However, the off-diagonal coefficient is not negligible when the coupled lines are very close. The two coupled R(12) lines of $F_2$ symmetry are an example of such a case.}

\textcolor{red}{Table \ref{table:wnn} also lists the values calculated for CH$_4$-air in \ccite{tra06a}, multiplied by 1.3 as done in \ccite{tra22a}. The qualitative agreement between the measured and calculated values is quite good in the P and R (up to $J'' = 9$) branches, although a few significant discrepancies are observed. The agreement is in contrast poorer in the Q branch and manifolds of the R branch with $J'' \geq 10$. In Table \ref{table:wnn_ow}, the present measurements are compared with 1.3 times the off-diagonal relaxation matrix coefficients measured in the Q branch of the $\nu_3$ band of CH$_4$ perturbed by air \ccite{dev18a} or N$_2$ \ccite{pin19a}. The corresponding calculated values \ccite{tra06a} are also listed.
\begin{table}[!htbp]
\centering
\begin{threeparttable}
\small
\caption{Off-diagonal relaxation matrix coefficients $W_{mn}$ (in $10^{-2}$ cm$^{-1}$atm$^{-1}$) for the $\nu_3$ band of $^{12}$CH$_4$ perturbed by CO$_2$ measured in the present work (TW) and \textcolor{red}{1.3 times the values measured \ccite{dev18a} or calculated \ccite{tra06a} for CH$_4$-air and measured for CH$_4$-N$_2$ \ccite{pin19a} ([0, $2d$, 2] model).}} \vspace{12pt}
\label{table:wnn_ow}
\begin{tabular}{llD{.}{.}{1.7}D{.}{.}{1.7}D{.}{.}{1.7}D{.}{.}{1.3}}
\toprule
Line $m$ & Line $n$ & \multicolumn{1}{c}{TW} & \multicolumn{1}{c}{\ccite{dev18a}} & \multicolumn{1}{c}{\ccite{pin19a}} & \multicolumn{1}{c}{\ccite{tra06a}} \\
\midrule
Q(12, $F_2$, 3, 46)          & Q(12, $F_1$, 3, 47)          & 1.706\,(64)  & 2.311\,(25) & 2.499\,(25) &       \\ % 1.778\,(19)
Q(12, $A_2$, 1, 17)          & Q(12, $A_1$, 2, 15)          & 3.217\,(120) & 3.556\,(59) & 3.331\,(51) & 1.540 \\ % 2.735\,(45)
Q(11, $F_2$, 3, 44)          & Q(11, $F_1$, 3, 42)          & 2.060\,(130) & 3.700\,(47) & 3.238\,(36) & 1.324 \\ % 2.846\,(36)
Q(10, $F_2$, 2, 37)$^{\dag}$ & Q(10, $F_1$, 1, 39)          & 0.263\,(12)  & 0.332\,(15) & 0.213\,(13) & 0.075 \\ % 0.255\,(11)
Q(10, $F_1$, 2, 40)          & Q(10, $F_2$, 3, 38)$^{\dag}$ & 1.862\,(69)  & 2.332\,(42) & 2.112\,(26) & 1.058 \\ % 1.794\,(32)
Q(9, $F_1$, 3, 35)           & Q(9, $F_2$, 2, 36)           & 1.267\,(48)  & 1.881\,(22) & 1.913\,(14) & 0.927 \\ % 1.447\,(17)
Q(7, $F_2$, 2, 29)           & Q(7, $F_1$, 2, 27)           & 1.091\,(40)  & 0.855\,(20) & 0.856\,(13) & 0.661 \\ % 0.658\,(15)
Q(6, $F_2$, 2, 23)           & Q(6, $F_1$, 1, 25)           & 0.940\,(35)  & 1.533\,(14) & 1.515\,(12) & 0.586 \\ % 1.179\,(11)
Q(5, $F_1$, 1, 19)           & Q(5, $F_2$, 1, 21)           & 0.266\,(11)  & 0.495\,(13) & 0.415\,(11) & 0.037 \\ % 0.381\,(10)
Q(3, $F_1$, 1, 12)           & Q(3, $F_2$, 1, 14)           & 0.584\,(22)  & 0.402\,(9)  & 0.331\,(8)  & 0.466 \\ % 0.309\,(7) 
Q(2, $F_2$, 1, 8)            & Q(1, $F_1$, 1, 5)            & 0.029\,(4)   & 0.564\,(13) & 0.354\,(11) & 0.373 \\ % 0.434\,(10)
\bottomrule
\end{tabular}
\begin{tablenotes}
    \item[$^{\dag}$] In \ccite{dev18a}, $\alpha' = 39$ (line $m$) and $\alpha'' = 1$ (line $n$). The assignments provided here are from the \texttt{HITRAN} database \ccite{gor20a}.
\end{tablenotes}
\end{threeparttable}
\end{table}
It is interesting to note that the agreement between the measurements is generally better than with the calculated values. Deviations between measured and calculated values are more apparent as $J$ increases, the present measurements being systematically higher than Tran \textit{et al.} \ccite{tra06a} (with one exception) and lower than the measurements \ccite{dev18a, pin19a} (with two exceptions).} These coefficients often result from different couplings so there is less reliability in these comparisons. This is a recurring problem \ccite{pin19a}, which highlights the issue of whether these parameters have a physical meaning or are just effective.

\textcolor{red}{Figure \ref{fig:tw_vs_obs} presents a room temperature spectrum of the $\nu_3$ band of CH$_4$ mixed with CO$_2$ at a total pressure of about 12.5 bars, recorded previously \ccite{tra22a}.
\begin{figure}[!htbp]
\centering
\includegraphics[width = \linewidth]{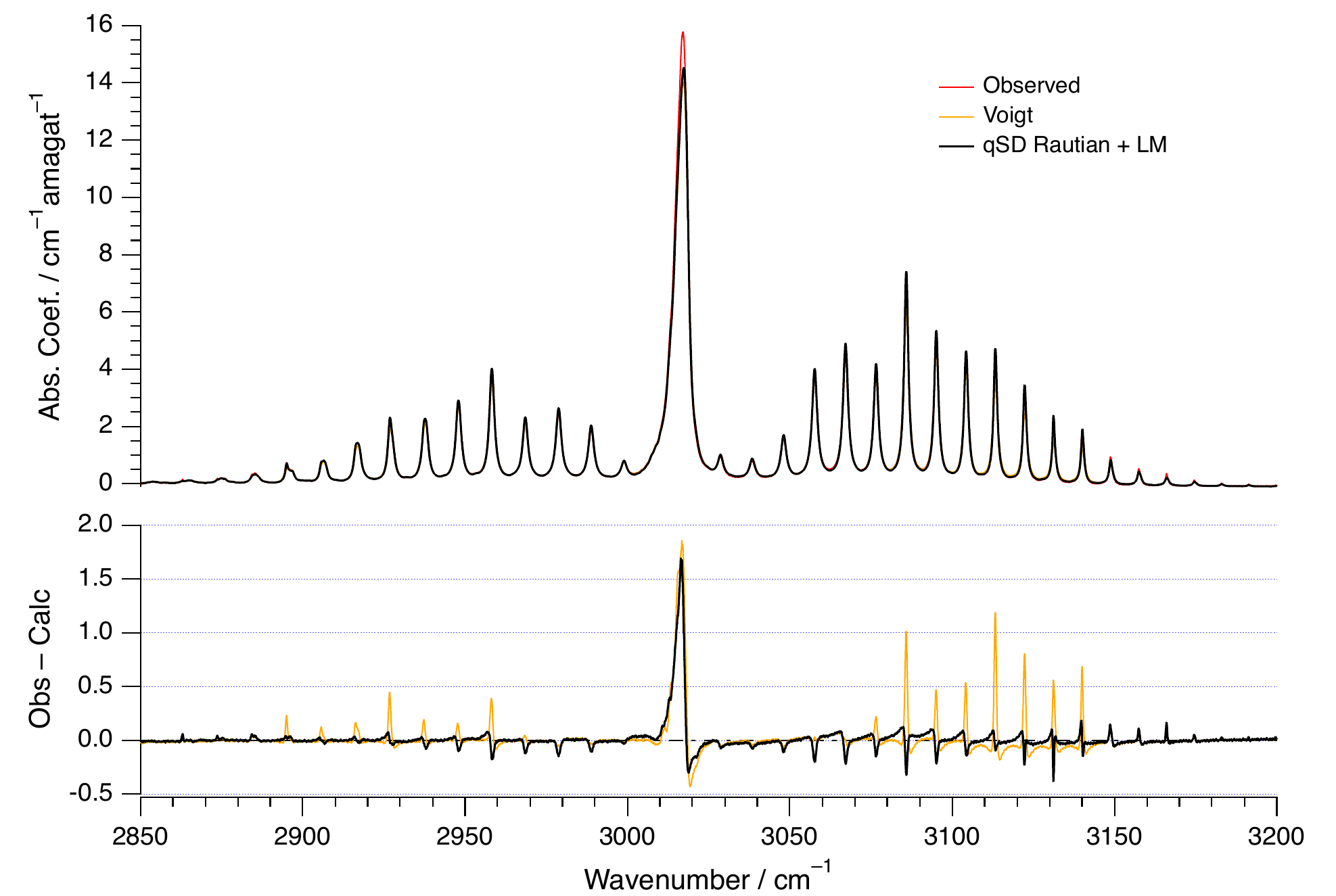}
\caption{Comparison between a spectrum of the $\nu_3$ band of CH$_4$ mixed with CO$_2$ at a total pressure of about 12.5 bars, recorded at room temperature (spectrum no. 2 in Table 1 of \ccite{tra22a}, red trace in the upper panel), and spectra calculated with the Voigt profile and neglecting line mixing effects (orange trace) and using the relaxation matrix formalism with all the parameters obtained in the present work (black trace). The corresponding residuals are presented in the lower panel. ``qSD'' and ``LM'' stand for ``quadratic speed dependent'' and ``line mixing,'' respectively.}
\label{fig:tw_vs_obs}
\end{figure}
Spectra calculated using the relaxation matrix formalism and the quadratic speed dependent Rautian profile with all the parameters obtained in the present work as well as using the Voigt profile and neglecting line mixing effects (called ``Voigt calculation'' here after) are also presented. Note that these calculations and the two mentioned here below actually involved fitting the 5 coefficients of the 4$^{\mathrm{th}}$ order polynomial expansion used to model the baseline of the spectra and the mole fraction of methane. As noted in \ccite{tra22a}, the fitted methane mole fraction was about 13 \% lower than the value at cell filling. If the residuals obtained considering line mixing are similar to those presented in Fig. 3 of \ccite{tra22a} for the P and R branches, they are close to those obtained with the Voigt calculation for the Q branch. Figure \ref{fig:tw_vs_tran} focuses on this Q branch. In addition to the spectra presented in Fig. \ref{fig:tw_vs_obs}, it includes two spectra calculated using the relaxation matrix formalism with the line parameters of the present work and 1.3 times some of the off-diagonal relaxation matrix elements reported for CH$_4$-air \ccite{tra06a} (all possible couplings in the $\nu_3$ band are actually considered in this work), as done in \ccite{tra22a}.
\begin{figure}[!htbp]
\centering
\includegraphics[width = \linewidth]{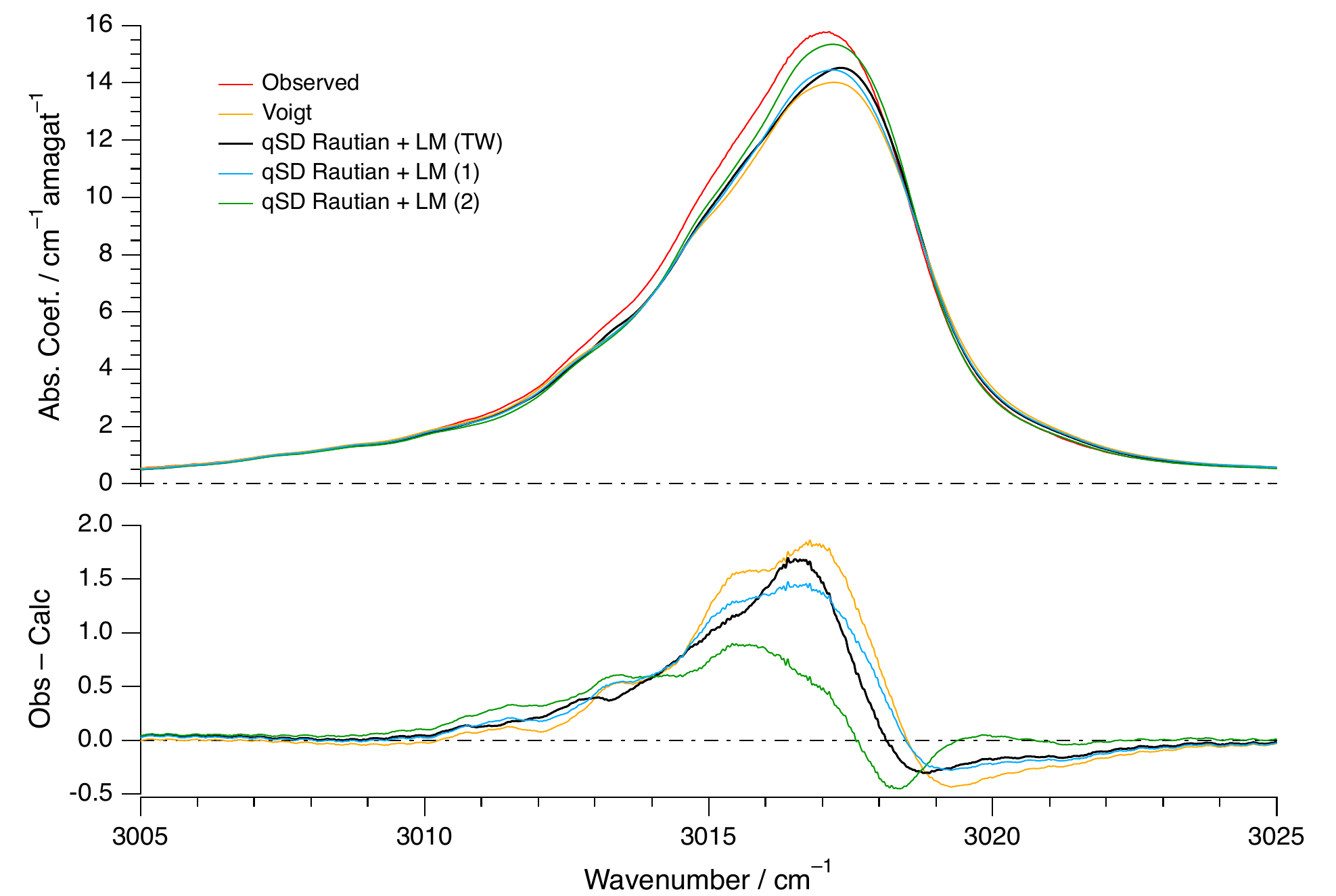}
\caption{Same as Fig. \ref{fig:tw_vs_obs}, focusing on the Q branch of the $\nu_3$ band of CH$_4$ (``qSD'' and ``LM'' stand for ``quadratic speed dependent'' and ``line mixing,'' respectively). The additional two spectra [identified by (1) and (2)] and their corresponding residuals were calculated using the relaxation matrix formalism with  the line parameters obtained in the present work and the values of the off-diagonal relaxation matrix elements reported in \ccite{tra06a}, multiplied by 1.3 as done in \ccite{tra22a}: the blue traces were generated neglecting all inter-$J$ and inter-branch couplings, while the green traces were calculated considering $\Delta J = \pm 1$ couplings. Both calculations included $\Delta J = 0$ couplings.}
\label{fig:tw_vs_tran}
\end{figure}
Both calculations include the $\Delta J = 0$ couplings. The blue traces correspond to the spectrum generated neglecting all inter-$J$ and inter-branch couplings, while the green traces were obtained also considering $\Delta J = \pm 1$ couplings. These two calculations show the importance of the $\Delta J = \pm 1$ couplings in the Q branch. Not considering them beyond $J''=4$ (see section \ref{sc:modeling_spectra}) therefore reduced the applicability to higher pressures of the parameters retrieved for the Q branch in the present work.} % Note that the off-diagonal relaxation matrix elements calculated for CH$_4$-air \ccite{tra06a} consider all possible couplings in the $\nu_3$ band. Only the $\Delta J = 0$ and $\pm 1$ were considered here to demonstrate the impact of the latter.

% As already noted in \ccite{dev18a}, the relaxation matrix modeled in \ccite{tra06a} includes inter-branch and inter-manifold couplings while the present study neglects such contributions, except for a few low $J$ Q branch lines. This may explain differences observed between the measured and calculated values of the $W_{mn}$ coefficients.

% Table \ref{table:wnn_ow} compares the off-diagonal relaxation matrix coefficients $W_{mn}$ measured in the present work with values obtained by multiplying by 1.3 \ccite{tra22a} the off-diagonal matrix coefficients calculated for CH$_4$-air collisions \ccite{tra06a}. The deviation between measured and calculated values is more apparent as $J$ increases, the present measurements being systematically higher (with one exception). Incidentally, applying this $1.3$ (or $1.2$ \ccite{ess21a}) factor to the measurements reported by Devi \textit{et al.} \ccite{dev18a} or Pine \ccite{pin19a} leads to the opposite observation. These coefficients often result from different couplings so there is less reliability in these comparisons. This is a recurring problem \ccite{pin19a}, which highlights the issue of whether these parameters have a physical meaning or are just effective.

\textcolor{red}{First order line mixing coefficients $Y_n^0$, calculated using Eq. \ref{eq:y_from_w}, are also listed in Table \ref{table:y0}. These calculated values agree within 20 \% with 44 of the 62 coefficients measured for P(12) to R(7). Significantly poorer agreement is observed for the more congested R(8) to R(12) manifolds, matching the emergence of signatures in the residuals of Figs. \ref{fig:fit_P} and \ref{fig:fit_R}, however one $J''$ higher. Inspection of the last column of Table \ref{table:y0} indicates that poorer agreement between observed and calculated values can also be attributed to missing measured off-diagonal relaxation matrix elements.}

%------------------------------------------------------------------------------
      
\section{Conclusion}\label{sc:conclusion}

CO$_2$ broadening and shift coefficients together with speed dependence of the broadening have been measured for lines belonging to $J''$ manifolds of the $P$ ($J'' = 1-13$), $Q$ ($J'' = 1-13$) and $R$ ($J'' = 0-13$) branches of the $\nu_3$ band of $^{12}$CH$_4$ observed near 3.3 μm. The CO$_2$ shift coefficients are reproted for the first time. These measurements were carried out using non-linear multispectrum least squares fitting techniques, applied to eleven high resolution Fourier transform spectra recorded at $296.5 \, (5)$ K. These consisted in one spectrum of pure methane at low pressure and 10 spectra of mixtures of methane and carbon dioxide at total pressures between 26 and 803 hPa. Methane lines were modeled using a hard collision speed dependent Rautian line shape model. This line shape model was chosen because it was found necessary to fit the observed shape of the R(0, $A_1$, 1, 3) and R(1, $F_1$, 1, 10) lines to the noise level. Although Dicke narrowing was included in the analyses, the measured Dicke narrowing coefficients were rather small and involved a possible correlation with the speed dependence of broadening. These observations were interpreted as an indication that the present measurements did not rely on enough spectra corresponding to total pressures below about 200 hPa. The measured narrowing coefficients were therefore not reported. As already reported in the literature for methane and other molecules \ccite{pin19a, dev15a, dev12a, devi07a, bui14a, oku23a}, the speed dependence of broadening coefficients is on average equal to about $\aw = 0.110$. Apart from the CO$_2$ broadening coefficients that exhibit a dependence on the tetrahedral symmetries already observed for other perturber gases, such a dependence was not observed for the other parameters measured in the present work.

Line mixing was considered in the analyses using the first order in pressure approximation or the relaxation matrix formalism. The work carried out demonstrated that first order line mixing can be sufficient to model absorption spectra of manifolds involving a low density of lines at total pressures up to 800 hPa. As $J$ increases, the density of lines in the manifolds increases and signatures indicating the failure of the first order approximation appear in the residuals. The use of the relaxation matrix formalism allowed bringing the residuals to the noise level. It however failed in the low $J$ part of the Q branch near 3020 cm$^{-1}$. More line couplings or a hybrid relaxation matrix / first order model accounting would probably be needed to be solve this problem. % Improvements of the measured parameters have been made in the R branch while the P branch showed mostly constant coefficients throughout the models

% Correlation arises on the measurement of Dicke narrowing and speed-dependence. A constant ratio of speed-dependent over non-speed-dependent width of $0.11$ seemed to be adequate and suggests that our $\beta $ in the other models are too low. Wether this constant ratio is sufficient to measure the real Dicke narrowing will require more intermediate pressures.

Prior to the present work, limited information was available in the literature on parameters characterizing the effects of the pressure of CO$_2$ on the absorption spectrum of the $\nu_3$ band of $^{12}$CH$_4$. Differences between the values of the parameters reported in the literature and measured in this work do not generally show any systematic discrepancies. % \textcolor{red}{Comment on the accuracies of the reported parameters...}

%------------------------------------------------------------------------------

\section*{Supplementary material}\label{sc:supmat}

The supplementary material provided with this article consists in two files listing the CO$_2$-induced collisional line parameters measured at $296.5\,(5)$ K for the $\nu_3$ band of $^{12}$CH$_4$ in fits F1 and F4 of this work.

%------------------------------------------------------------------------------

\section*{Acknowledgments}

T. Bertin thanks the \textit{Fonds pour la formation \`a la Recherche dans l'Industrie et dans l'Agriculture} (FRIA, Belgium) for a PhD fellowship. This work was financially supported by the \textit{Fonds de la Recherche Scientifique}--FNRS (Belgium, contract J.0217.20).

%------------------------------------------------------------------------------

\bibliographystyle{elsarticle-num}
% \bibliography{ch4-co2}

\end{document}